\documentclass[sigconf]{acmart}

\AtBeginDocument{%
  }

\copyrightyear{2026}
\acmYear{2026}
\setcopyright{cc}
\setcctype{by}
\acmConference[CHI '26]{Proceedings of the 2026 CHI Conference on Human Factors in Computing Systems}{April 13--17, 2026}{Barcelona, Spain}
\acmBooktitle{Proceedings of the 2026 CHI Conference on Human Factors in Computing Systems (CHI '26), April 13--17, 2026, Barcelona, Spain}
\acmDOI{10.1145/3772318.3790311}
\acmISBN{979-8-4007-2278-3/2026/04}

\usepackage{listings}
\usepackage{graphicx}
\usepackage{subcaption}
\usepackage{xcolor}
\usepackage{cancel} 
\usepackage{float}

\begin{document}

\title{Counting How the Seconds Count: Understanding Algorithm-User Interplay in TikTok via ML-driven Analysis of Video Content}

\author{Maleeha Masood}
\affiliation{%
\department{Siebel School of Computing and Data
Science}
  \institution{University of Illinois Urbana-Champaign}
  \city{Urbana, IL}
  \country{United States}}
\email{maleeha2@illinois.edu}

\author{Shreya Kannan}
\affiliation{%
\department{Siebel School of Computing and Data
Science}
  \institution{University of Illinois Urbana-Champaign}
  \city{Urbana, IL}
  \country{United States}}
\email{shreya28@illinois.edu}

\author{Zikun Liu}
\affiliation{%
\department{Siebel School of Computing and Data
Science}
  \institution{University of Illinois Urbana-Champaign}
  \city{Urbana, IL}
  \country{United States}}
\email{zikunliu@illinois.edu}

\author{Deepak Vasisht}
\affiliation{%
\department{Siebel School of Computing and Data
Science}
  \institution{University of Illinois Urbana-Champaign}
  \city{Urbana, IL}
  \country{United States}}
\email{deepakv@illinois.edu}

\author{Indranil Gupta}
\affiliation{%
\department{Siebel School of Computing and Data
Science}
  \institution{University of Illinois Urbana-Champaign}
  \city{Urbana, IL}
  \country{United States}}
\email{indy@illinois.edu}

\renewcommand{\shortauthors}{Masood et al.}

%%%% abstract.tex starts here %%%%

\begin{abstract}
Short video streaming systems such as TikTok, YouTube Shorts, Instagram Reels, etc., have reached billions of active users worldwide. At the core of such systems are  (proprietary) recommendation 
algorithms which recommend a sequence  of videos to each user, in a personalized way. We aim to understand the temporal evolution of recommendations made by such algorithms, as well as the interplay between the recommendations and user experience. While past work has studied recommendation algorithms using {\it textual} data (e.g., titles, hashtags, etc.) as well as {\it user studies and interviews}, we add a third modality of analysis---we perform {\it automated analysis of the videos themselves}. To perform such multimodal analysis, we  develop a  new {HCI measurement approach that starts with our new tool called VCA (Video Content Analysis)  that leverages recent advances in Vision Language Models (VLMs). We  
apply VCA on a trifecta of HCI methodologies---real user studies, interviews, and data donation.} 
This allows us to understand
{\it temporal} aspects of how well TikTok's recommendation algorithm is perceived by users, is affected by user interactions, and aligns with user history; how users are sensitive to the order of videos  recommended; and how the algorithm's effectiveness itself may be predictable in the future. While it is not our goal to reverse-engineer TikTok's recommendation algorithm, our new findings indicate behavioral aspects that {the TikTok user community can benefit from.}
\end{abstract}

% \ccsdesc[500]{Human-centered computing~Empirical studies in HCI}
% \ccsdesc[300]{Human-centered computing~Social media}
% \ccsdesc[300]{Information systems~Recommender systems}



% \keywords{{Algorithmic Experience; Social Media Feeds; Video Analysis}}

\maketitle

%%%% introduction.tex starts here %%%%

\section{Introduction}
\label{sec:intro}

{We are seeing an explosion in the popularity of short video streaming applications over the past few years.
TikTok, one of the leading short
video platforms, reports more than 1 billion users worldwide, with an estimated 82 million active users daily in the US in 2025 \cite{1bil}. On average, users on TikTok spend
nearly an hour per day on the app \cite{explodingtopics}. Other major social media platforms have also 
released short video streaming services like Instagram Reels, Facebook Watch and YouTube Shorts. As a result, the overall adoption of short video applications continues to grow worldwide \cite{poptiktok, poptiktok2, poptiktok3}.}

{On TikTok, the “For You Page” is an ongoing algorithmic experience between the short video streaming platform and a user, consisting of a \textit{sequence of videos} appearing to be uniquely personal.
Central to this experience is a recommendation algorithm,  
which generates the user-specific 
 video recommendations. A user watches these videos in the order shown, although they can always skip to the next video by swiping.} These algorithms require little explicit input about user preferences, but rather learn about interests through interactions.  The recommendation algorithms play a key role in the success of such platforms, but these algorithms remain proprietary.

{Existing knowledge about how recommendation algorithms function and interact with users has been largely derived from qualitative accounts of users' personal experiences with short video streaming systems and ultimately, user‑constructed folk theories. 

Prior human-computer interaction research \cite{facebooktheories, likeithideit, selfrepresentation, folktheories, changingplatforms, marginalizedtheories} has found that users develop improvised  folk theories (explanatory frameworks, often partial) to make sense of the recommendation algorithm. These user-generated explanations, in turn, shape everyday practices: users attempt to engage in “training” the algorithm by strategically liking, skipping, sharing, searching and commenting on videos, and the user interprets subsequent changes in their feed as personalized responses to specific actions or attributes. Such practices not only encapsulate users’ beliefs (mistaken or otherwise) about the system, but they also loop back 
into the platform, further entangling user behavior, algorithmic inference, and perceived control.}

{In this work, we {\it precisely} 
examine how the recommendation system behaves as it interacts with a user over time, by shifting focus from the \textit{qualitative} experiences of a user to the \textit{quantitative} behavior of the recommendation system itself. We aim to \textit{understand the evolution of recommended content over time and the interplay between the algorithm’s output (i.e., recommended videos) and the user experience}.
{Rather than proposing or deepening HCI theories, we offer a practical HCI measurement approach that is  immediately usable by HCI researchers in this fast-moving field. Our approach 
enables analysis of video content at scale, which is difficult to achieve with existing methods of working with multimodal data. We also show how our HCI measurement approach enables us to unearth several findings that empirically test or challenge widely held folk theories about recommender systems, including the role of user interactions and sequence continuity.

}

We answer the following research questions (RQs):
}
\begin{itemize}
    \item \textbf{RQ1} How does the content of recommendations made by TikTok’s recommendation algorithm evolve over time at short and long timescales?
    \item {\textbf{RQ2} What is the relationship between user interactions (likes and shares) and recommendations made by the algorithm?}
    \item \textbf{RQ3} Do users engage more with recommendations that align with their historical video consumption patterns? 
    \item {\textbf{RQ4} Beyond the videos recommended, does the continuity of the recommended video sequence matter (in terms of user satisfaction)?} 
    \item \textbf{RQ5} To what extent can we predict whether a user will watch a recommended video using our video analysis pipeline?
\end{itemize}

\noindent{\textbf{\textit{Interpreting the RQs:}}}  
{These RQs cover all three entities involved---provider, user, video sequence. Concretely, we study the content of video recommendations from the provider (RQ1), user interactions and preferences (RQs 2-3), and the {\it sequence} of videos recommended (RQs 4-5). }

Alternately, these RQs can be seen as focusing on both past and present histories (RQs 1-4), and also looking forward to the future (RQ5). {A third way to see these RQs is that together they trace the life cycle of recommendations: from how they evolve over time (RQ1),  how they respond to user behavior (RQ2–RQ4), and whether they affect engagement (RQ5). Other important questions (e.g., fairness or creator outcomes) are orthogonal to our scope. } 

\noindent{\textbf{\textit{Existing Work, and Need for a New Approach:}}}  
Recent work has attempted to analyze the emergent behaviors of TikTok's recommendation algorithms through several 
techniques. First, some studies \textit{manually} analyze video content~\cite{TTclimatechange2022, covidvaccine2021, ALLEY2022102611} or conduct in-person or online interviews to understand folk theories with respect to the recommendation algorithm ~\cite{folktheories, algorithmiccrystal, devito2022transfeminine}. For e.g., manual analysis of 100 videos was used to study the content of climate change-related videos~\cite{TTclimatechange2022}. 
Manual analysis requires extensive effort and cost, and is hard to scale beyond a few hundred videos. Other  work~\cite{urman2022personalfactors,devito2022transfeminine} relies on automatic analysis of TikTok browsing histories to conduct large-scale empirical analysis of user interactions in TikTok (e.g., finding that users more likely to `\textit{like}' content posted by creators whom they `\textit{follow}' \cite{gummadi2024donation}). This analysis relies on the interactions but does not go deeper towards analyzing video content.

To answer our RQs, we need an HCI measurement approach that has three properties: (i) ability to analyze videos at {\it scale}, suggesting the need to develop an {\it automated} way to analyze video content, (ii) {\it multimodal} analysis combining user studies, interviews, data donation, and automated video analysis, in scientific ways, and (iii) deep analysis of the {\it content} inside the videos, as opposed to ``surface'' properties like hashtags which are often misleading.

\begin{figure}[h]%[btph]
    \centering
    \begin{subfigure}{0.48\linewidth}
        \centering
        \includegraphics[alt={Alternative text}, height=5cm]{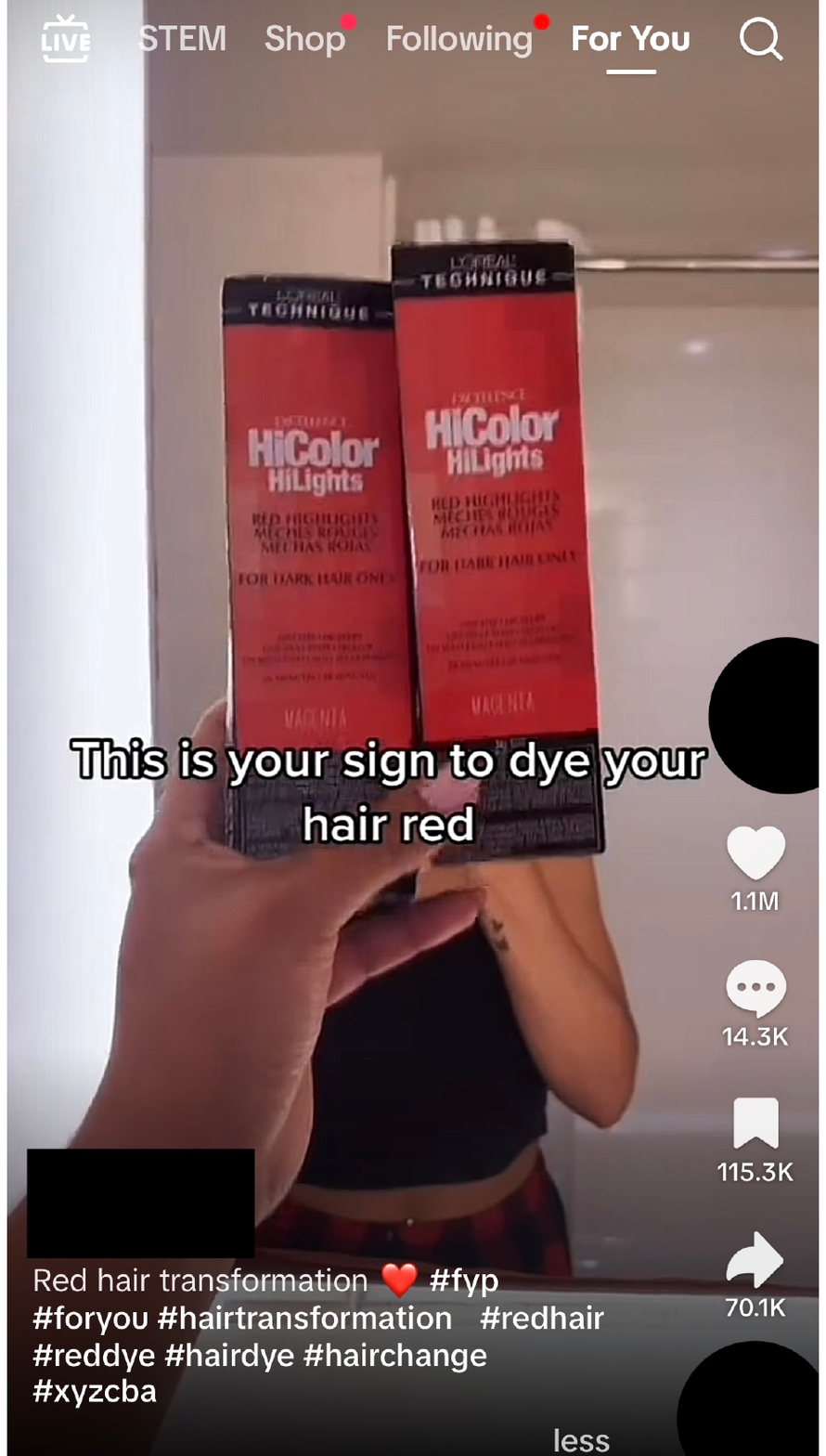}
        \Description{This subfigure contains a screenshot from a video of woman who is talking about red hair transformation. She is standing in front of the mirror and is displaying two red hair dyes to the audience. There is white bold text on screen which says, “This is your sign to dye your hair red”. This video is labeled with the hashtag “\#reddye''.}
        \caption{\centering \#reddye video \cite{tiktok_reddye_video_1} about red hair dye.}
    \end{subfigure}\hfill
    \vspace{0.6em}
    \begin{subfigure}{0.48\linewidth}
        \centering
        \includegraphics[alt={Alternative text},height=5cm]{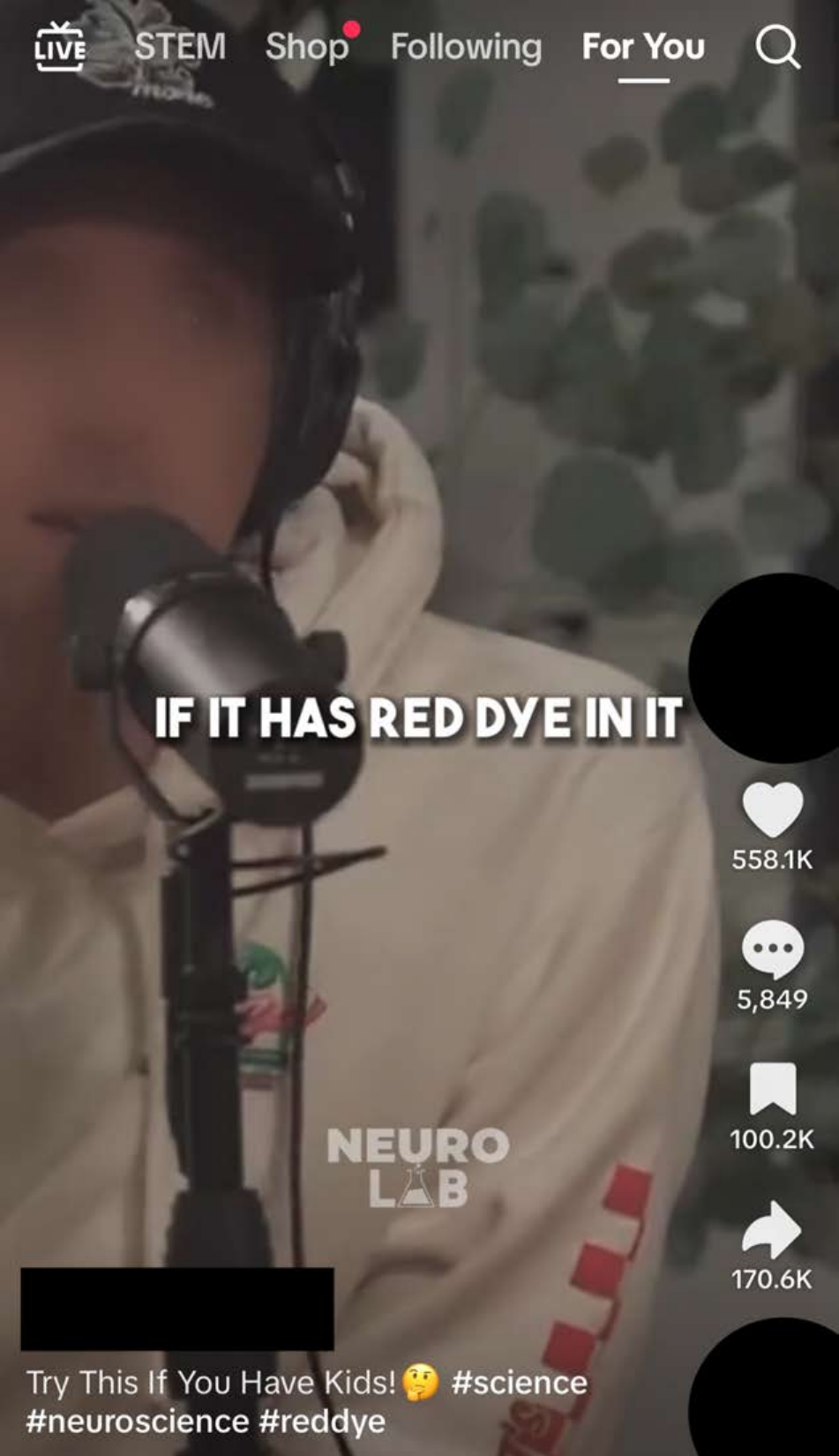}
        \Description{This subfigure contains a screenshot from a video of a man talking into a microphone, in what appears to be a podcast. The man is wearing headphones, a cap and is talking about red dye which is used in food. The video is labeled with the hashtag reddye.}
        \caption{\centering \#reddye video \cite{tiktok_reddye_video_2} about the red dye used in food.}
    \end{subfigure}\hfill
    \begin{subfigure}{0.48\linewidth}
        \centering
        \includegraphics[alt={Alternative text},height=5cm]{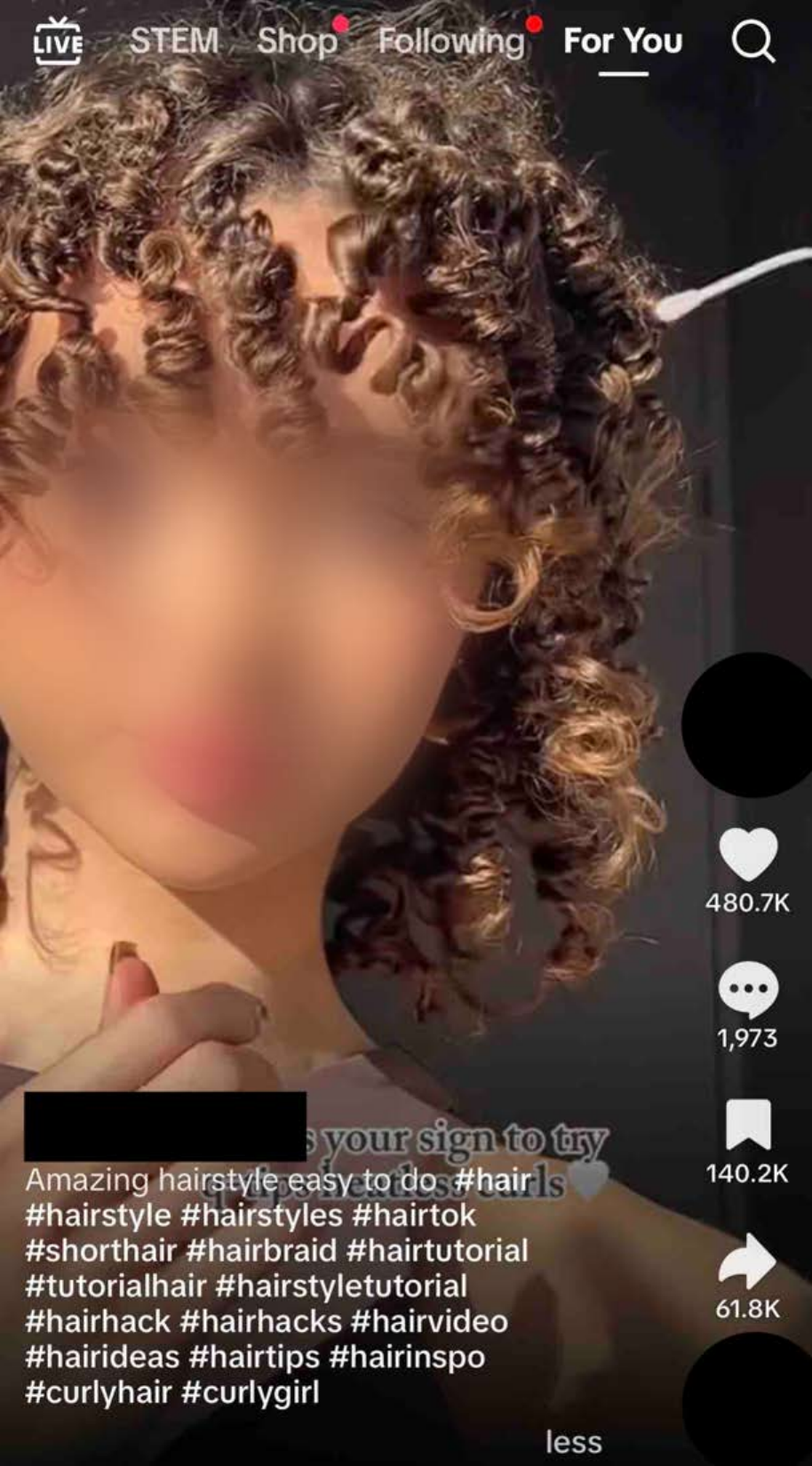}
        \Description{In this subfigure, we have a screenshot of a video containing a woman who is talking about curling hair. She is facing the camera and showing her curls. The text displayed on the video contains the words “This is your sign to”.}
        \caption{\centering Similar video to (a) using VCA \cite{tiktok_reddye_video_3}.}
    \end{subfigure}\hfill
    \begin{subfigure}{0.48\linewidth}
        \centering
        \includegraphics[alt={Alternative text},height=5cm]{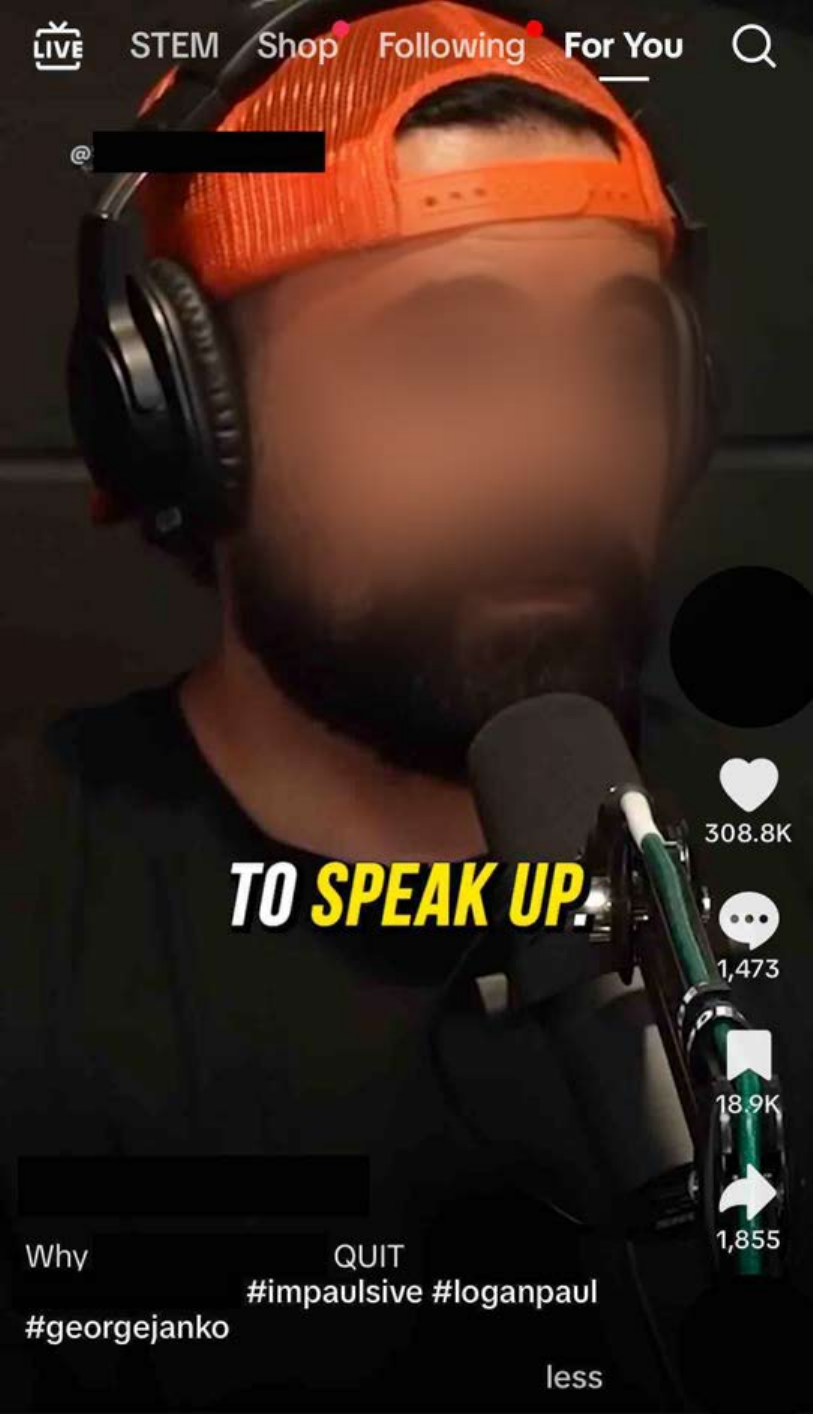}
        \Description{In this subfigure we have a screenshot of a video of a man talking into a microphone at a podcast show. The man is wearing headphones, a cap and the text “To Speak Up” is displayed on the screen.}
        \caption{\centering Similar video to (b) using VCA \cite{tiktok_reddye_video_4}.}
    \end{subfigure}\hfill
    \Description{In Fig. 1, we have 4 subfigures, which are each a screenshot of a public TikTok video.}
    \caption{\centering Two TikTok videos that share the hashtag \#reddye but refer to different topics: (a) is a video about dyeing hair red while (b) talks about red food dye. The most similar videos to both, found using VCA, are also shown. Note that similar videos do not share the same hashtags but are similar in content. Source: TikTok creators (a) notyourleidy, (b) neurolab\_, (c) bestviral5, (d) socialmediamoney.}
    \label{fig:samehashtagexample}\vspace*{-10pt}
\end{figure}

Analysis of video content at scale is challenging. Some past work has analyzed (i) objects present in videos (e.g., using deep learning models like ResNet)~\cite{lu2019object}, and (ii) video description texts and by using {\it the video's hashtags}~\cite{gummadi2024audit} as a proxy for video content. We find both these approaches 
present a limited view of a video's content. Object detection does not necessarily capture the content of the videos. Additionally, videos containing very different hashtags may talk about the same content. Fig.~\ref{fig:samehashtagexample} shows an example of videos with identical hashtags but vastly different content. Hashtags are also unreliable as they are often missing, misspelled, or misleading (i.e., do not match the video content) for a video.  Even when available, hashtags may not often capture implicit reasons why a video is attractive, e.g., sarcasm, comedy, etc. Because a user sees a video (and doesn't merely read hashtags), for deeper insights we need to go beyond hashtag-based analysis to analyze the video itself.   

\noindent{\textbf{\textit{A New Video Content Analysis Tool:}}}  
To perform large-scale automated analysis of video content, 

{we introduce an automated measurement tool called Video Content Analysis (VCA) that enables content-aware measurement of short video recommendation experiences at scale. We deploy VCA in a {\it mixed-methods study} involving controlled browsing sessions and large-scale data donation. This enables us to quantify how recommendation content evolves over time, how user actions relate to subsequent recommendations, and how sequence continuity affects user experience.}
VCA is fueled by recent advances in audio-visual querying capabilities in Vision Language Models (VLMs). Specifically, language models such as Video-LLaMA~\cite{zhang2023video} use transformer models to create an embedding of a video (and its accompanying audio). The embedding enables a language model to a) understand the video and b) interact with users about that video. Such embeddings allow VLMs to not just reason about objects in the video, but also their audio-visual dynamics. For instance, using these embeddings, VLMs can generate descriptions like `a person is holding a microphone and asking questions to passersby'. Given the rich content captured by these embeddings, they are a natural contender to capture audio-visual dynamics. We exploit these embeddings created by pre-trained transformer models as a mathematical representation of the video content. The {\it key novelty inside our VCA \it measurement tool} lies in the application of VLMs to analysis of user behavior (rather than the development of new models via training). 

\noindent{\textbf{\textit{Applying our VCA Tool to Understand TikTok-User Behavior:}}}  
{With our VCA analysis pipeline, we produce human-interpretable {\it clusters} that can be used in user-facing explanations, design probes, and audits. We validate the clustering produced with users and confirm that our VCA analysis is grounded in how people perceive content (described in \S\ref{sec:rq1new}).} {VCA allows us to {\it augment} existing ways of studying user behavior, leading to richer insights than if we used any of these approaches in isolation.} 
{More broadly, we view VCA as a generalizable tool for conducting human-driven analysis of videos (or other multimodal data) at scale, complementing traditional user studies. By summarizing long video histories into interpretable clusters, it allows researchers to explore and validate hypotheses about user behavior without always needing to recruit participants or conduct interviews.}

We combine our new VCA methodology with two modalities. (1) We conduct a user study of 68 users, conducted in a mix of in-person and online settings. In these user studies, users browse TikTok in a web browser on macOS computers for ten-minute sessions and rate their level of engagement, content quality, and network/interface quality. (2) We compile a large-scale dataset of browsing histories, via a data donation program, comprising over 2.65 million videos. These  
techniques, used in tandem, allow us to study the {\it implications} of the TikTok algorithm across different notions of {\it time}: both short-term (e.g., do small changes in the recommender algorithm behavior affect user engagement?), and long-term (e.g., how does TikTok's recommendation evolve over months, for a given user?).

{Our findings include both validation of existing folk theories about TikTok, as well as discovery of {new findings that go against folk theories.} Our key findings are:}

\begin{enumerate}

\item {We find that a user spends a large fraction of their time every day on videos that only cover a {\it small} number of topics. Further,  we find that the set of topics of interest for a user changes frequently.}

\item {Our analysis shows that recommended video feeds do not converge toward trending content and recommendations become increasingly individualized over time.}

\item  Contrary to the folk theory that current interactions immediately reshape future recommendations, we find that recommendations do not reflect just-performed actions. 

\item We find that users like seeing change in the topics of videos they are presented, i.e., a user's interest never ``converges''. {This suggests that session novelty improves perceived content quality.}
\item We find that recommendation sequence matters: dropping a small number of videos from the recommended sequence (unbeknownst to the user) {quickly} degrades user experience. 
\item We find that we can use VCA to predict whether a user will watch more than 10\% of a video, with an accuracy of 70\%. {This simultaneously demonstrates the benefits of a content-based representation (through improved prediction accuracy) and the limitations of our ability to predict user interactions which exhibit randomness and agency.}\enlargethispage*{16pt}

\end{enumerate}

%%%% related.tex starts here %%%%

\section{Related Work}

Understanding algorithm–user interplay on TikTok requires linking insights in and across four  areas: 
(i) how users perceive recommendations, (ii) how researchers have analyzed the TikTok recommendation algorithm, (iii) how researchers have analyzed video at scale, and (iv) Vision Language Models. We discuss these areas and contrast our approach to each area.

\subsection{User's Perception and Awareness of Recommendation Systems}
Past work has tried to reason about "black-box" recommender systems that govern their social media feeds \cite{likeithideit, facebooktheories, invisible, ruineverything}. These papers  present folk theories or perceptions that users develop about their interactions with and presence within social media networks. 

Multiple user studies have been conducted that specifically aim to learn about theories by TikTok users.
A user study including 24 interviews was conducted to study the interactions between TikTok's recommendations and the user's concept of self \cite{algorithmiccrystal}.  Similarly, 15 US-based TikTok users were interviewed using a semi-structured technique to understand their folk theories about TikTok's recommendation algorithm by combining the sense of person and social presence \cite{folktheories}. 
Vera et al. \cite{controllingunwanted} conduct qualitative interviews to explore TikTok users’ perceptions of recommended content on their For You Page (FYP), especially focusing on unwanted recommendations.
Because the recommendation algorithms that drive social media feeds today are mostly private, it is hard for users to obtain concrete reasoning about them or validate their existing ones. Our work links interactions directly to TikTok's recommendation algorithm and offers users insights that can help them gain awareness and shape their perceptions about their interactions with TikTok.

\subsection{Existing Analysis of TikTok's Recommendation Systems}

Several papers have tried to understand TikTok and its users. A data donation program was conducted to gather historical browsing and interaction data from TikTok users to analyze different statistical aspects of user habits such as a user's watch time over an extended timespan (few months), and the relationship between their likes and follows \cite{gummadi2024donation}. Building on this work, browsing histories were combined with bot accounts to identify features of the recommendation algorithm such as the trade-off between exploration and exploitation of videos on a user's feed~\cite{gummadi2024personalization}. Other authors conducted a mixed-method study including 28 qualitative interviews with TikTok users and a large-scale analysis of TikTok videos using their hashtags and comments to identify the impact of factors like hashtags and posting time on video popularity \cite{userassumptionTT}. An audit study \cite{gummadi2024audit} analyzed the reasoning that TikTok provides on why it recommends a certain video. This reasoning can be accessed via TikTok itself and by using different flavored sock puppet accounts, the paper shows that a significant number of explanations provided by TikTok are illogical and there is a need for more accurate and detailed explanations for the end users. Existing work does not link the longitudinal evolution of a user to their recommendations, nor do they quantify the impact of interactions or modifying the recommendations on the user experience.

\subsection{Existing Methodology for Video Analysis}
Any work that tries to understand TikTok has to work with large amounts of video data. A category of past work exists that tries to gain insights about video data but is severely limited because their methodology involves authors annotating videos. For e.g., to judge the quality and exposure of news about climate change on TikTok, authors manually go through 100 videos \cite{TTclimatechange2022}. Another work found 100 trending videos with the hashtag \#covidvaccine and analyze videos to understand misinformation and sentiments associated with the topic \cite{covidvaccine2021}. Similarly, 200 videos from a corpus of 25000 relevant videos on Douyin were analyzed to understand propagation and interactions of content related to Intangible Cultural Heritage \cite{chi2024culturalheritage}.

There is also some research that uses tools to perform automated analysis of video using machine learning techniques in limited settings, largely using object recognition or activity recognition. For e.g., the cultural differences between TikTok and Douyin (the Chinese variant of TikTok) have been studied using techniques for object detection \cite{TTvsDouyinContent}. Other tools use thumbnails of YouTube videos across 10 countries instead of metadata to discover cultural preferences~\cite{YTThumbnails}. Both these works use Faster-RCNN for object detection. In another work, authors develop a video comprehension tool that assists dance learners through automatic dance move identification \cite{videodancetool}. 
Our video content analysis tool combines the depth and content-driven insights of manual analysis with the scale of history-based analysis. In doing so, it removes the scale limitation of manual content-based analysis, while providing rich insights about each video watched by a user. To the best of our knowledge, we are the first to use large generative models for analyzing user recommendations and preferences on short-form video. Our use of large generative models to interpret the entire video provides richer insights compared to analysis provided by object recognition or thumbnails alone.

\subsection{{Vision Language Models for Video Analysis}}
{
Recent advances in Vision Language Models (VLMs) have enabled reasoning over multimodal data like videos. VLMs such as Video-LLaMA~\cite{zhang2023video} use transformer models to create an embedding of a video (and its accompanying audio). Video-LLaMA is built on top of MiniGPT-4 and is composed of two core components: (1) Vision Language (VL) Branch and (2) Audio Language (AL) Branch. The video and audio branches create video and audio embeddings respectively and enable the language model to understand and interact with users about that video. The embeddings allow VLMs to not just reason about objects in the video, but also their audio-visual dynamics (e.g., using these embeddings, VLMs can generate detailed descriptions about videos as seen in Fig. \ref{fig:chatexample}).

Prior work has used such models for tasks such as video captioning, retrieval, chatbots \cite{valor}; we instead use them as a tool to represent TikTok videos when studying algorithm-user interplay.
}

\section{Methodology: Data Collection and Measurement Design}\label{sec:curation}

To retrieve data from TikTok and its users for our analysis, we conduct a user study on a university campus and through online participants. TikTok users mainly interact with the app's recommendation algorithm via their "For You Page" (FYP). Our user study observes a user's interactions with videos recommended on their FYP (\S\ref{sec:study}) and also, has an additional (optional) component that allows participants to donate historical activity data that is available to download via their TikTok accounts (\S\ref{sec:bhstudy}). The user study and historical data collection effort was reviewed by our Institutional Review Board and found to be exempt from a full IRB review. Details about the study are listed in the remainder of the section and further, in the appendix. We note that all our data collection and analysis were completed prior to both (i) TikTok's acquisition by the US entity, and (ii) the new terms of service that went into effect on January 22, 2026.

{We also describe our measurement tool called Video Content Analysis (VCA) that we use to convert each video into a compact vector capturing its audio-visual content in \S\ref{sec:design}.}

\subsection{User Study: Browsing Two TikTok Sessions}
\label{sec:study}

\subsubsection{Sessions and Questions}
Our user study was a 30-minute activity where participants saw their own TikTok For You Page (FYP), the default page that a user is at on TikTok, at \url{https://www.tiktok.com/foryou}. The participants used TikTok in \textit{two sessions of 10-minutes each} using Google Chrome on a macOS computer. In one of these two sessions, the users saw their original TikTok feed, as recommended by the TikTok algorithm. In the other session, a small fraction of the videos are dropped from the user feed, i.e., a small fraction of randomly selected videos recommended by TikTok's algorithm are hidden from the users. The order of these sessions was randomly assigned.

To enable this study, we built a Google Chrome plugin. Participants are not made aware of the existence (and nature) of the modifications to their FYP we make in the sessions either before the study or after, as approved by IRB protocol. 
For each session, the participants are expected to scroll through their FYP as they normally would, i.e., they don't have to see all videos entirely. We limit participants from spending time reading comments or browsing outside the FYP.

After the two sessions, participants complete a survey form with questions about demographics, including questions on age, occupation, TikTok join date, and details about their platform usage, and their experiences with both the sessions in terms of their content quality, network quality and their overall engagement levels using Likert scales. In additional questions, an emphasis was also placed on the relative experience between the sessions and we asked questions that ranked the two sessions based on interest, frustrations, historical similarity, and overall satisfaction. The questions we asked are detailed further in the appendix.

To be able to participate in our study, we required an active TikTok account that is regularly used to browse content. We asked the following questions at the time of recruitment and only an affirmation to both questions led to participation.

\begin{enumerate}
    \item Do you have a TikTok account that is at least a few weeks old?
    \item Do you use TikTok regularly?
\end{enumerate}
%\mc{Add in the exact questions we ask in a list.} 
Each participant received \$15 as compensation to participate.

\subsubsection{Chrome Extension}
Our browser extension handles the dropping of videos by modifying the HTML of the TikTok page. To drop videos during the modified sessions, the Chrome extension monitors when TikTok loads new batches of video links (typically 5 to 11 at a time) as users scroll through their FYP. For each batch of newly loaded videos, a  percentage of videos were randomly selected, and their HTML was removed, making them hidden to the user. 

\begin{figure*}[h]
\centering

% --- Top large image ---
\begin{subfigure}{\columnwidth}
  \includegraphics[width=1\linewidth]{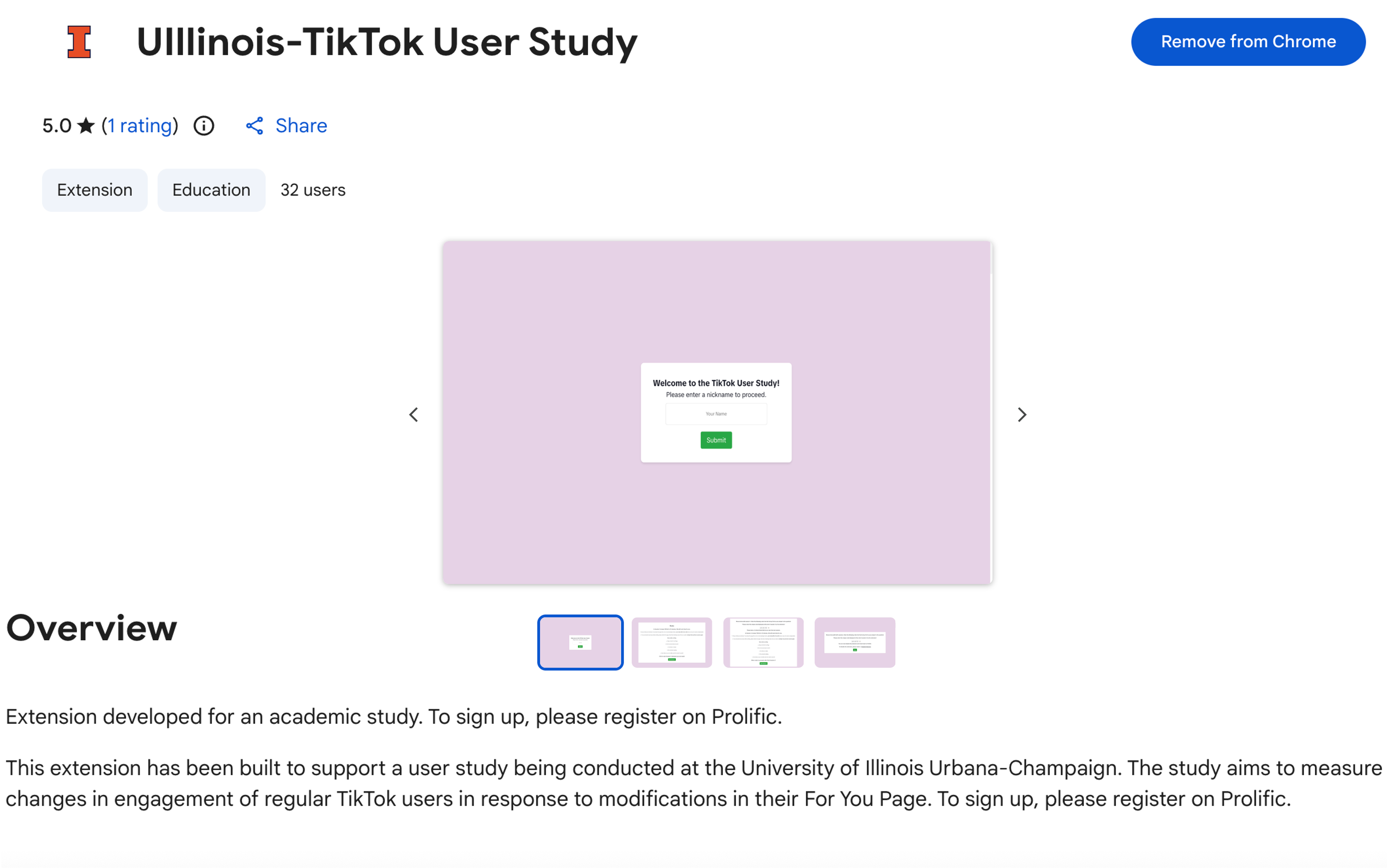}
  \Description{Screenshot of the Chrome Web Store page for our extension, showing its overview, ratings, and installation button.}
  \caption{\centering Our extension page on the Chrome Web Store.}
\end{subfigure}

\vspace{6pt}

% --- Bottom two images ---
\begin{subfigure}{0.7\columnwidth}
  \includegraphics[width=\linewidth]{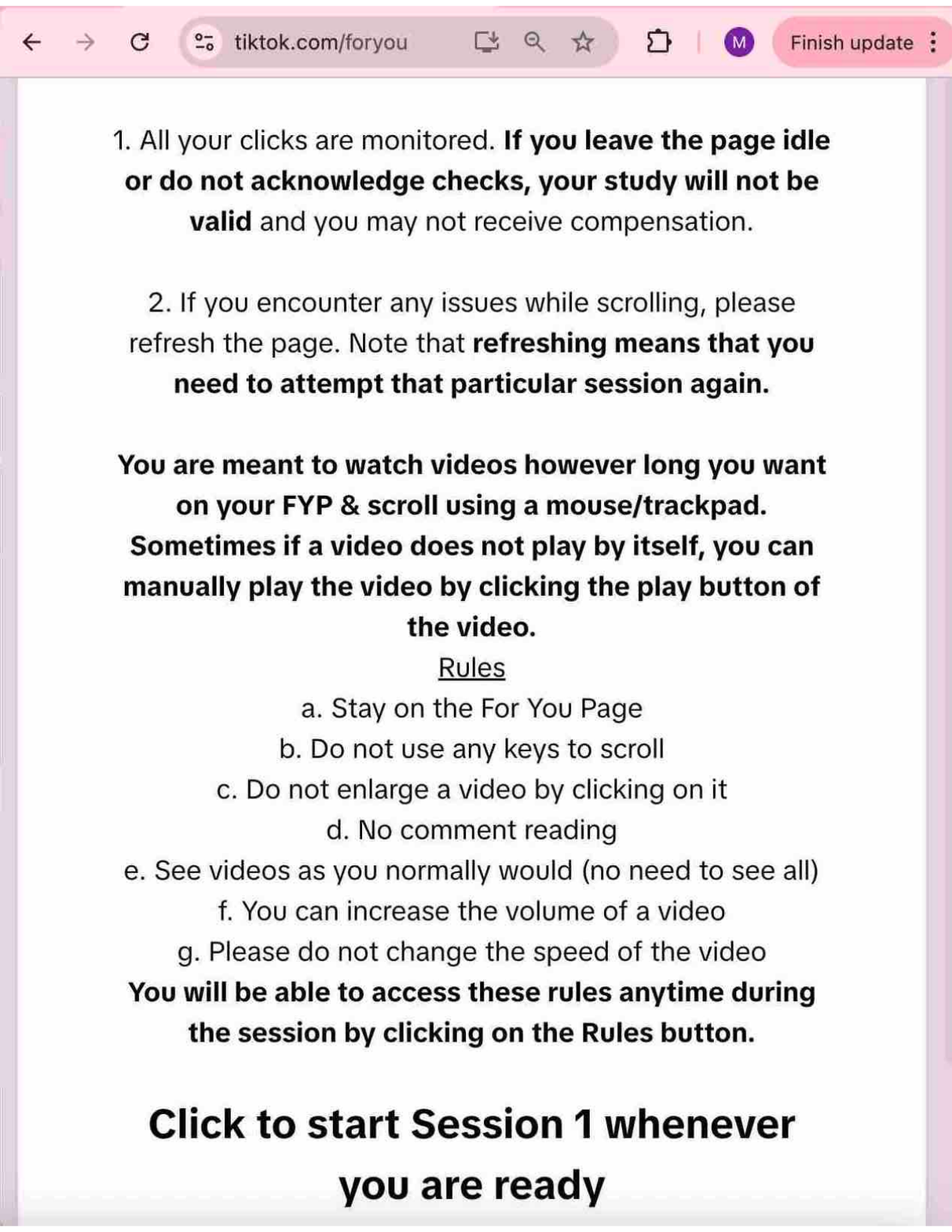}
  \Description{Rules page shown to participants before beginning a study session, outlining guidelines and attention checks.}
  \caption{\centering Rules page shown before a session begins.}
\end{subfigure}
% \hfill
\hspace{40pt} 
\begin{subfigure}{0.7\columnwidth}
  \includegraphics[width=\linewidth]{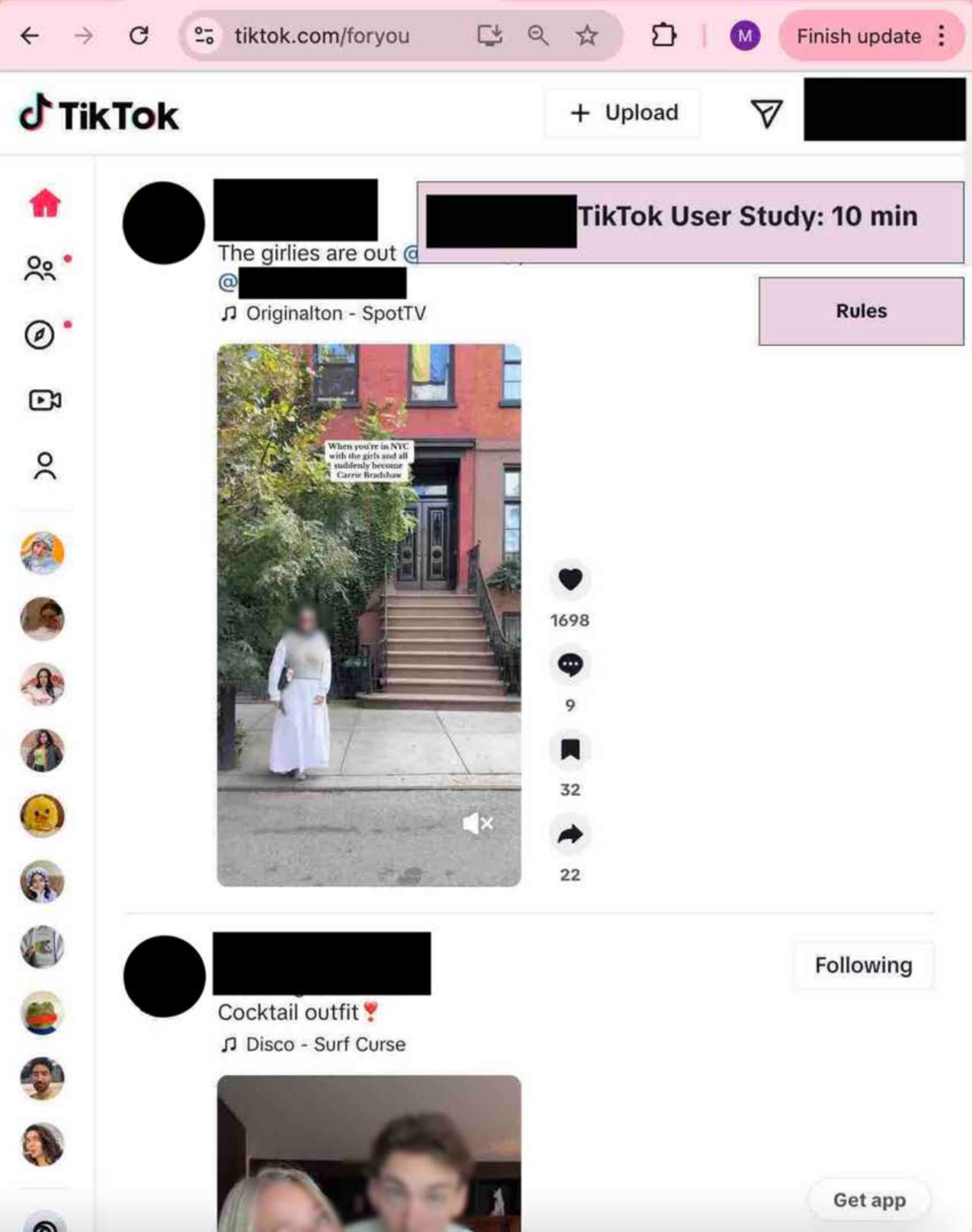}
  \Description{User study interface during a session, showing a timer and access to the rules page.}
  \caption{\centering User interface during a study session.}
\end{subfigure}\vspace*{-5pt}

\Description{Screenshots of the Chrome extension used during the online user study.}
\caption{\centering Chrome extension used during the online study. A similar local version was used for the in-person study. Video appearing in (c) by TikTok creator: loren.asad \cite{tiktok_fig2}.}
\label{fig:chromeextension}\vspace*{-5pt}
\end{figure*}

The extension also extracts video IDs of the videos users actively watched during the two sessions. It does so by querying specific HTML selectors and transmitting the collected data to a back-end server. Our repository of the user study data is the back-end server that allows us to know the videos seen by the participants as well as additional information like video watch time. Participants with unusually low video view times, significant discrepancies, prolonged engagement with a single video, failure to respond to periodic attention checks, or extended periods of inactivity on the FYP were rejected and excluded from the dataset. Some screenshots of the extension are shown in Fig.~\ref{fig:chromeextension}.

\subsubsection{Platform of Study}

The same study was conducted (1) in-person on a university campus and (2) online via Prolific \cite{prolific}. The style of the study and questions asked were the same for both the platforms.

In-person participants were recruited through flyers and posts on social media platforms. At the time of the study, participants used a designated laptop provided by us, with the extension pre-installed. Online participant recruitment was handled by Prolific. Participants used their own macOS laptops and were guided to download the Chrome extension from the Chrome Web Store. 
% The name of the extension reveals our identity and to maintain anonymity, it is not shared in this version of the paper. {
Once the study was completed, we instructed online participants who had to download the Chrome extension themselves to uninstall it. We share instructions with them on how to do so. In-person participants need not uninstall the extension since they were using a device designated and set up by us.

We conducted the in-person study between November 2023 and May 2024. The online study was conducted between May and September 2024.

\subsubsection{Variants}

\begin{table*}[t]
\centering
\begin{tabular}{|c|c|c|c|}
\hline
 & \multicolumn{3}{c|}{\textbf{Study Condition}} \\
\cline{2-4}
 & \textbf{In-person} & \textbf{Prolific} & \textbf{Prolific} \\
 & \textbf{(30\% Drop)} & \textbf{(5\% Drop)} & \textbf{(30\% Drop)} \\
\hline
\textbf{No. of Valid Participants} & 13 & 21 & 34 \\
\hline
\textbf{No. Participants who saw modified session first} & 8 & 10 & 18 \\
\hline
\textbf{No. Participants who saw unmodified session first} & 5 & 11 & 16 \\
\hline
\textbf{Total No. of Unique Videos} & 543 & 1321 & 1900 \\
\hline
\textbf{Average No. of Videos per Session} & 22.31 & 31.45 & 27.94 \\
\hline
\end{tabular}
\vspace{5pt}
\caption{\centering Statistics from the variants of in-person and Prolific user studies.}
\label{table:userstudyvariants}\vspace*{-20pt}
\end{table*}

The user study had three variations depending on the platform and drop percentage we use. We explain the reasoning for the drop modification in \S\ref{sec:rq4}. We choose two drop percentages: 5\% (low drop rate) and 30\% {(moderate drop rate).} Details of each variant are shared in Table \ref{table:userstudyvariants}. 

\begin{enumerate}
    \item In-person (30\% Drop): 30\% of videos were randomly dropped in exactly one of the sessions.
    \item Prolific (5\% Drop): 5\% of videos were randomly dropped in exactly one of the sessions.
    \item Prolific (30\% Drop): 30\% of videos were randomly dropped in exactly one of the sessions.
\end{enumerate}

{The actual number of videos that were dropped is decided at runtime based on the number of videos that are recommended as participants scroll. For e.g., assuming 20 videos were recommended at a time, if the drop setting was set to 5\%, 1 video was dropped and if drop setting was set to 30\%, 6 videos were dropped. Note that this does not decide the total number of videos seen by each participant as they can always scroll further to request more videos from the TikTok server.}

For all these variants, the other session (at a random order) was the original TikTok FYP. {We don't inform users about the sessions they appeared in - our protocol was approved by IRB.}
We conducted more user studies than those listed in Table \ref{table:userstudyvariants} but they are not considered due to concerns like data not recording as expected, set up not done correctly, and invalid variants. %and only use the part 2 materials to increase our dataset of videos. Extra details about the study are provided in the appendix.

\subsection{Historical Activity Files Donation}
\label{sec:bhstudy}

\begin{figure}[h]
\centering
\begin{lstlisting}[language=python]
Date: 2023-12-12 12:12:12
Link: www.tiktokv.com/share/video/7313716511095442693/

Date: 2023-12-12 12:12:24
Link: www.tiktokv.com/share/video/7374043158331657504/

Date: 2023-12-14 00:00:07
Link: www.tiktokv.com/share/video/7466011678879010066/
\end{lstlisting}
\Description{Snippet of a browsing history file showing timestamps and corresponding TikTok video URLs.}
\caption{\centering Contents of a browsing history file. This example user watched video 7313716511095442693 for 12 seconds and watched video 7374043158331657504 next for an unknown amount of time. We also do not know the watch time of video 7466011678879010066 from the snippet shared. The like list and share list files also look similar to the browsing history file but only have the links to videos that were liked or shared.}
\label{fig:bh-example}\vspace*{-10pt}
\end{figure}

\begin{table*}[t]
\centering
\begin{tabular}{|c|c|}
\hline
\textbf{No. of Valid Participants} & 100 \\
\hline
\textbf{No. of Unique Videos} & 3.9M \\
\hline
\textbf{No. of Videos with Embeddings and Metadata} & 2.65M \\
\hline
\textbf{Data Time Range} & 170--450 Days \\
\hline
\textbf{Average No. of Unique Videos in a Browsing History File} & 39K \\
\hline
\textbf{No. of Valid Participants with Like List} & 66 \\
\hline
\textbf{No. of Valid Participants with Share History} & 54 \\
\hline
\textbf{Average No. of Videos Seen in a Day} & 332 \\
\hline
\end{tabular}
\vspace{5pt}
\caption{\centering Statistics from the browsing history files we collected.}
\label{table:bhfilesstats}\vspace*{-10pt}
\end{table*}

We requested all participants in our user study to optionally donate their TikTok activity data for an additional \$15 compensation. {We had a total of 100 valid donations.} 
% - almost everyone who participated in the study chose to donate their data. 
TikTok activity data contains files like the browsing history file, like list, and share list. These files contain a record of videos a user has watched or interacted with (typically in the last 6 months) in separate files, along with the date and time each video was viewed, as illustrated in Fig. \ref{fig:bh-example}. The browsing history file has an exhaustive list of all the videos seen, and like lists and share lists contain the time and URLs of liked and shared videos respectively. We require at least 1000 videos in the browsing history files to consider an activity folder as valid. We borrow this data collection technique from \cite{gummadi2024donation} to build a database of user profiles. Sometimes, the activity folder misses files, which is beyond both our and the participant's control. We note that the donation activity has more participants because the session preference data from some users was rejected due to reasons mentioned above (failed attention checks, incorrect extension variants, etc.). The historical activity data is valid in spite of these issues, and is therefore retained for these participants.

\subsubsection*{The Information that Activity Files Provide}

Activity files are important to completely understand a user and their interests, and are central to reaching our findings.
Using the browsing history file, we can collect key information such as the date and time a user watched a video and the video URL to view the video. We can infer the watch time of most videos (an example is shown in Fig. \ref{fig:bh-example}) and, by comparing this time to the video duration, we can get the video watch percentage. Intuitively, metrics like watch percentage are positively correlated to user interests. Like lists and share lists contain the time and URLs of liked and shared videos and these videos are especially important to define user preferences. However, in our data analysis, we note that like lists and share lists are not exhaustive (also noted in \cite{gummadi2024donation}). We still use these lists in \S\ref{sec:rq2} to address folk theories surrounding likes and shares.

For each video, we gather metadata information about them using an open-source unofficial TikTok API \cite{Teather_TikTokAPI_2024}. 
While collecting the data, some videos were either deleted or not publicly available - we ignore these videos. We also use the videos and a VLM (Video-LLaMA, discussed in \S\ref{sec:design}), to generate video content embeddings. On average, we successfully collect metadata and embeddings for about 70\% of the videos in each user's browsing history. We share some facts about the data donation part of the study in Table \ref{table:bhfilesstats}.

\begin{table*}[t]
\centering
\begin{tabular}{|c|c|c|c|c|}
\hline
 & \textbf{Category} 
 & \textbf{User Study (Browsing Session)} 
 & \textbf{Activity Data Donation} 
 & \textbf{TikTok} \\
\hline
\textbf{Age} & \textbf{18--24} & 58 & 60 & 36.2 \\
\hline
 & \textbf{25--34} & 33 & 36 & 33.9 \\
\hline
 & \textbf{35--44} & 9 & 4 & 15.8 \\
\hline
 & \textbf{45--54} & 0 & 0 & 7.9 \\
\hline
 & \textbf{55+} & 0 & 0 & 6.2 \\
\hline
\textbf{Gender} & \textbf{Male} & 41 & 32 & 45 \\
\hline
 & \textbf{Female} & 59 & 67 & 55 \\
\hline
 & \textbf{Non-Binary} & 0 & 1 & 0 \\
\hline
\end{tabular}
\vspace{5pt}
\caption{\centering Demographic distribution of the user study participants. The last column shows the distribution of TikTok's general user base. All numbers are percentages.}
\label{table:demographic}
\vspace{-5pt}\end{table*}

The demographic distribution of our participants is illustrated in Table \ref{table:demographic}. Approximately 72\% of participants watch videos in English, 8\% in Spanish, while the remaining view content in various other languages. About 45\% of our participants identify as White/Caucasian, 23\% as Asian/Pacific Islanders, 15\% as Black/African American, and 12\% as Hispanic, with the remaining identifying as other ethnicities. The distribution of participants makes a good effort to mirror that of TikTok’s general user base ~\cite{tiktokdemo},  particularly in terms of age (with skew toward a younger audience) and gender, making our collected data a close representation of the platform's audience. 
\enlargethispage*{16pt}
\subsection{Ethics}

Both our study and data collection effort received an Institutional Review Board's exemption from our university's Office for Protection of Research Subjects under Category 3 (Benign Behavioral Interventions).

At the beginning of our study, all participants were  informed of the process of the study and the activity data we were requesting, and were asked to explicitly opt in and consent. Participants were not coerced to share data they did not want to and could change their consent at any time. We showed participants how to download their data from TikTok and then left it up to the participants to analyze the contents of the data and submit it later if they were still comfortable doing so. At the end of the user study session, participants were instructed to log out of their TikTok accounts.

Our goal was to create abstract user profiles that conjoined the user study with the activity data donated - we did not try to obtain more information than what was required (video IDs of videos seen, video watch time, video timeline, liked and shared videos) and were not interested in who the user is in real life. The data we collected was handled only by IRB-approved researchers who had completed relevant training and was stored in a way that facilitated our goal without revealing information using alphanumeric IDs for labeling. We did not attempt to link these IDs to any identifiable information and did not contact participants once we had received the data for our study. All of our data is stored on our lab's server in our university and an additional AWS machine that only we have access to.

We limit ourselves to publicly available videos on TikTok and do not try to access private videos seen by participants. We plan to store all the data we collected for two years after % the study 
the paper is published, without sharing it with anyone, and deleting it permanently thereafter.

{
\subsection{Video Content Analysis Tool}
\label{sec:design}
We now discuss the technical details of how we convert the content of a video to a VCA vector, which is the mathematical representation of the content of a video. This tool enables content-aware measurement of short-video recommendations and, together with user studies and data donation, forms our HCI measurement approach.\enlargethispage*{16pt}

\subsubsection{Capturing Video Content}

\begin{figure}[h]
\centering

% --- Top (centered, 0.7 column width) ---
\begin{subfigure}{0.7\columnwidth}
  \centering
  \includegraphics[width=\linewidth]{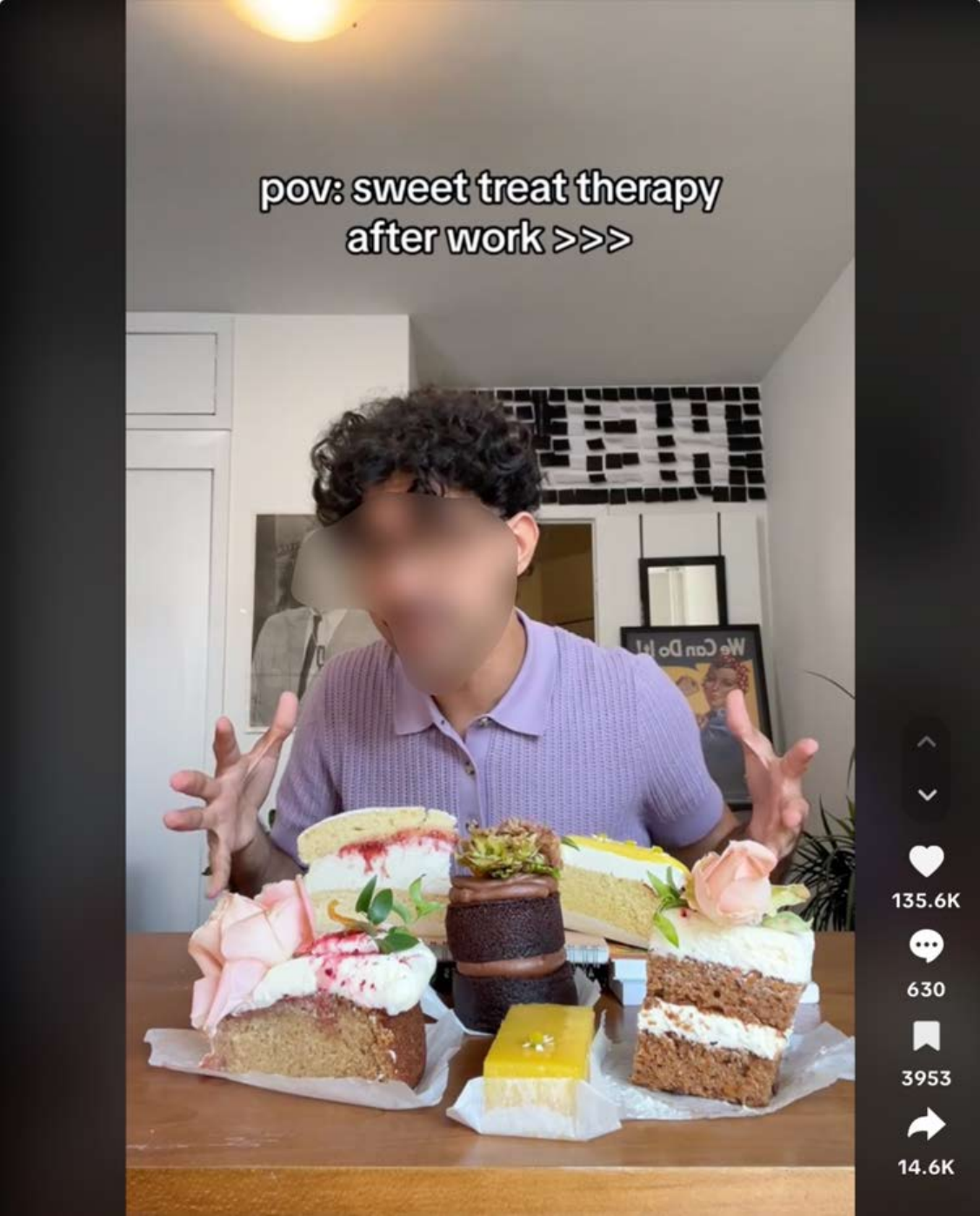}
  \Description{Screenshot of a TikTok video showing a man sitting in front of multiple cake slices, with on-screen text reading ``pov: sweet treat therapy after work >>''.}
  \caption{\centering Snapshot of a TikTok video \cite{sodcake} in which a man expresses his love for cake slices after a day at work.}
\end{subfigure}

\vspace{6pt}

% --- Bottom (full column width) ---
\begin{subfigure}{\columnwidth}
  \centering
  \includegraphics[width=\linewidth]{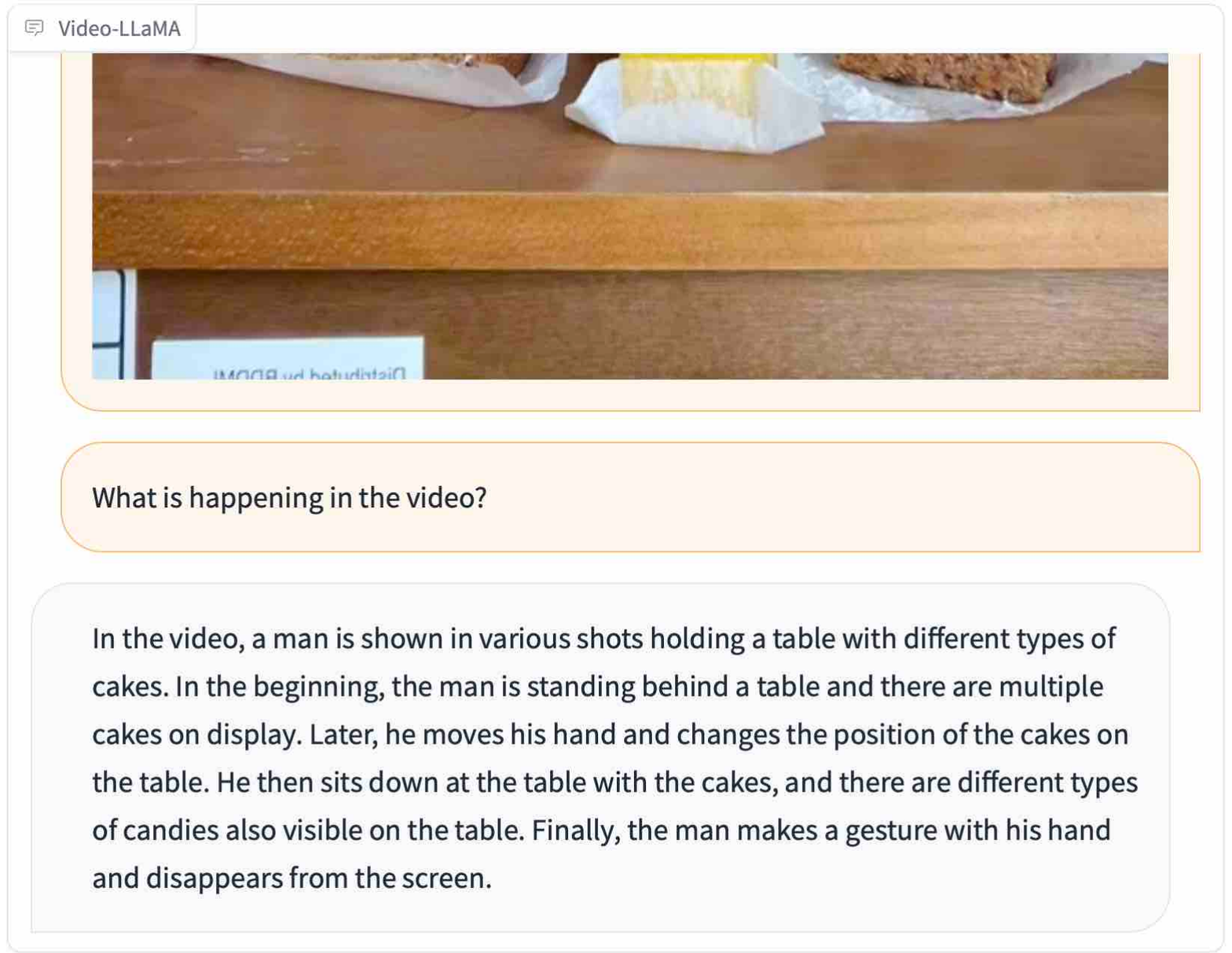}
  \Description{Video-LLaMA generated description after processing the video shown above in response to the prompt ``What is happening in the video?''.}
  \caption{\centering Description generated by Video-LLaMA when prompted with ``What is happening in the video?'' after the video in (a) is uploaded to the chat. The response is reasonably accurate and detailed.}
\end{subfigure}

\Description{Example illustrating a TikTok video and the corresponding description generated by Video-LLaMA.}
\caption{\centering Example TikTok video and the description produced by Video-LLaMA for the same video. Video by TikTok creator sodakhtar.}
\label{fig:chatexample}\vspace*{-10pt}
\end{figure}

Our method of understanding videos is inspired by the ``upload video to chat'' functionality of some language models \cite{li2024llms, ataallah2024minigpt4}. VLMs like Video-LLaMA convert videos to their embedding representation to reason about it.

Given the rich content captured by these embeddings, they are a natural contender to capture the audio-visual dynamics and hence, the content of a video. We exploit these embeddings created by pre-trained transformer models as a mathematical representation of the video content.

Video-LLaMA is an open-source VLM \cite{opensource}. 

Since we are able to access its codebase to intercept and record embeddings, it is the language model we use to collect our embeddings. We use the pretrained model available with 7 billion (7B) parameters. Note that there are other much larger VLMs available: Video-LLaMA itself has a 13B version available, DeepMind's Flamingo VLM consists of 80B parameters etc. Since we work with a pretrained model, collecting our embeddings was not computationally intensive and we were able to do so using Virtual Machine resources available to us at a university campus. An example machine we use to collect embeddings ran Ubuntu 22.04.5 LTS equipped with 1 NVIDIA A40 with memory 48GB. On this machine, we could run 2 parallel instances of the pretrained transformer model. Collecting the embedding of one video took 2 seconds.

Video-LLaMA has separate audio and visual branches that convert audio and video into embeddings of size 8, 4096 and 32, 4096 respectively. We stack these two embedding tensors into a single vector for each video. 

Each vector then gets normalized using feature-wise min–max normalization, mapping each feature within the vector to the 
[0,1] range. The resulting vector captures the content as well as the context of the video. The number of features in the resulting vector is very large. To obtain a more compact representation, we applied Principal Component Analysis (PCA) \cite{WOLD198737} and reduced the vector size to 100 dimensions. We chose 100 components as a practical trade-off between keeping enough variance in the data and computational cost, that constrains how large an embedding we can store and process efficiently. We refer to this resulting 100-dimensional representation as the VCA vector. 

The VCA vector needs to be computed only once for each video. This computation is similar in scale to other common video processing techniques. For e.g., video streaming systems, like TikTok,\enlargethispage*{16pt} already perform compression at video upload time to create different versions of the video for different network conditions and screen types~\cite{li2023dashlet}. 

\begin{figure}[h]
\centering
\footnotesize

% \mc{set one7377356449468534034, set two is 7248981972222004507, 7392259053235326251, 7393082396419165471, set 3 is 7225971694245104939, 7333773698161265963, 7254752968451362054}

% -------- Cluster 39 --------
\begin{subfigure}{0.31\columnwidth}
  \includegraphics[width=\linewidth]{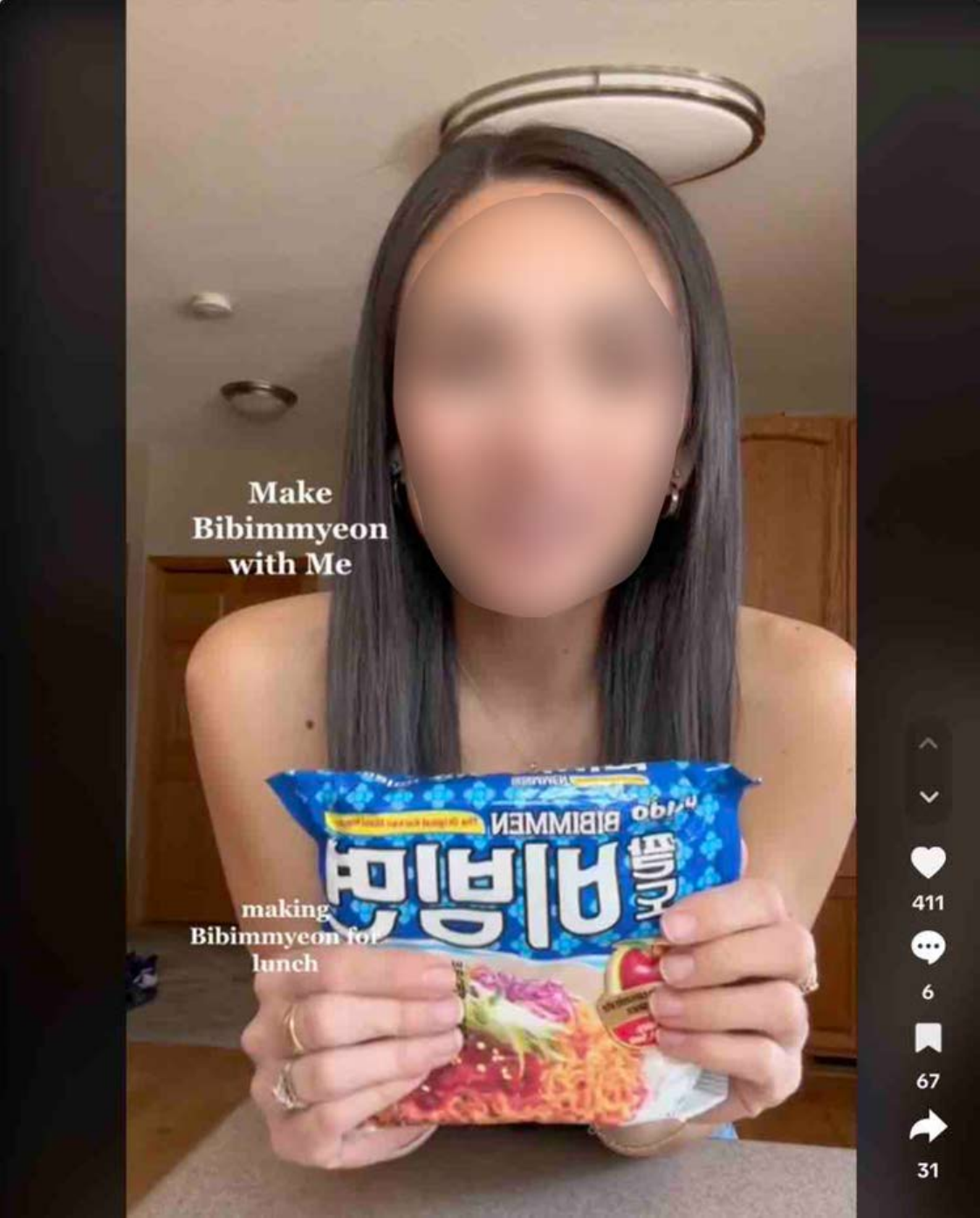}
\end{subfigure}\hfill
\begin{subfigure}{0.31\columnwidth}
  \includegraphics[width=\linewidth]{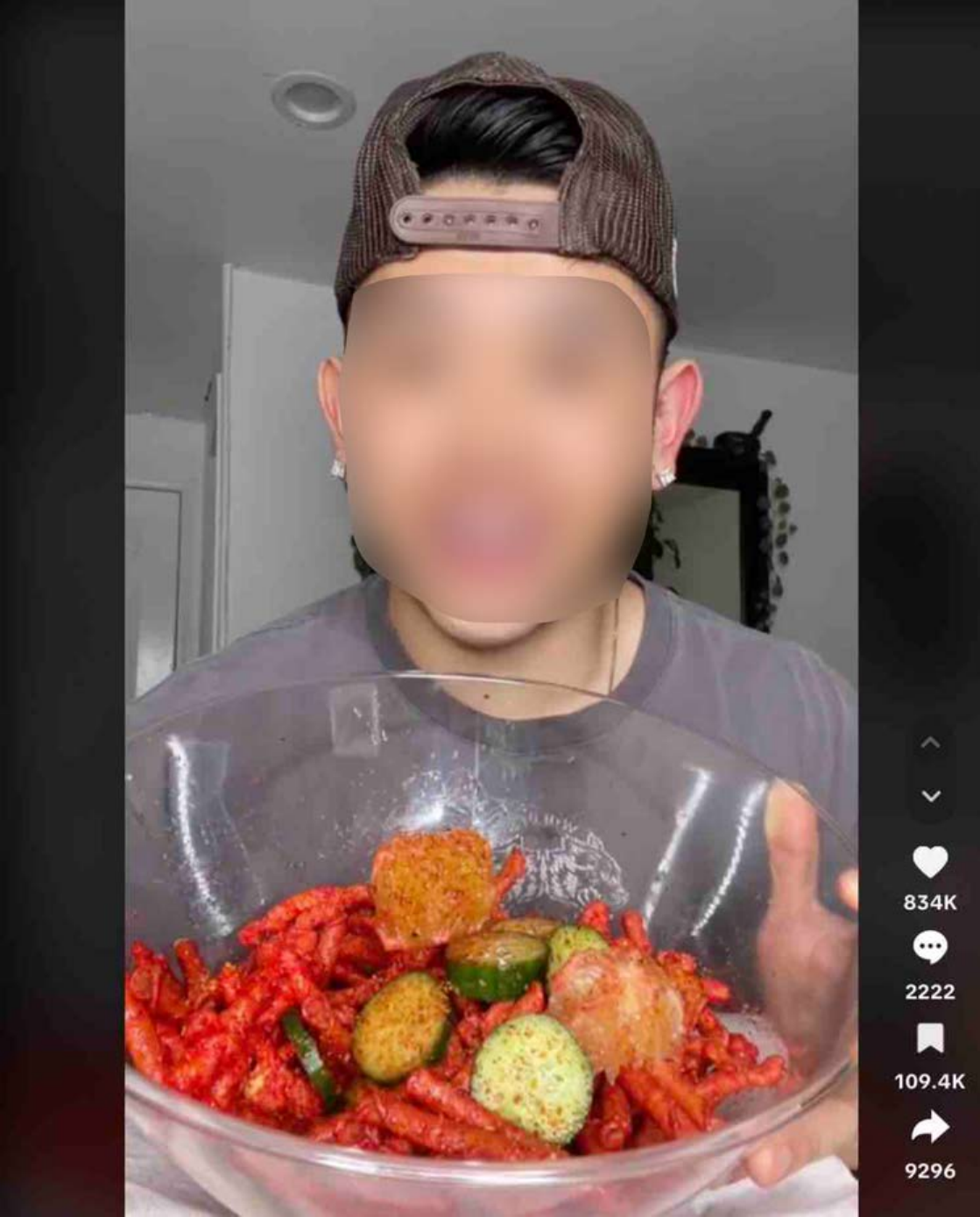}
\end{subfigure}\hfill
\begin{subfigure}{0.31\columnwidth}
  \includegraphics[width=\linewidth]{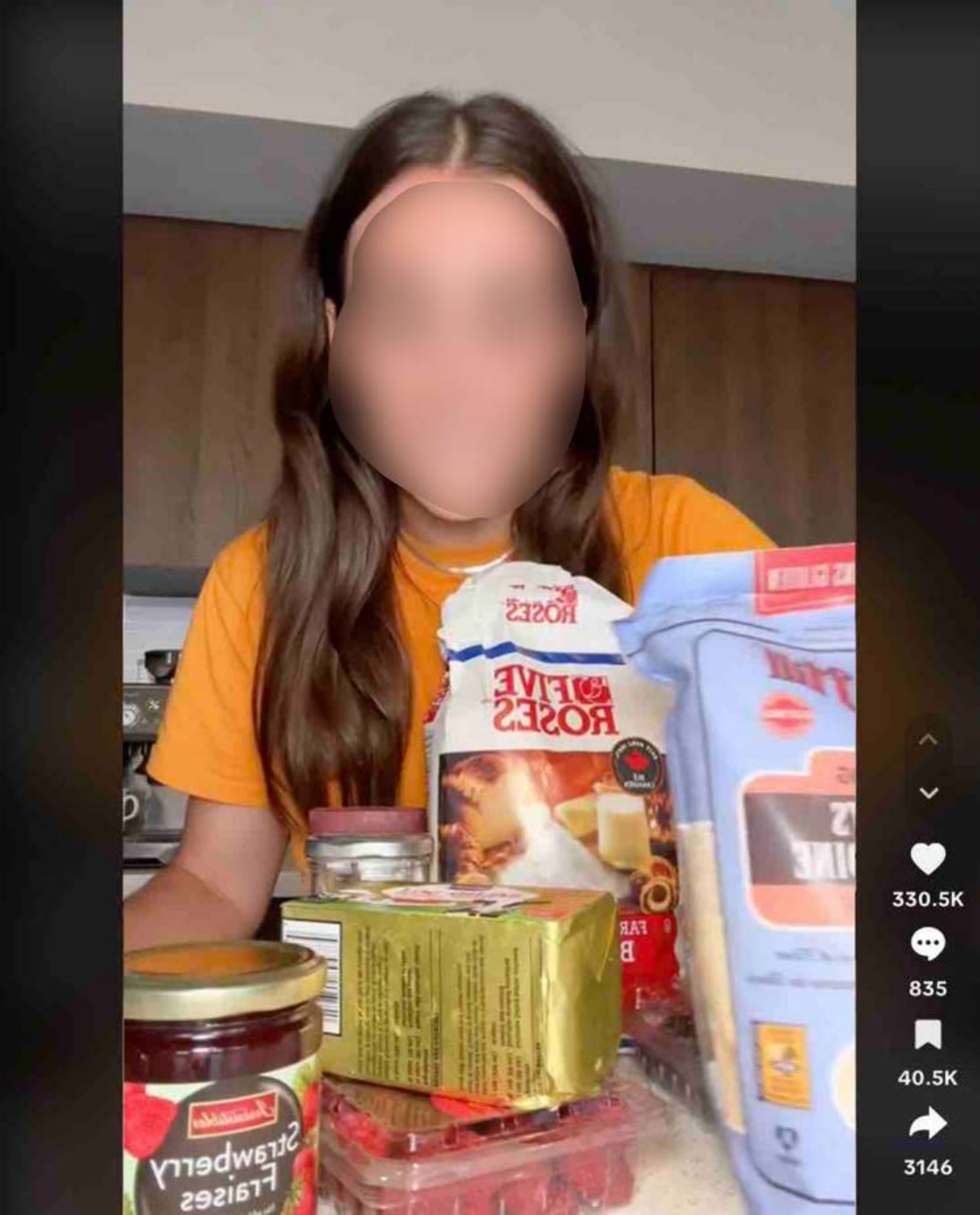}
\end{subfigure}
	
\vspace{4pt}
\Description{Screenshots of three videos from cluster 39 showing people cooking while speaking to the camera.}
\caption*{\centering (a) Cluster 39: Videos of people talking to the camera while cooking. Videos by TikTok creators: kindasortasimple \cite{food_1}, thereal.ceyana \cite{food_2}, hey.itsbel \cite{food_3} (Left-Right).}

\vspace{6pt}

% -------- Cluster 47 --------
\begin{subfigure}{0.31\columnwidth}
  \includegraphics[width=\linewidth]{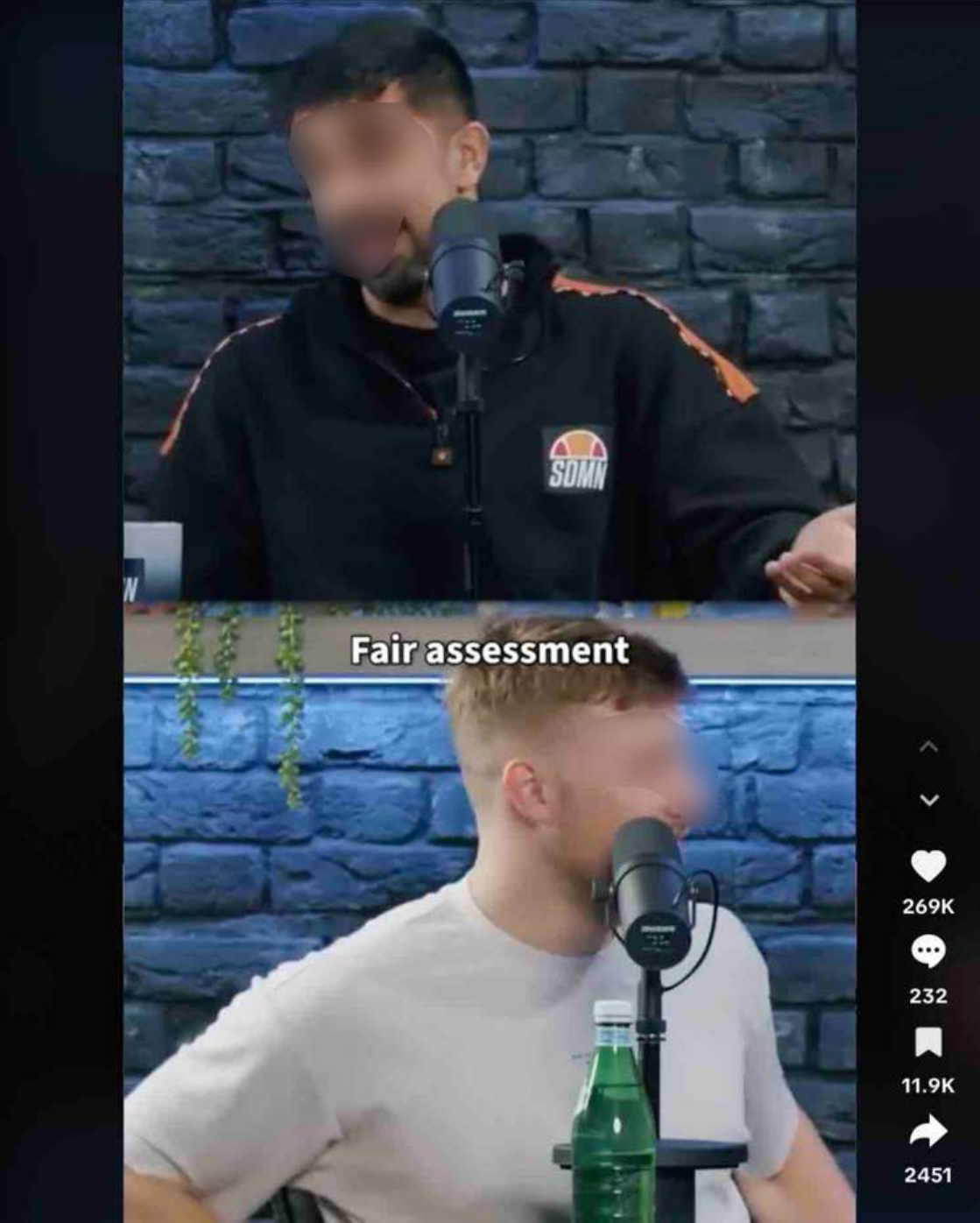}
\end{subfigure}\hfill
\begin{subfigure}{0.31\columnwidth}
  \includegraphics[width=\linewidth]{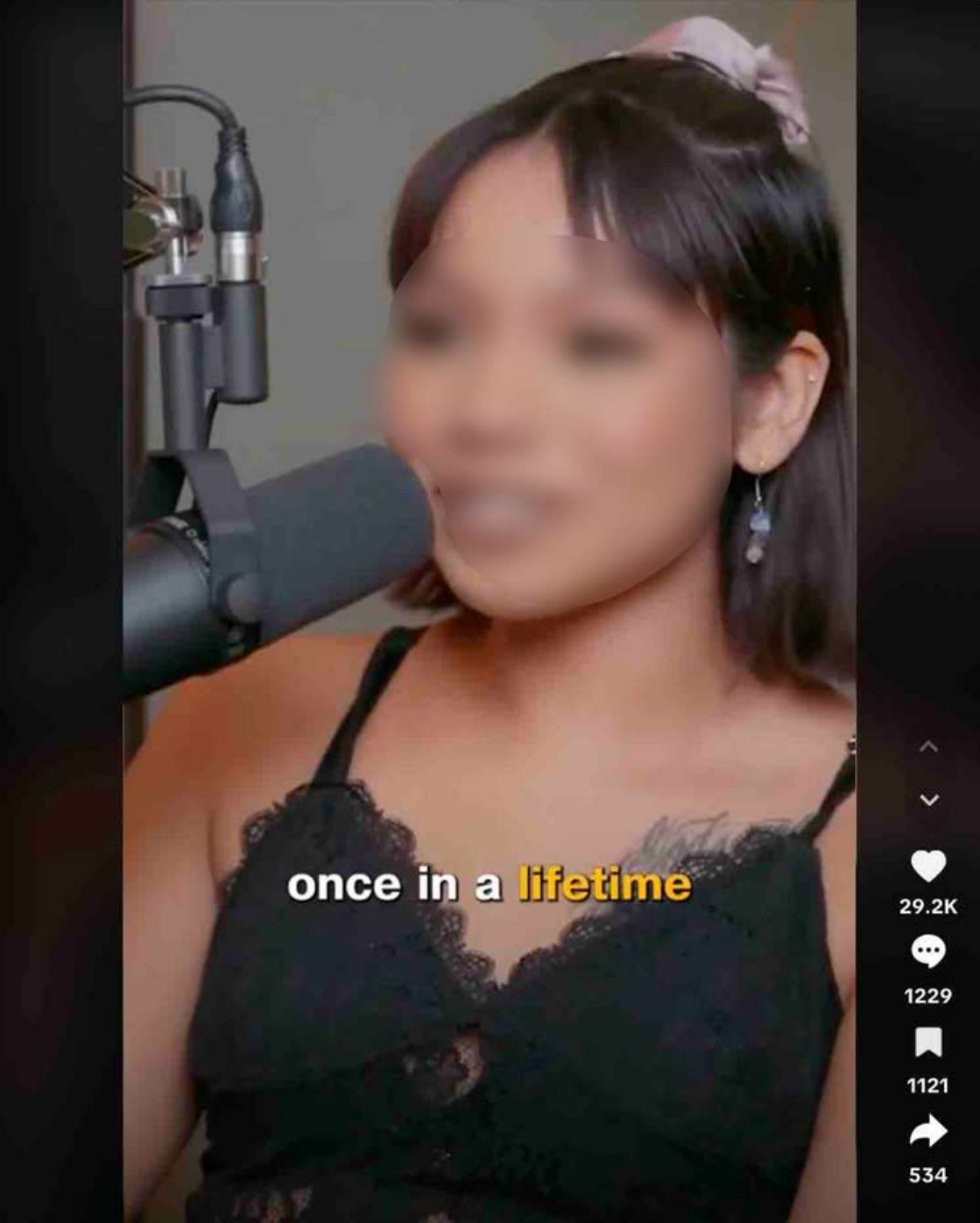}
\end{subfigure}\hfill
\begin{subfigure}{0.31\columnwidth}
  \includegraphics[width=\linewidth]{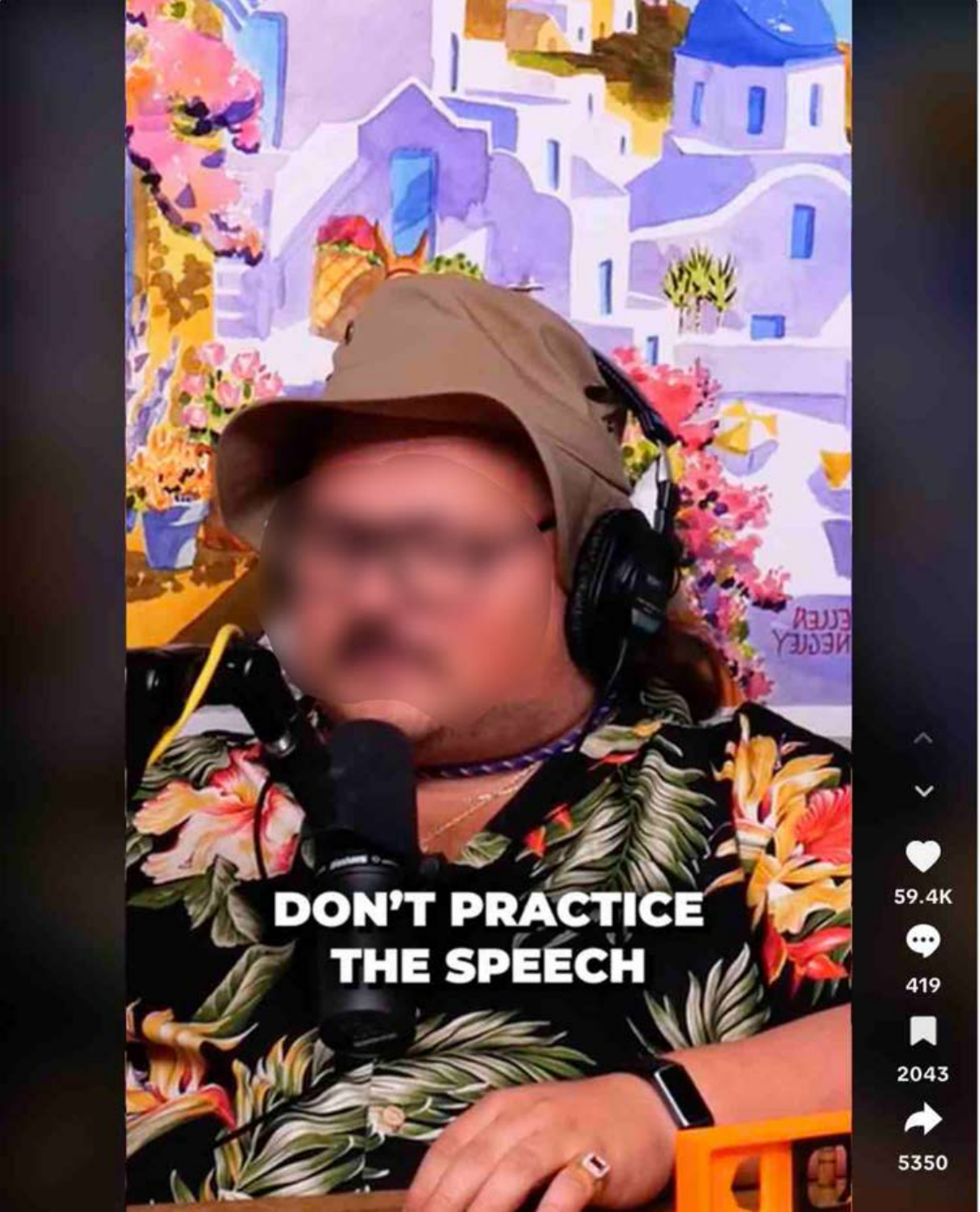}
\end{subfigure}

\vspace{4pt}
\Description{Screenshots of three videos from cluster 47 featuring people speaking into microphones, commonly in podcast settings.}
\caption*{\centering (b) Cluster 47: Videos of people talking into microphones. Videos by TikTok creators: joinsideplus \cite{micro_1}, calebhammercomposer \cite{micro_2}, stavvysworld \cite{micro_3} (Left-Right).}

\vspace{6pt}

% -------- Cluster 48 --------
\begin{subfigure}{0.31\columnwidth}
  \includegraphics[width=\linewidth]{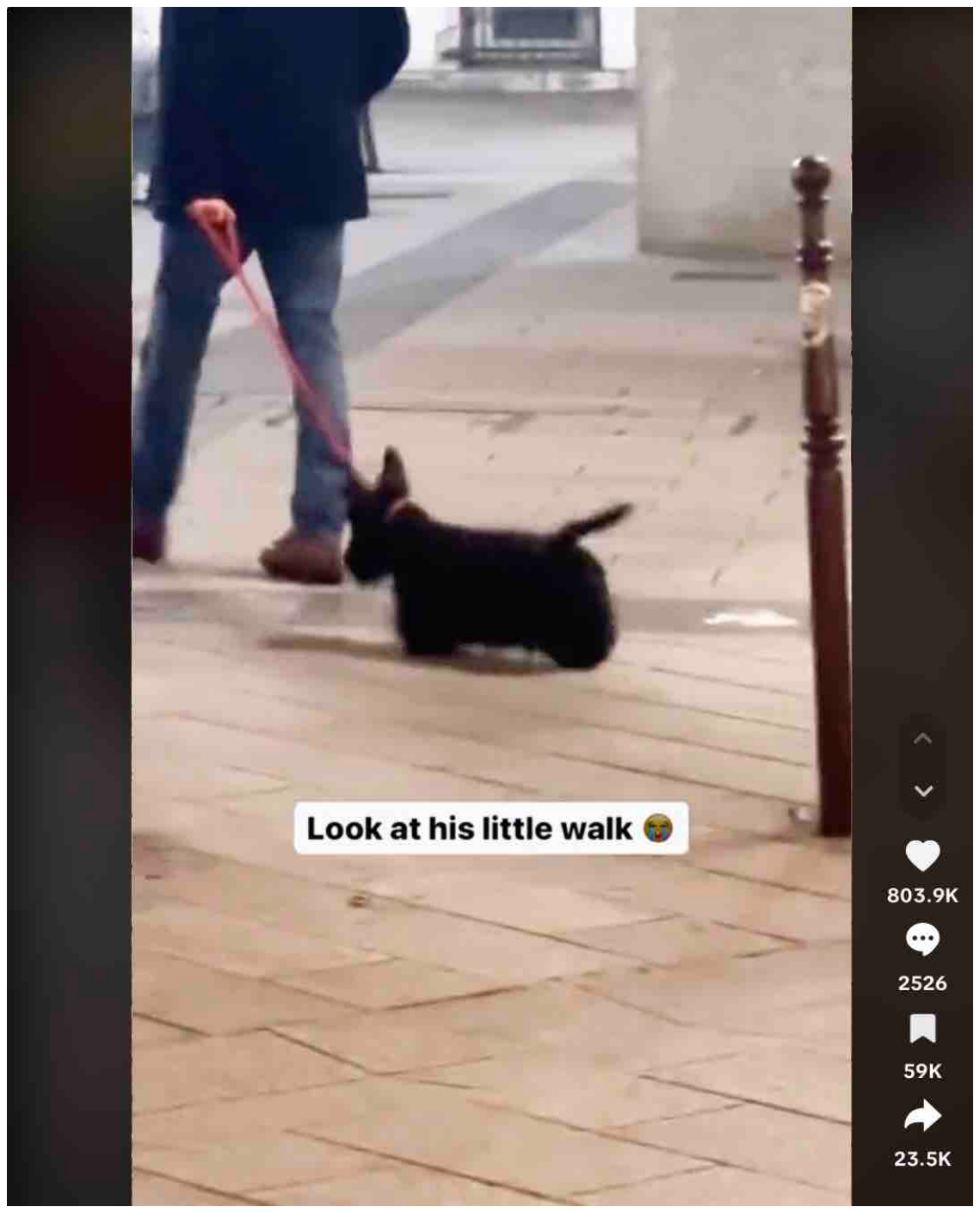}
\end{subfigure}\hfill
\begin{subfigure}{0.31\columnwidth}
  \includegraphics[width=\linewidth]{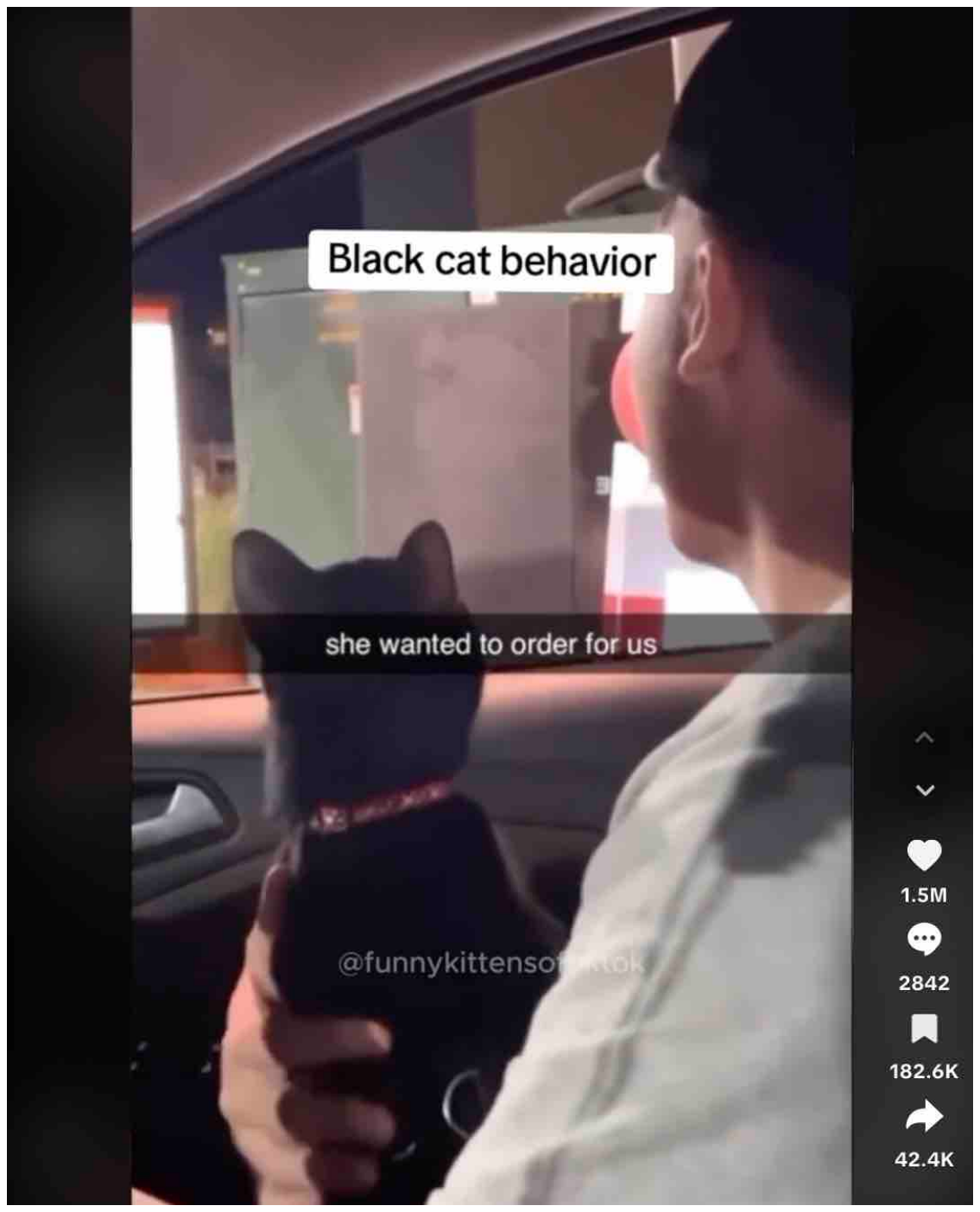}
\end{subfigure}\hfill
\begin{subfigure}{0.31\columnwidth}
  \includegraphics[width=\linewidth]{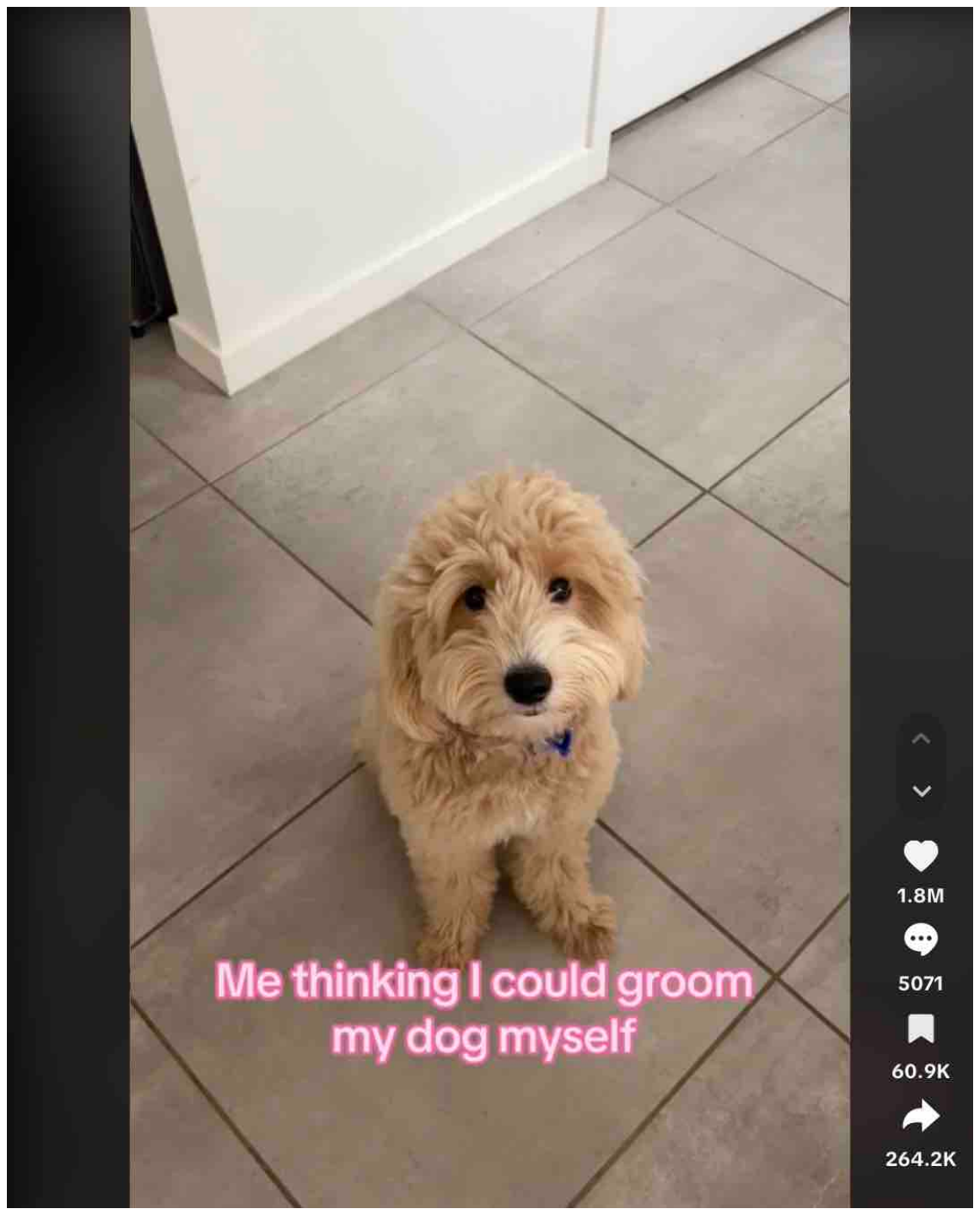}
\end{subfigure}

\vspace{4pt}
\Description{Screenshots of three videos from cluster 48 featuring pets interacting with their owners.}
\caption*{\centering (c) Cluster 48: Videos featuring pets. Videos by TikTok creators: ryflow\_people \cite{pet_1}, funnykittensoftiktok \cite{pet_2}, samaraspaulding \cite{pet_3} (Left-Right).}

\Description{Snapshots of videos grouped into the same semantic clusters. Each row shows representative examples from one cluster.}
\caption{\centering Examples of videos grouped into the same cluster.}
\label{fig:videossamecluster}
\end{figure}

\subsubsection{Clustering Videos}
\label{sec:kmeans}

Given the VCA vectors of a large number (millions) of videos, we need to identify different mechanisms to analyze this data. We choose to perform KMeans clustering~\cite{na2010research}. KMeans is an unsupervised learning algorithm that partitions a dataset into K clusters, where each data point belongs to the cluster with the nearest mean. For our use case, each VCA vector was a datapoint. We chose an unsupervised clustering approach because (a) it is non-trivial to define a taxonomy of social media videos, and (b) it requires substantial manual effort to manually label millions of videos with clusters.

\begin{figure}[h]
    \centering
    \includegraphics[width=\linewidth]{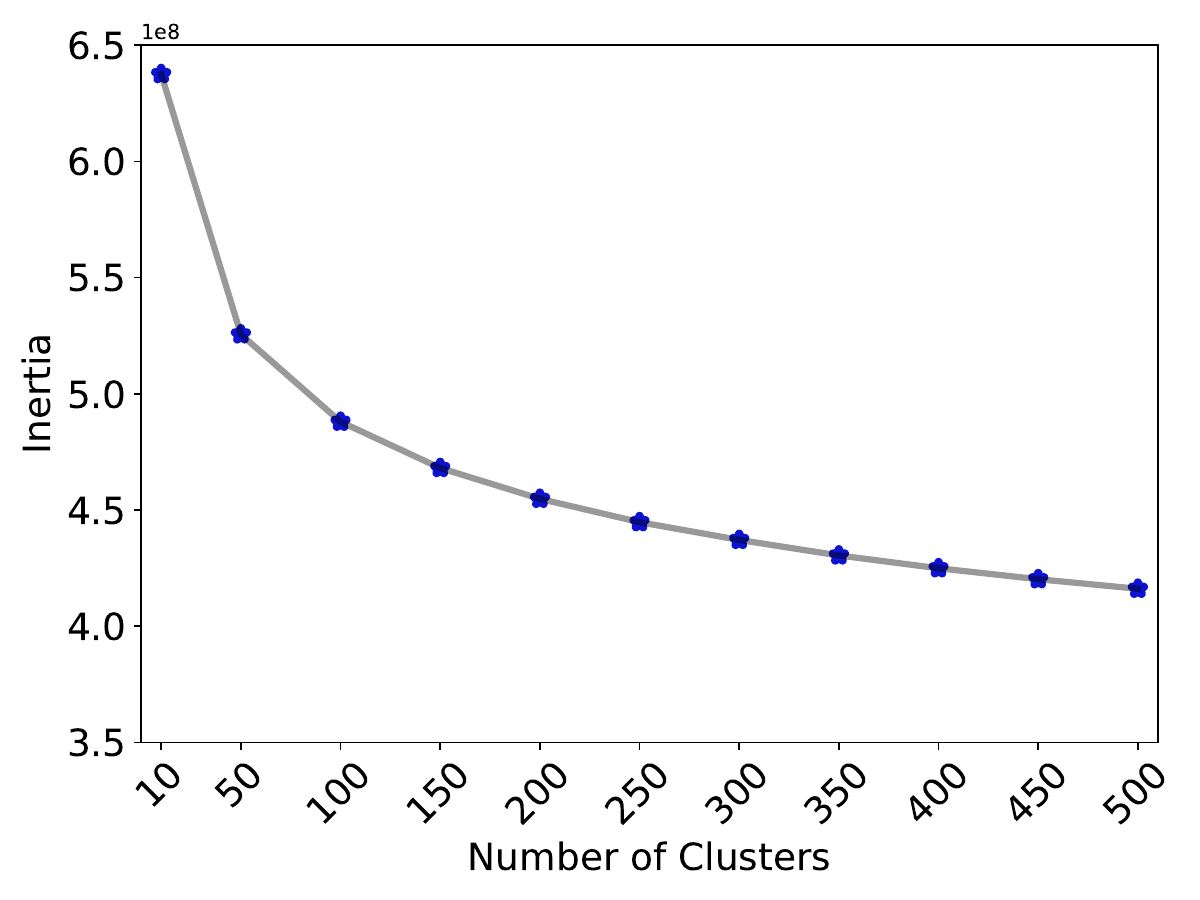}
    \Description{Line plot showing K-means inertia versus number of clusters. The x-axis ranges from 10 to 500 clusters, and the y-axis shows inertia values on the order of 10⁸. Inertia decreases steeply from K=10 to around K=100, then declines more gradually beyond K≈100, forming an elbow that indicates diminishing returns from adding more clusters.}
    \caption{\centering Elbow plot of K-means inertia as a function of the number of clusters 
K. The curve shows an elbow around 
$K=100$, after which additional clusters yield only marginal reductions in inertia.}
    \label{fig:elbow}
\end{figure}

 We selected the number of clusters using the elbow method on a subsample of 500,000 VCA vectors. We plot K-means inertia (i.e., sum of squared distances of samples to their closest cluster center - a lower inertia value suggests better clustering) as a function of 
K in Fig. \ref{fig:elbow}. We limit K between 10 and 500 to ensure that the final number of clusters is manageable by humans. The curve shows an elbow around 
$K=100$, where additional clusters yield only marginal reductions in inertia. We therefore chose 
$K=100$ as a trade-off between model fit and complexity.

With $K=100$, we find that the clustering scheme yields meaningful clusters. Some example clusters and representative videos are highlighted in Fig.~\ref{fig:videossamecluster}. Most clusters in our dataset appear coherent, except for 12 clusters. Some examples of identifiable and unclear clusters are in Table~\ref{table:imp_clusters}. All clusters are listed in the appendix. \enlargethispage*{16pt}

\begin{table*}[t]
\centering
\begin{tabular}{|c|c|}
\hline
\textbf{Identifiable Clusters} & \textbf{Topic} \\
\hline
1 & Videos of influencers interviewing other people on the streets \\
\hline
18 & Videos of people unboxing products and reviewing them; no person is seen on screen \\
\hline
44 & Make-up and hair tutorials by female influencers \\
\hline
72 & Clippings from talk shows and news rooms \\
\hline
\textbf{Unclear Clusters} & \textbf{Description} \\
\hline
52 & Unsure; videos focus on couples and games like Minecraft \\
\hline
65 & Unsure; some videos focus on travel, others focus on ASMR videos \\
\hline
\end{tabular}
\vspace{5pt}
\caption{\centering We classified the videos into 100 clusters. Of these, 12 clusters have topics that are not clearly identifiable, while the remaining clusters are well-defined and easily classified. We provide some examples in the table.}
\label{table:imp_clusters}
\end{table*}
}

\section{Results and Findings}

\begin{table}[h]
\centering
\begin{tabular}{|c|c|}
    \hline
    \textbf{RQ} & \textbf{Data} \\
    \hline
    \textbf{0} & VCA Vectors \\
    \textbf{1} & VCA Vectors, Browsing History \\
    \textbf{2} & VCA Vectors, Browsing History \\
    \textbf{3} & VCA Vectors, Browsing History, User Study Sessions \\
    \textbf{4} & User Study Sessions \\
    \textbf{5} & VCA Vectors, Browsing History \\
    \hline
\end{tabular}

\caption{\centering The data sources we used to conduct our analysis for each RQ.}
\label{table:toolbox}\vspace*{-10pt}
\end{table}

We address each RQ defined in \S\ref{sec:intro} sequentially by summarizing our key findings first and then explaining the methodology used to conduct our analysis. {At the end of each RQ, we also discuss a few implications of our results on the broader HCI domain.}  

Table \ref{table:toolbox} outlines the data sources used for the analysis of each RQ. The VCA vectors form an integral component in answering the RQs and in order to verify the alignment of the VCA vector with the video content, we conduct an additional user study that we describe in \S\ref{sec:rq1new}.

\subsection{Ecological Validity (RQ0): Are videos within a VCA cluster similar in content?}
\label{sec:rq1new}

To validate that the VCA clusters meaningfully group similar videos, we conducted a user study in which participants compared video pairs drawn from within and across clusters.

\noindent {\bf Key Takeaway:} Participants, on average, can differentiate between two videos belonging to the same VCA cluster vs two randomly sampled videos 84.9\% (95\% CI = [80.5\%, 89.2\%]) of the time. 

\begin{figure}[h]
    \centering
    \includegraphics[width=\linewidth]{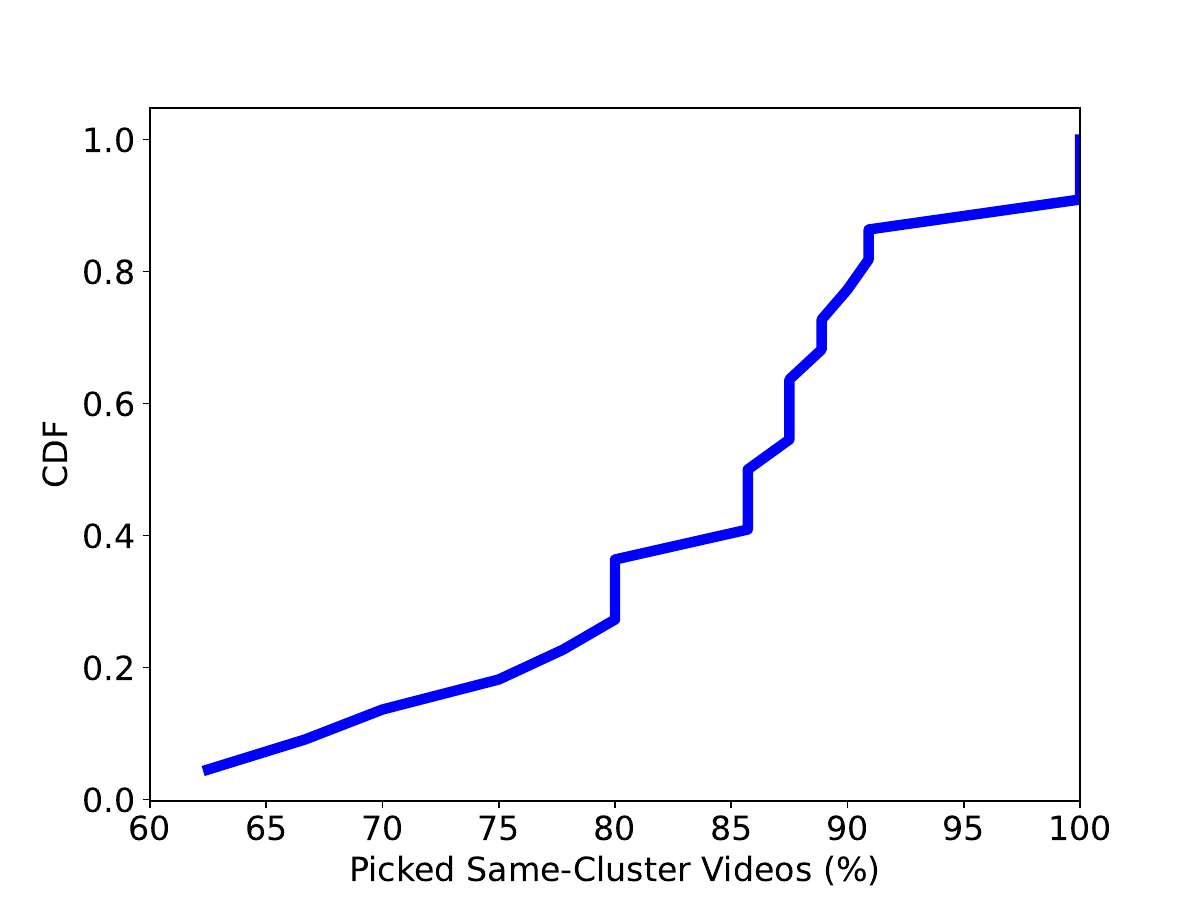}
    \caption{\centering The plot shows the CDF of the percentage of rounds where participants selected videos that were sampled from the same cluster as similar -- higher curve indicates better accuracy. Participants can verify that similar videos are in the same cluster on average 85\% ($\sigma=10\%$) of the time. 
    }
    \Description{Cumulative distribution function (CDF) plot of the percentage of rounds in which participants selected videos from the same cluster as similar. The x-axis ranges from approximately 60\% to 100\%, and the y-axis shows the cumulative fraction of participants. The curve increases gradually up to about 80\%, then rises steeply between roughly 85\% and 90\%, indicating that most participants selected same-cluster videos in a high fraction of rounds.}
    \label{fig:similarity_us}
\end{figure}

\subsubsection{Method}

We verify that our clustering approach creates meaningful groups of videos by conducting in-person proctored interviews on a university campus (N=22). We recruited participants by advertising the study among students and anyone who watched short videos was eligible.  Each participant was shown 44 videos in rounds of 4. The 4 videos were shown in sets of 2 - exactly one of the sets had videos sampled from the same cluster; the other set had videos from different clusters. After watching both sets of videos, participants were asked to pick whether they thought that set 1 had more similar videos or set 2 (we randomly picked which set had similar videos ahead of time). Participants were instructed to identify videos they considered similar, reflecting the way they would judge similarity while watching videos for e.g. similar audio, similar topic, same individuals, etc.

If any video was unavailable (video was deleted or private), users could skip the round. {We used all 100 clusters in this experiment. Had we excluded the "unclear" clusters from Table \ref{table:imp_clusters}, we expect our results may have had 
even higher 
accuracy.}

Fig. \ref{fig:similarity_us} plots 
a CDF of the accuracy of each participant, measured as the percentage of rounds in which they selected the correct set with same-cluster videos. 

We conduct a one-sample t-test comparing participant accuracy against the chance level of 0.5. Participants identified the same-cluster videos with a mean accuracy of 0.85 ($\sigma=0.10$, 95\% CI = [80.5\%, 89.2\%]), which was  statistically significant, t(21) = 16.18, p < 0.001.

\subsection{RQ1: How does the content of recommendations made by TikTok's recommendation algorithm evolve over time at short and long timescales?}
\label{sec:rq1}

Understanding of the behavior of TikTok's recommendation algorithms for each individual user across time has remained limited. Our Video Content Analysis (VCA) measurement tool in combination with watch history allows us to answer a key insight: for a given user, how does the content of TikTok's recommendations change over varying time scales of days and months?

\noindent {\bf Key Takeaway:} 
\textbf{\textit{User watch time is  clustered in the short-term, but over long-term these clusters change.}}
 On any single day, about 50\% of watch time is focused on the user's top 5 clusters even though about 52.5 clusters ($\sigma=8.42$) are recommended daily. However, the identity of these top 5 clusters shifts daily. Across six months, users spend only 22\% of their time in their top 5 clusters, indicating that while the system exploits heavily each day, the set of exploited clusters rotates over time, producing long-term exploration.

\subsubsection{Method}

RQ1's analysis is based on the browsing history files that each contain videos recommended to a user over time as well as the watch duration of the video by the user (example in Fig. \ref{fig:bh-example}). For each user, we truncate the length of the browsing history to 160 days because about 20\% of the users do not have histories beyond 160 days. We also discard all videos with watch time more than 3 times the duration of each video to remove bias (e.g., due to a user moving away from the app). On average, this left us with about 36,000 videos seen by each user.

By clustering the videos in the watch histories using our measurement tool, we are able to identify the top interests of a user by accumulating the time spent watching videos in each of the VCA clusters. To identify a user's top N clusters, we identify the N clusters that they spent the most time watching videos from. For e.g., a user's top 1 cluster is the VCA cluster they spent the most time watching videos from. To obtain the ratios of time spent in these top N clusters, we divide the time spent watching videos in the cluster by the total time spent watching all videos for each user. 
We find patterns by looking at the clusters recommended to each user and the top clusters of each user at both daily and six-month time scales.

\subsubsection{Results at the granularity of a day}
\label{subsubsec:daily}

\begin{figure*}[h]
    \centering
    \begin{minipage}[b]{0.5\textwidth}
        \centering
        \begin{subfigure}[b]{\textwidth}
            \includegraphics[width=\linewidth]{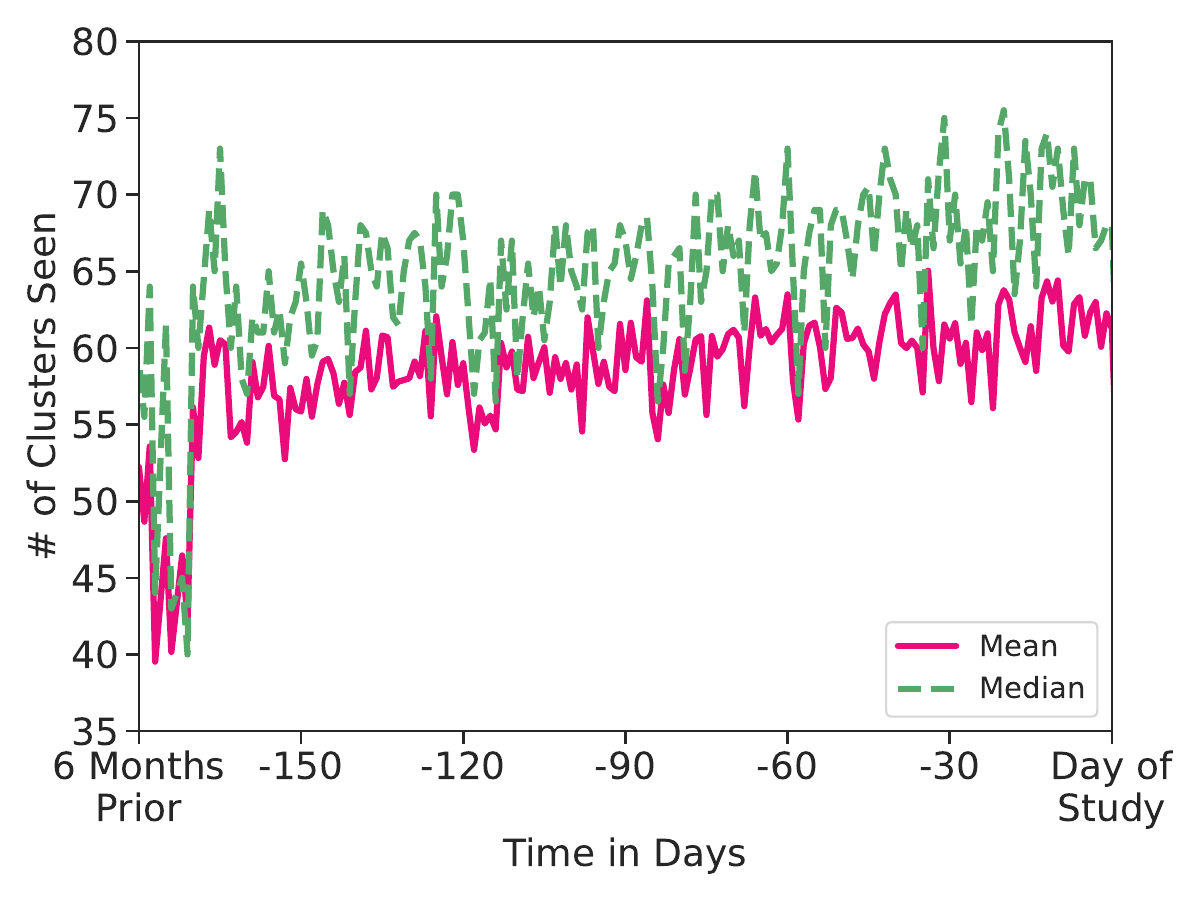}
            % \vspace{-20pt}
            \Description{In this figure, we describe the average number of clusters recommended to users by TikTok on a day-to-day basis. The figure is a plot with one line for the mean number of recommended clusters and one for the median. The interquartile range of the average number of clusters is between 44 to 60. The mean is 52.5 clusters and the median is 56.}
            \caption{\centering 
            % An average user will be recommended around 60 clusters every day by TikTok.
            The number of clusters recommended to TikTok users in our study every day.}
            \hfill
            \label{fig:day_avg_num_cluster}
            \vspace{1cm}
        \end{subfigure}
    \end{minipage}
    \hfill
    \begin{minipage}[b]{0.45\textwidth}
        \centering
        \begin{subfigure}[b]{\textwidth} 
            \includegraphics[width=\linewidth]{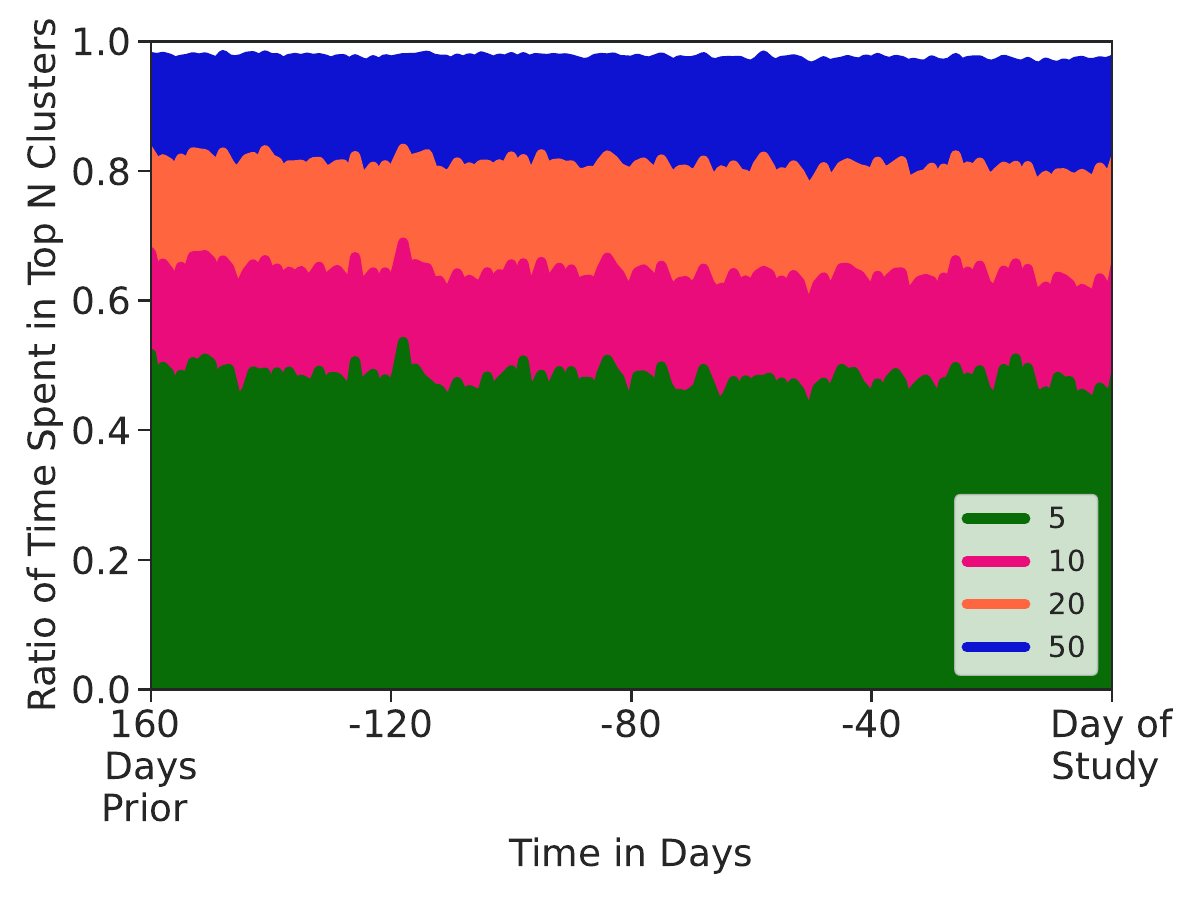}
            \vspace{-17pt}
            \Description{In this figure, we illustrate that the time spent in top N clusters recommended remains consistent across time for an average user. The X axis represents the number of days leading up to the user study, from 160 days prior to the day of the study. The Y axis is  the ratio of time spent on top N clusters, where N is 5, 10, 20 and 50. The Y axis range from 0 to 1.0.}
            \caption{\centering Ratio of time spent by TikTok users in their top N clusters every day.}
            \hfil
            \label{fig:day_time_in_top_cluster}
        \end{subfigure}
        % \hfill
        \begin{subfigure}[b]{\textwidth}
            \centering
            \begin{tabular}{|c|c|}
                \hline
                \textbf{Top N Clusters} & \textbf{Avg No. of New Clusters per Day} \\
                \hline
                5 & 3.96 ($\sigma=0.11$) \\
                % 3.9597554598303297 0.1095172885126284
                10 & 7.22 ($\sigma=0.20$) \\
                % 7.220878214374222 0.19641830484593442
                20 & 12.30 ($\sigma=0.39$) \\
                % 12.300354255393133 0.3867751677170055
                50 & 18.04 ($\sigma=0.77$) \\
                % 18.048183500253838 0.769866215590278
                \hline
            \end{tabular}
            \caption{\centering The average churn of labels in the top N clusters recommended is between 36\% (N=50) to 79\% (N=5).}
            \label{table:day_churncluster}
        \end{subfigure}
    \end{minipage}
    % \hfill
    
    \Description{In Fig 8, we have three subfigures which describe findings related to the number of clusters recommended to users on a day-to-day basis.}
    \caption{\centering TikTok recommends users a variety of content on a day-to-day basis.}
    % How an average TikTok user interacts with 100 clusters 
    \label{fig:day_clusters}
\end{figure*}

We find that TikTok's recommendations for each user spans a large number of clusters. On average, in a single day, TikTok recommends videos spanning 52.5 clusters to each user with a standard deviation of 8.42 clusters. The median number of clusters recommended was 56 (IQR = 45–60) (Fig. \ref{fig:day_avg_num_cluster}). However, as shown in Fig.~\ref{fig:day_time_in_top_cluster}, a user spends a large fraction of their time watching videos from just a few clusters. % (likely representing their interests). 
About half the time is spent on videos from just five (that user's top 5) clusters, and 80\% of the time is spent on videos from the user's top 20 clusters. For all values of N, regression slopes were effectively zero (|$\beta$| < 0.0004), indicating that the proportion of time spent in the top N clusters stayed consistent over time.

At the same time, users' recommendations experience a lot of churn. In Fig.~\ref{table:day_churncluster}  we find that (for a user) on average, about 4 out of the top 5 clusters are different even between consecutive days. Similarly, an average 70\% of top 10 clusters and 62\% of top 20 clusters change across any two days.

\subsubsection{Results on a 6-month scale}
\label{subsubsec:entirebh}

    \begin{figure}[h]
    % #{0.45\textwidth}
        \centering
        \includegraphics[width=\linewidth]{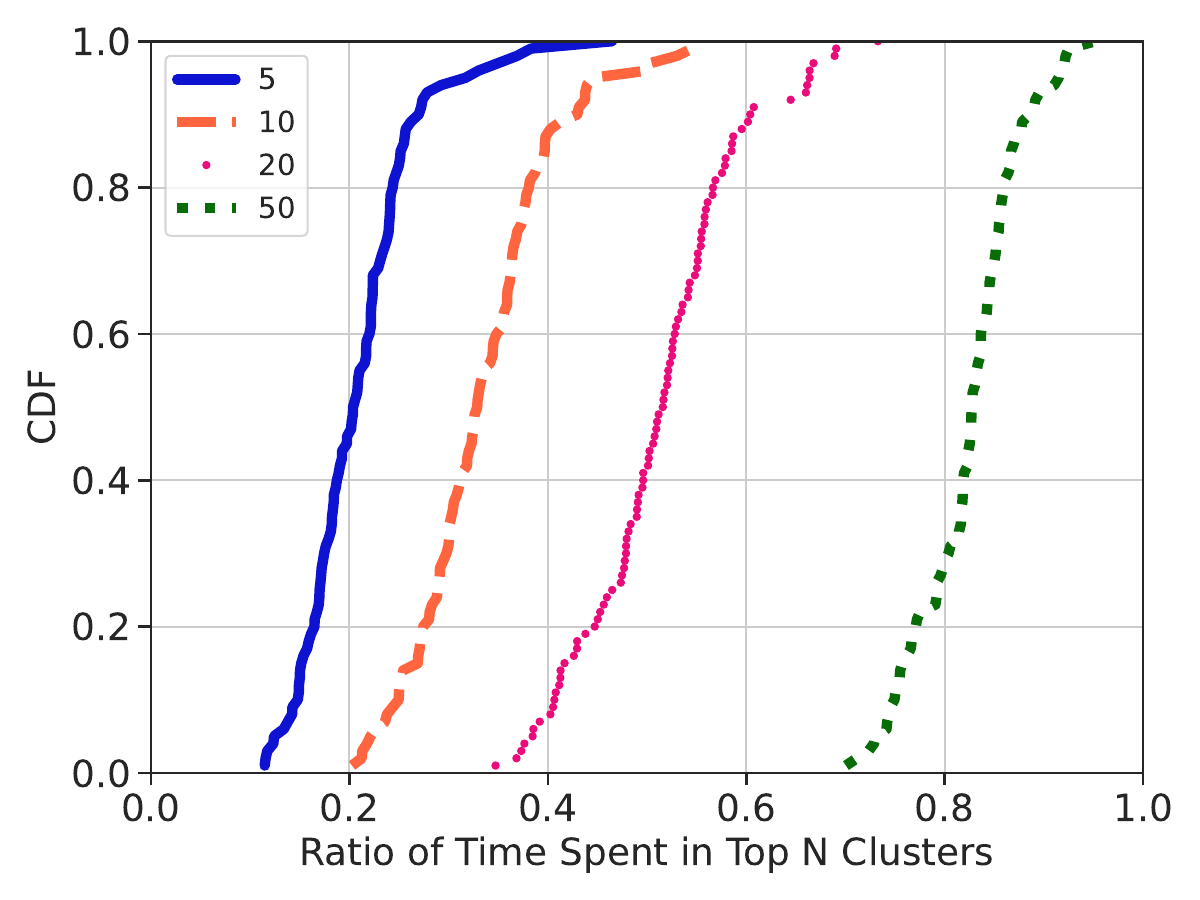}
        \Description{In this figure, we depict the ratio of time spent by a user in their top N clusters, where N takes the value of 5, 10, 20 and 50. The X axis is the ratio of time spent on the top N clusters. The values on the X axis range from 0 to 1.0. The Y axis represents the CDF, with values from 0 to 1. Each of the curves is upward sloping. As N increases in number, the curves shift rightward. }

    \caption{\centering Ratio of time spent by a user in their top recommended clusters over the course of their entire browsing history.}
    \label{fig:overall_clusters}
\end{figure}

We repeat the same temporal analysis of recommendations across a longer period of 6 months to analyze the time spent by users in their top N recommended clusters across 6 months. The median user spends 21\% of their time in videos from top 5 clusters---see Fig. \ref{fig:overall_clusters}. 
Similarly, a median user spends 33\%, 52\%, and 83\% of their time watching videos from top 10, 20, and 50 clusters, respectively. Comparing these ratios to the day-to-day churn, we observe the time spent in the top 5 clusters is considerably lower in the 6-month study. This substantiates our observation of churn in the video topics that TikTok recommends to users. 

\subsubsection{An Extended Insight: TikTok's algorithm tailors recommended videos over time}

\begin{figure*}[h]
\centering

\begin{subfigure}[b]{0.49\textwidth}
  \centering
  \includegraphics[width=\linewidth]{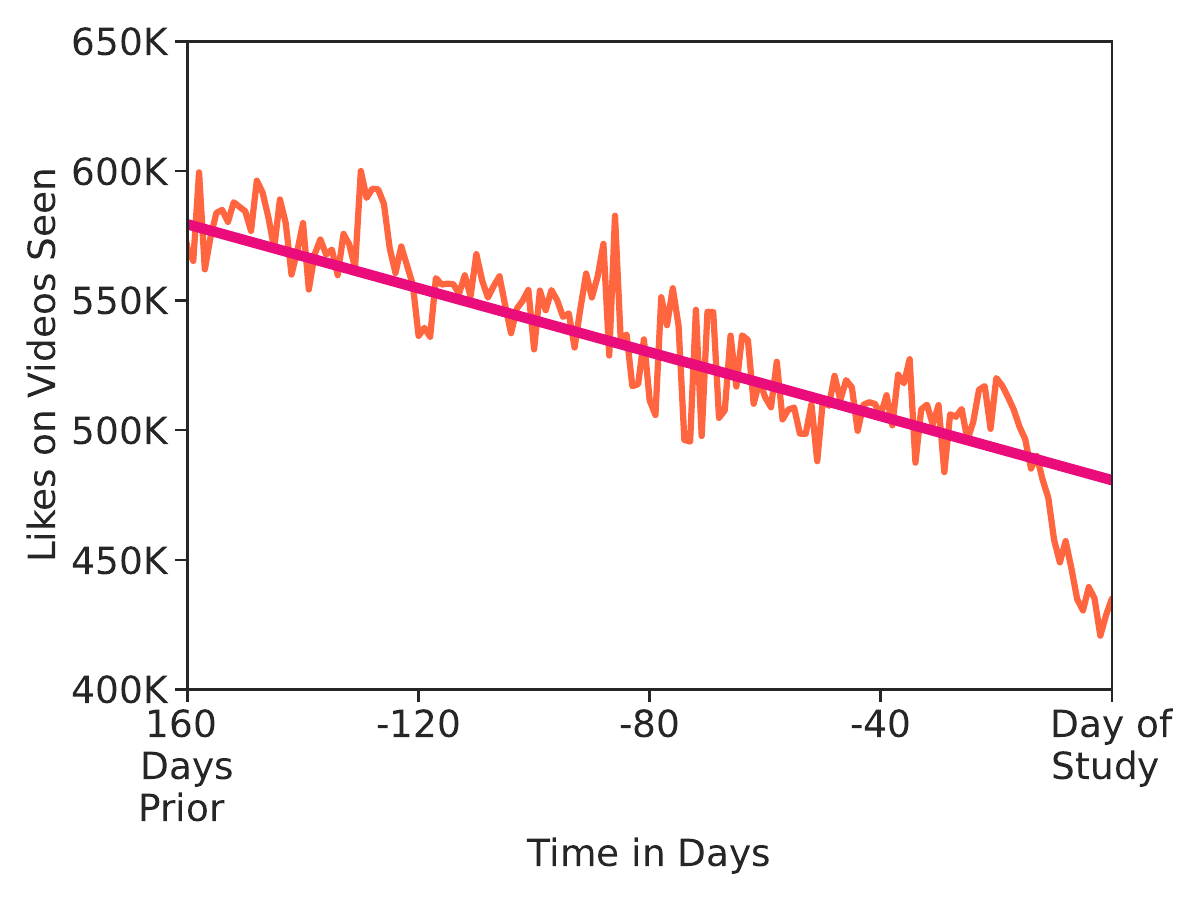}
  \Description{Relationship between time and video likes. The x-axis shows days leading up to the study, and the y-axis shows the number of likes. The trend decreases over time.}
  \caption{\centering The average likes on videos seen each day decrease over time for all users. Linear regression confirms a statistically significant negative relationship (slope = -617 likes/day, $p < .001$).}
  \label{fig:day_likes_in_a_session}
\end{subfigure}
\begin{subfigure}[b]{0.49\textwidth}
  \centering
  \includegraphics[width=\linewidth]{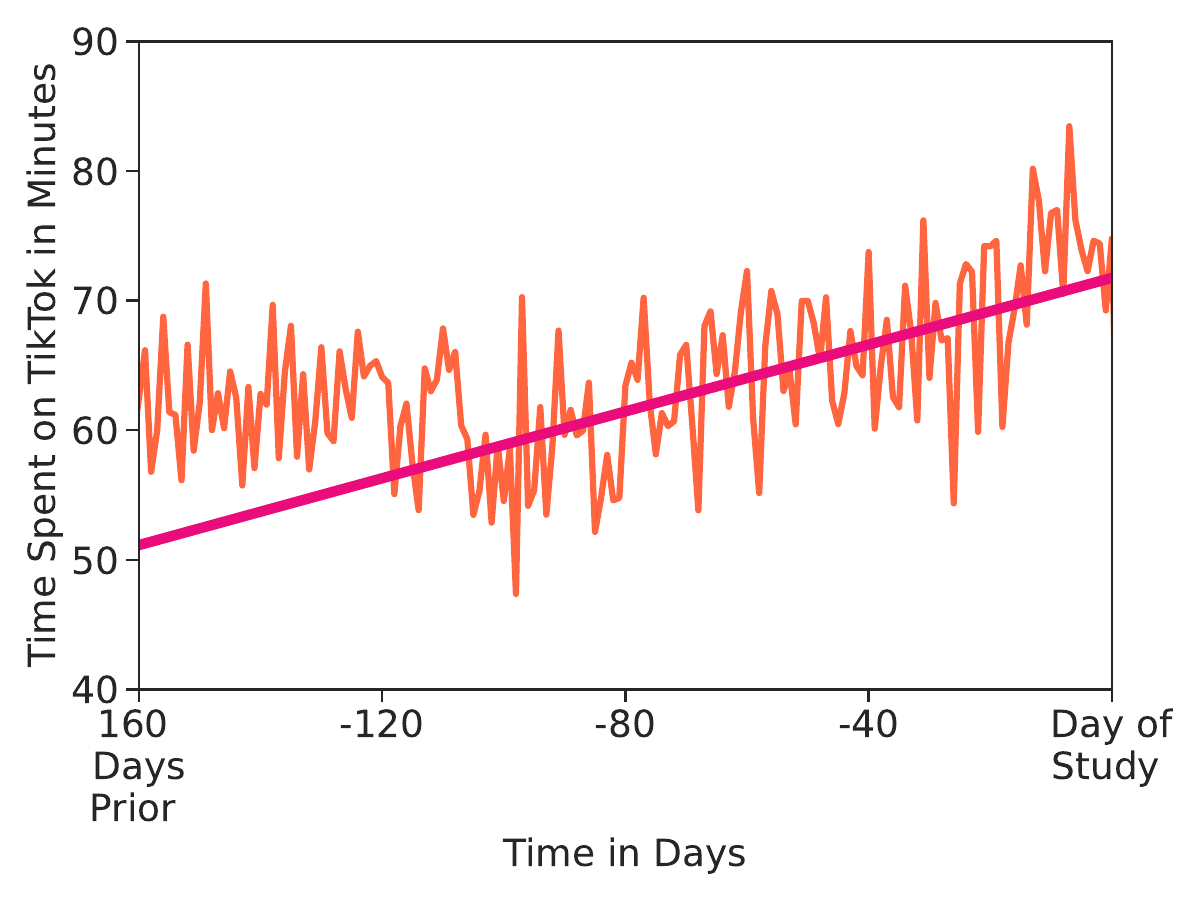}
  \Description{Relationship between time and duration spent on TikTok. The x-axis shows days leading up to the study, and the y-axis shows time spent in minutes. The trend increases over time.}
  \caption{\centering The average time spent watching videos increases over time for all users. Linear regression shows a statistically significant positive relationship (slope = 0.128 minutes/day, $p < .001$).}
  \label{fig:day_timespent}
\end{subfigure}

\Description{Two trends over time: average likes on viewed videos decrease, while total time spent watching videos increases.}
\caption{\centering Over time, TikTok users are recommended less popular videos even as their total time spent on the app increases.}
\label{fig:day_likes_time}
\end{figure*}

Any recommender algorithm like TikTok's is aimed at learning about a user over time \cite{herrman2019tiktok}. Across all users, a trend clearly visible in Fig. \ref{fig:day_likes_in_a_session} is that the average likes received by videos that are being recommended,  (i.e., video popularity) decreases consistently with time. We believe this indicates that as TikTok {\it gets to know a user more, it recommends less popular but niche videos}. At the same time 
(despite watching less popular content) 
the average time spent on TikTok actually {\it increases} over time (Fig. \ref{fig:day_timespent}), indicating stable or increased engagement with the app. Alternatively, TikTok's recommendations become more fine-tuned to the user over time, recommending less popular but more engaging (for the user) content.

\subsubsection{HCI Implications}
{Our analysis shows that recommendations are highly clustered over small timescales i.e., a lot of watch time is spent in only a few recommended clusters. However, the identity of these clusters changes over time and so, long-term behavior spans a much broader set of topics. This observation shows that a user's short video algorithmic experience is inherently multi‑timescale: users may inhabit narrow “micro‑bubbles” in a given session while still being exposed to diverse content across time. This pattern complicates the usual understanding of “filter bubbles” by revealing both short‑term bubbles and long‑term diversity in content. This further motivates the need for a) metrics that account for temporal dynamics of diversity, and b) methods that help users understand and navigate how their sequence of video recommendations evolve over time,  for e.g., using visualizations of interests across time.}

\subsection{{RQ2: What is the relationship between user interactions (likes and shares) and recommendations made by the algorithm?}}
\label{sec:rq2}

TikTok users commonly assume that interacting with a video sends a stronger signal to the algorithm (than, say, merely watching it). 
This section attempts to unearth this key insight: when a user interacts with a video (like or share): (i) does the interaction affect immediately recommended videos, and (ii) is the video interacted with similar to recently shown videos? 

\noindent {\bf Key Takeaway:} Contrary to popular perception, we find that after a user interacts with a video (like or share), there is little change in future recommendations. 
% over both short and long time spans
However, we find that \textbf{\textit{recently viewed content tends to influence a user's inclination to like or share a video}}.

\subsubsection{Method}

\begin{figure}[h]
\centering

\begin{subfigure}{0.33\columnwidth}
  \centering
  \includegraphics[width=\linewidth]{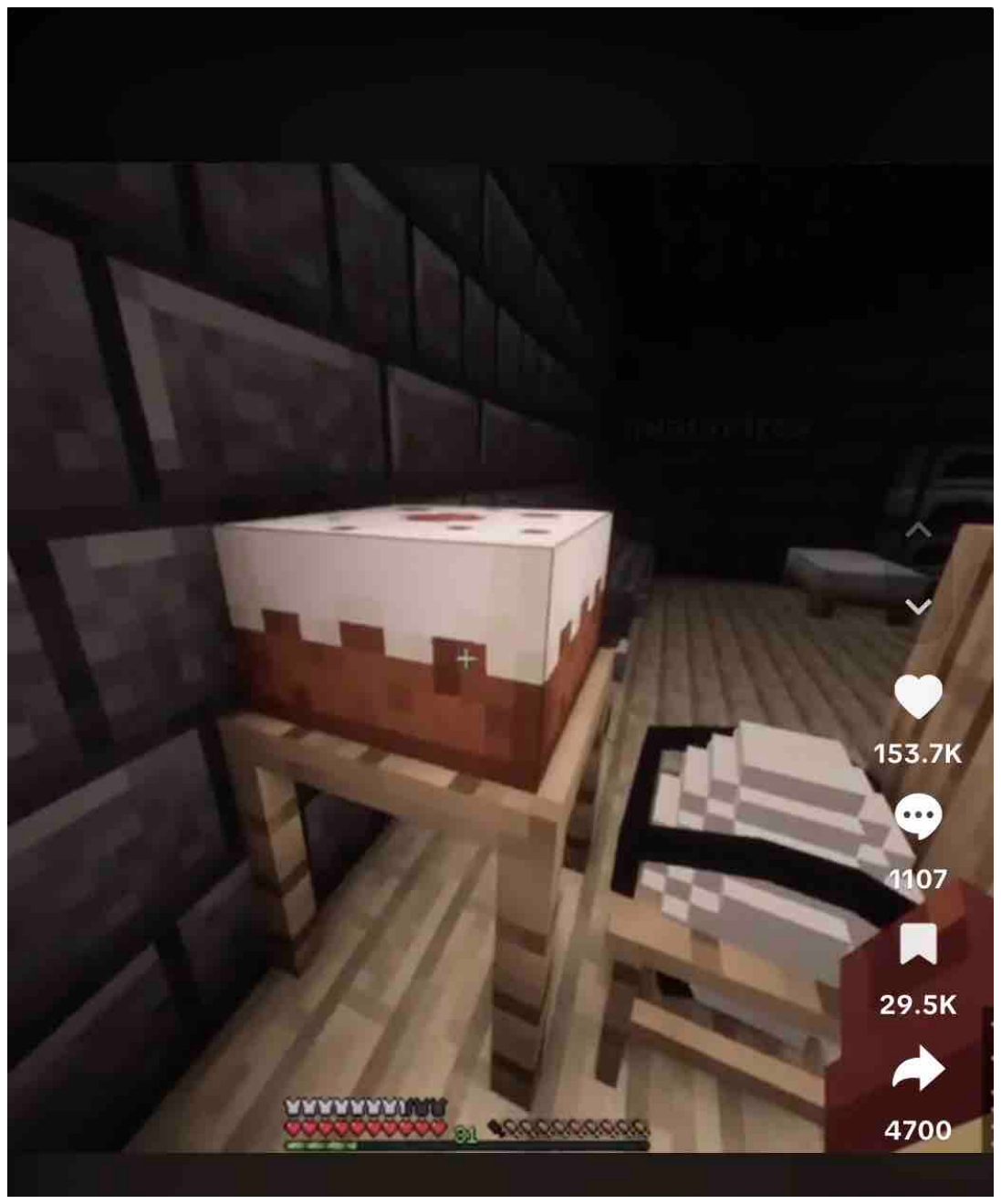}
  \caption{\centering An example video \cite{tiktok_mine_1}.}
  \label{fig:mine_a}
\end{subfigure}\hfill
\begin{subfigure}{0.33\columnwidth}
  \centering
  \includegraphics[width=\linewidth]{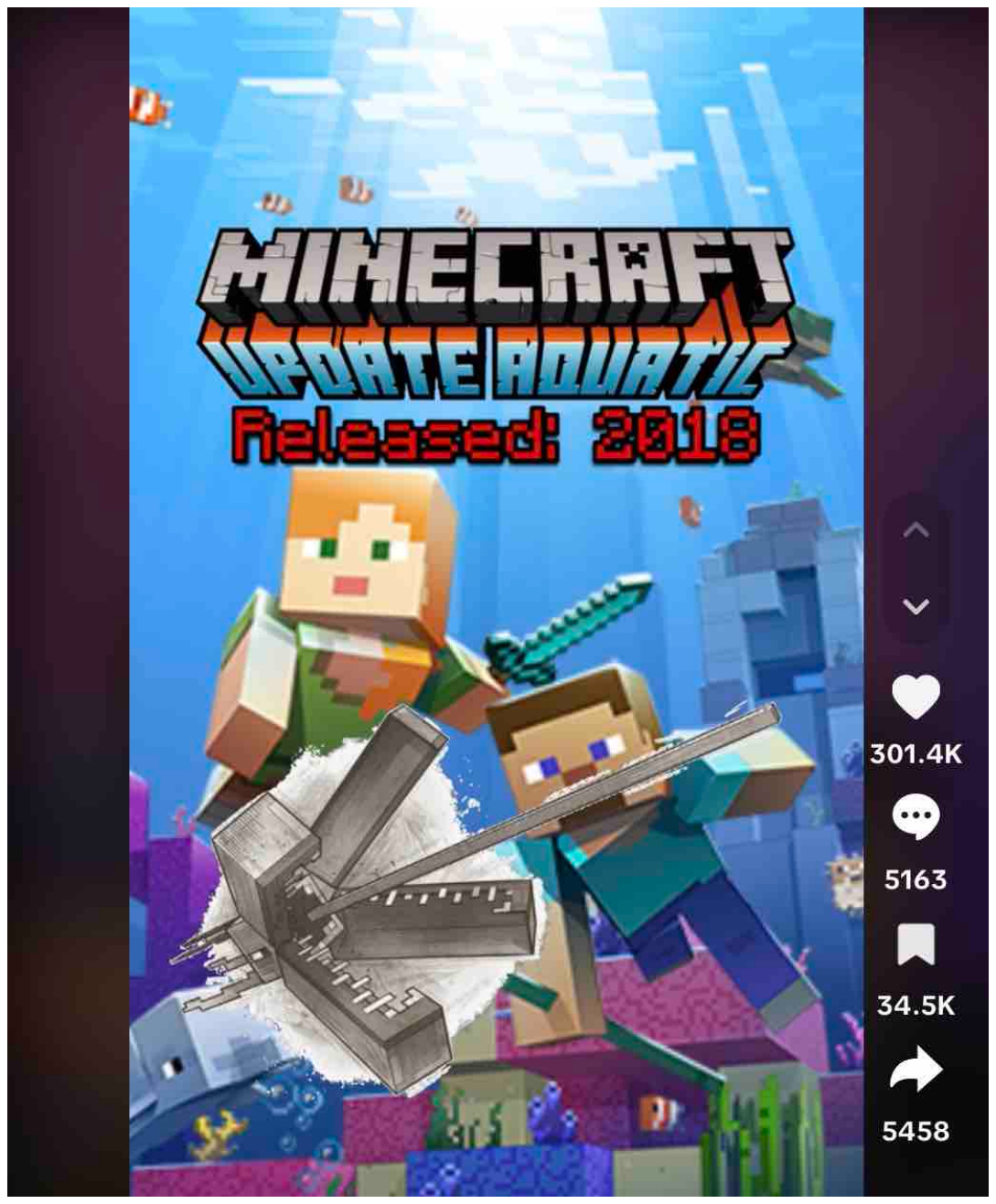}
  \caption{\centering A video \cite{tiktok_mine_2} similar to (a).}
  \label{fig:mine_b}
\end{subfigure}\hfill
\begin{subfigure}{0.33\columnwidth}
  \centering
  \includegraphics[width=\linewidth]{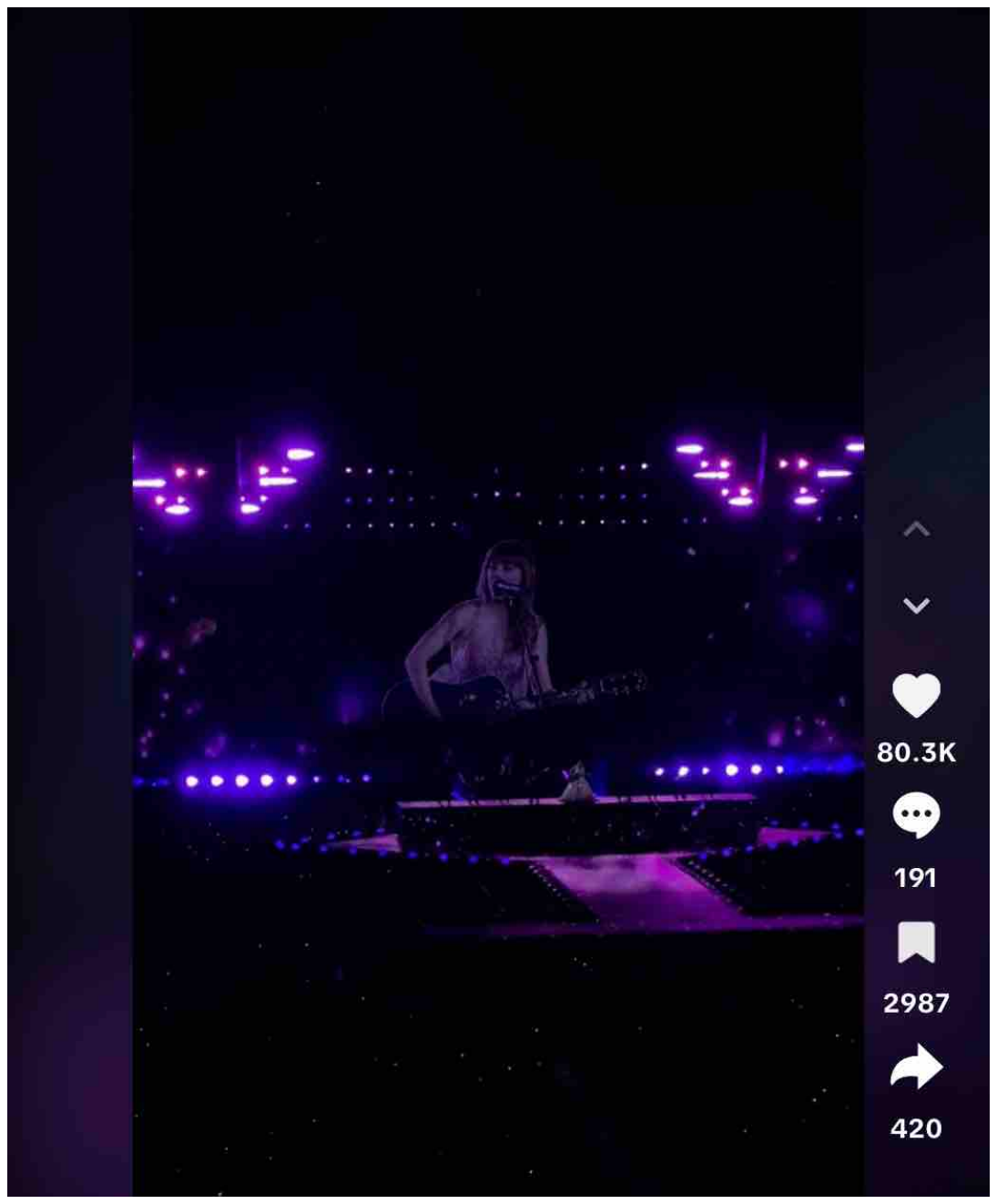}
  \caption{\centering A video \cite{tiktok_dissimilar} not similar to (a).}
  \label{fig:concert}%\vspace*{20pt}
\end{subfigure}

\Description{Screenshots of three public TikTok videos. Videos (a) and (b) depict Minecraft gameplay, while (c) shows a concert performance with stage lighting and a musician playing guitar.}
\caption{\centering Three example TikTok videos. Videos (a) and (b) are video gaming related, whereas video (c) is from a musical event. The Euclidean distance between the VCA vectors of (a) and (b) is nearly half that between either gaming videos and the event video. Videos by TikTok creators: (a) \_miaundrea, (b) nightwing7974, (c) csyrd.}
\label{fig:exampless}
\end{figure}

We use the Euclidean Distance metric to quantify distances between VCA vectors, i.e.,  we measure differences between videos by using our mathematical representation of their contents. A larger distance between videos means that they are further apart in content. Similar videos have a smaller distance between them. Fig. \ref{fig:exampless} shows three examples of videos---the first two which are similar yield a distance of 28.01 while the third video which is quite different is at respective distances of 58.03 and 57.29. We calculate the distance of a liked (or shared) video with videos recommended right after the interaction (\textit{future}) or the videos recommended right before the interaction (\textit{past}). To get the distance of a video with its neighboring k videos, we calculate the distance of the video with each of its k neighbors that were available.

\begin{figure}[h]
    \centering
    \includegraphics[width=\linewidth]{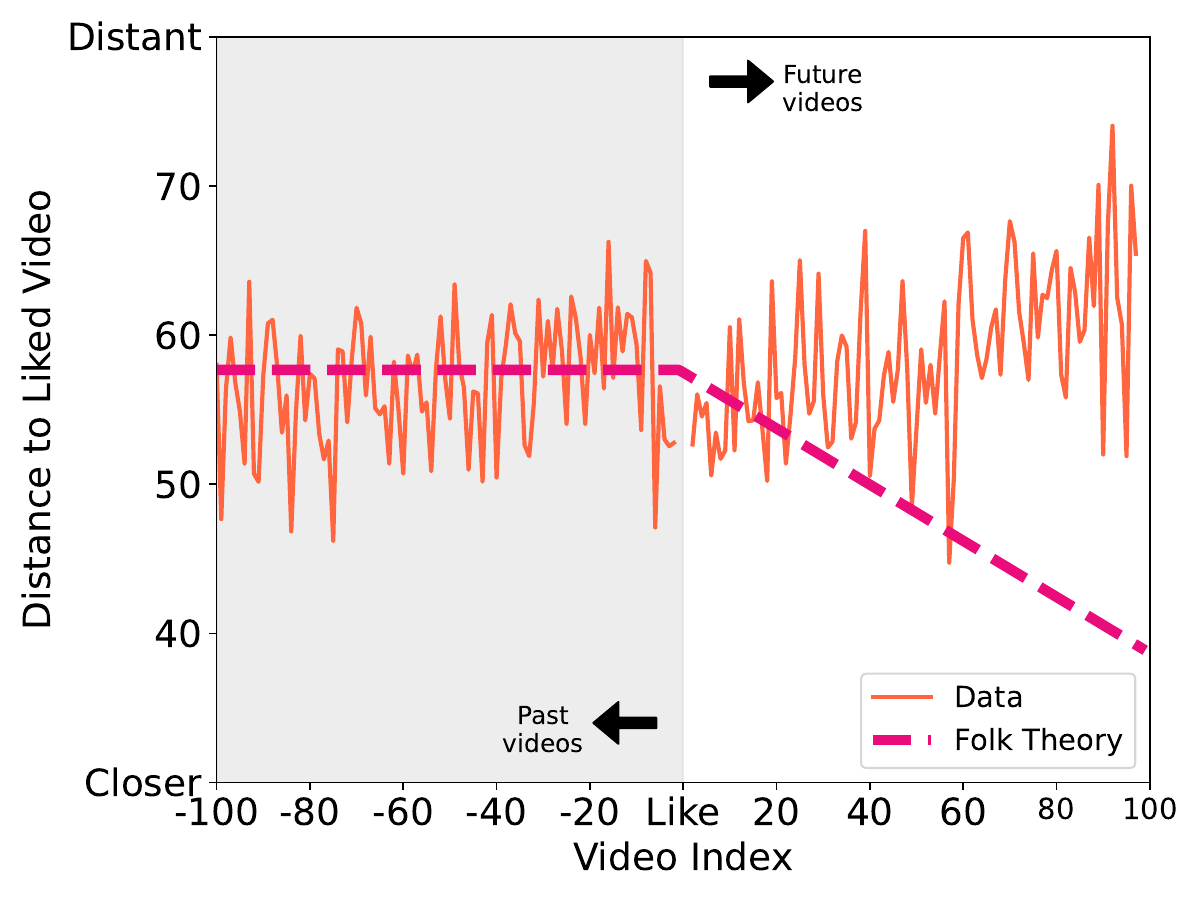}
    \Description{In this figure, the X axis represents the video index representing the 100 neighboring videos in the past and future. The X axis values range from -100 to 100, where the positive values represent videos in the future, and negative values represent videos in the past. The Y axis represent the distance to the liked video. The Y axis values range from 40 to 70. We draw a comparison between the data and the folk theory. The data shows that the relationship between increase in video index and distance to liked video is largely positive, whereas folk theory implies the relationship after a point is downward sloping.}
    \caption{\centering Distance between a real user's liked video and its 100 neighboring videos in the past and future. The expected pattern, based on folk theories, is also shown.} 
    
    \label{fig:like_theory_example}
\end{figure}

\begin{figure*}[t]
\centering

\begin{subfigure}{0.49\textwidth}
  \centering
  \includegraphics[width=\linewidth]{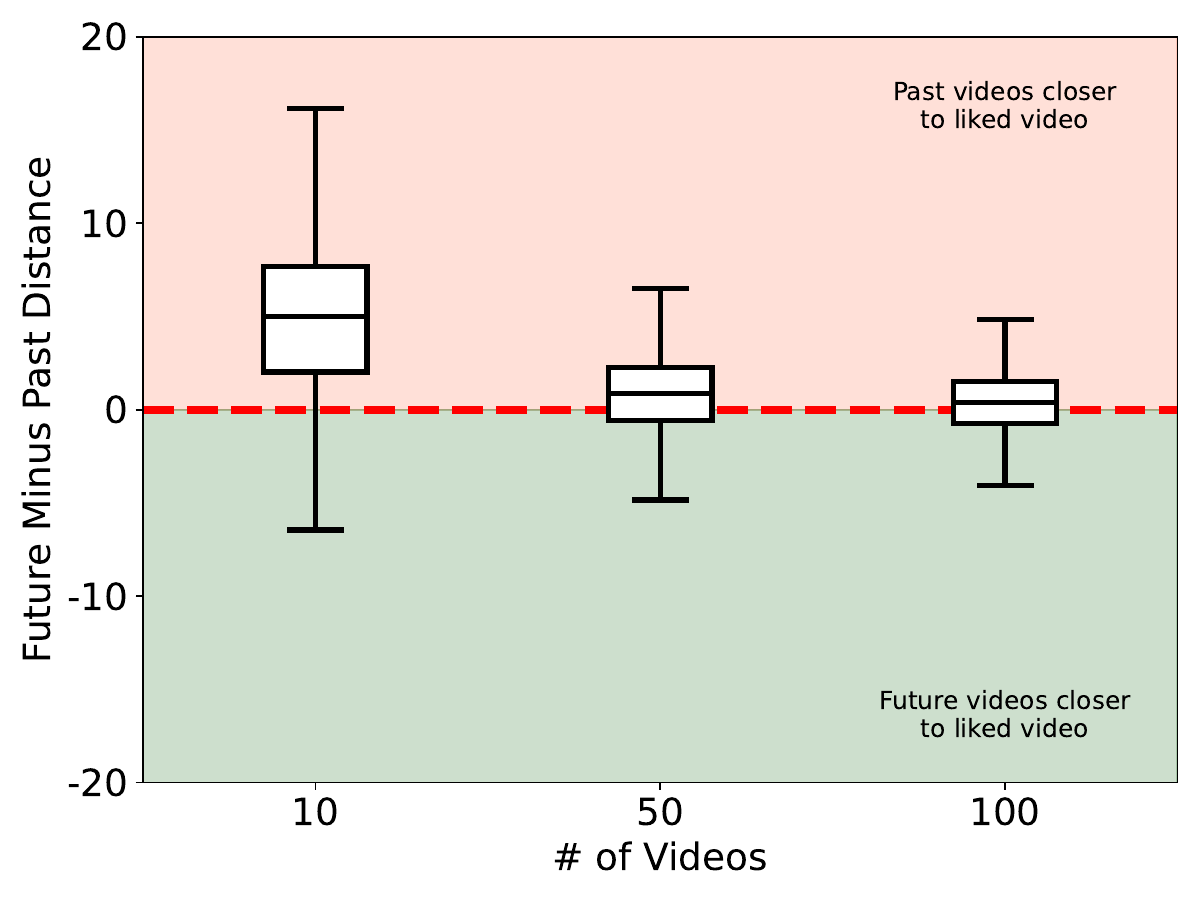}
  \Description{Future–past distance distribution for videos neighboring a liked video. The x-axis shows the number of neighboring videos (10, 50, 100), and the y-axis shows the future–past distance.}
  \caption{\centering Users see videos more similar to a video they \emph{liked} in the past than in the future.}
  \label{fig:likedist}
\end{subfigure}
\begin{subfigure}{0.49\textwidth}
  \centering
  \includegraphics[width=\linewidth]{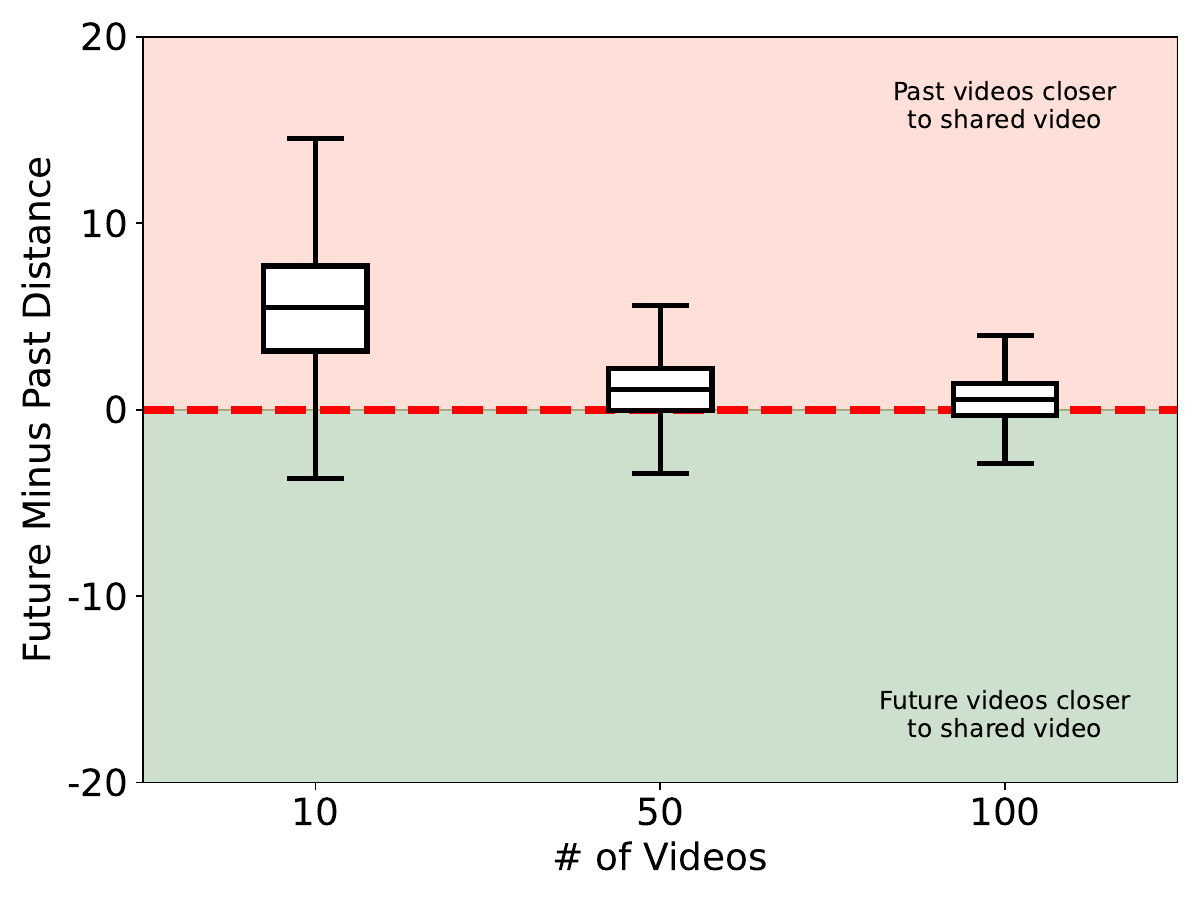}
  \Description{Future–past distance distribution for videos neighboring a shared video. The x-axis shows the number of neighboring videos (10, 50, 100), and the y-axis shows the future–past distance.}
  \caption{\centering Users see videos more similar to a video they \emph{shared} in the past than in the future.}
  \label{fig:sharedist}
\end{subfigure}

\Description{Box plots showing the distribution of future–past distances for neighboring videos. Boxes represent the interquartile range with the median marked, and whiskers extend to 1.5 times the interquartile range.}
\caption{\centering Once a video is liked/shared, the distance between X (X = 10, 50, 100) videos recommended after does not decrease; i.e., further video recommendations are not close to the liked/shared video. In fact, the past X recommendations are closer to the liked/shared video. In the box plot, the box represents one quartile below and above median, and the whiskers extend to 1.5 the interquartile range.}
\label{fig:like_share}
\end{figure*}

\subsubsection{Folk Theory}
\label{sec:shorttermfolk}

We have observed it is common for TikTok users to assume that \textit {if they interact with a video such as when they like, share, or comment on it, their future recommendations are immediately influenced by the contents of the video} \cite{tiktokrecom1, tiktokrecom2}. We call this the \textit{ ``Post Interaction Anticipation''}  (PIA) Folk Theory. {Past research supports the PIA folk theory qualitatively by reporting user beliefs about interactions with recommendation systems immediately shaping future recommendations \cite{personalizedteens, likeithideit, selfrepresentation}.}

We represent the PIA folk theory using the dashed line in Fig. \ref{fig:like_theory_example}. Before a video is liked (negative video indices), according to the folk theory, the distance of the videos with the liked video should stay relatively constant. However, once a video is liked, users believe that the action impacts their recommendations almost immediately, and subsequent future videos are closer to the liked video i.e., the distance between the videos shrinks. PIA folk theory anticipates that the distance with respect to the \textit{future} videos is smaller than the distance with respect to the \textit{past} videos, showing that recommendations after the interaction were closer to the liked video.

\subsubsection{Results}

We find evidence {\it against} the Post Interaction Anticipation (PIA) folk theory. Interacting with a video, we find, does {\it not} influence immediate recommendations. Fig.~\ref{fig:like_theory_example} shows a clear contrast between the dotted PIA line (expected to drop) and the actual behavior (stable, or increasing). The distances  between future videos and the liked video do not shrink, unlike the expectation.

At the same time, we find that the distance of the videos immediately {\it prior} to the like or share interaction is closer (smaller) to the liked video. This is visible in Fig.~\ref{fig:like_theory_example}, and we also see evidence in further detailed analysis of other liked videos in our dataset. 
Fig. \ref{fig:likedist} contains the range of average distance of the set of future videos with the liked video minus the average distance of the past videos to the liked video. The figure shows that majority of the time, videos recommended right before a like event are more similar to the liked video than the videos recommended after a like event. We conduct this analysis with 10, 50, and 100 past and future neighboring videos\footnote{We observe that typically most users view about 300 videos in a day.} and conduct a one-sided t-test to test the statistical significance of our result. On average, past videos were closer to the liked video than the future ones. For 10 neighboring videos, the mean future - past distances is 4.60 (95\% CI [4.55, 4.63], t(83,186) = 255.2, p < 0.001); for 50 neighboring videos, 0.78 (95\% CI [0.76, 0.80], t(83,186) = 84.1, p < 0.001); for 100 neighboring videos, 0.35 (95\% CI [0.34, 0.37], t(83,186) = 48.3, p < 0.001). Positive values of the future - past distances indicate that the past videos are closer to the liked video compared to those seen after the interaction and a one-sided t-test confirms that at all sizes of neighboring videos, the past video distances are significantly smaller than the distances of the future videos.

We found this  observation  also holds for share interactions---see  Fig.~\ref{fig:sharedist}. The average future - past distance for 10 neighboring videos is 5.34 (95\% CI [5.30, 5.38], t(40,156) = 291.8, p < 0.001); for 50 neighboring videos, 1.07 (95\% CI [1.05, 1.08], t(40,156) = 118.9, p < 0.001); for 100 neighboring videos, 0.54 (95\% CI [0.52, 0.55], t(40,156) = 76.5, p < 0.001).
This leads us to two possible hypotheses: (a) user interactions are influenced by consistent exposure to a content type---a sequence of similar videos leads to a user interacting with a video of that type, and/or (b) the TikTok recommendation algorithm views an interaction as a peak of engagement and tends to become more exploratory after these events. This allows the algorithm to increase engagement. We demonstrate in \S\ref{sec:rq3} that users find \textit{fresher} content to be more engaging. 

\subsubsection{HCI Implications}

{Our analysis quantifies the distance within sequence of videos recommended to a user before and after a like or share interaction to find proof against the PIA folk theory. This suggests a mismatch between users’ perceived alignment with the recommendation system and its actual responsiveness: likes and shares influence further recommendations very little, even though past research shows that users treat them as tuning knobs \cite{personalizedteens, likeithideit, selfrepresentation}. We envision researchers working on algorithmic self‑presentation to benefit from our analysis. By quantifying the impact of like and share events on the similarity between pre‑ and post‑interaction recommendation sequences, we provide a concrete metric for validating whether interactions truly shift alignment that does not just rely on human feedback. This opens space for empirically comparing alternative feedback designs (for e.g., session‑level controls) and evaluating how well they support users in constructing a more faithful self‑presentation.}

\subsection{RQ3: Do users engage more with recommendations that align with their historical video consumption patterns?}
\label{sec:rq3}

Recommendation algorithms are often based on the assumption that past (history) is a good indicator of the future. 
This section attempts to unearth this key insight: Do user preferences ever converge? When presented with a session that either aligns with  or deviates from past videos they have been recommended, which one do they prefer? 

\noindent {\bf Key Takeaway:} 

{Our analysis shows that users rate  sessions higher when they have \textbf{\textit{different content compared to their past histories}}, rather than similar content. We find a small but statistically significant positive association between session-level novelty and perceived content quality, suggesting that diversity in recommended content can enhance perceived quality.}

\begin{figure}[h]
    % \centering
    \begin{subfigure}{0.45\textwidth}
        \centering

        \includegraphics[width=\textwidth]{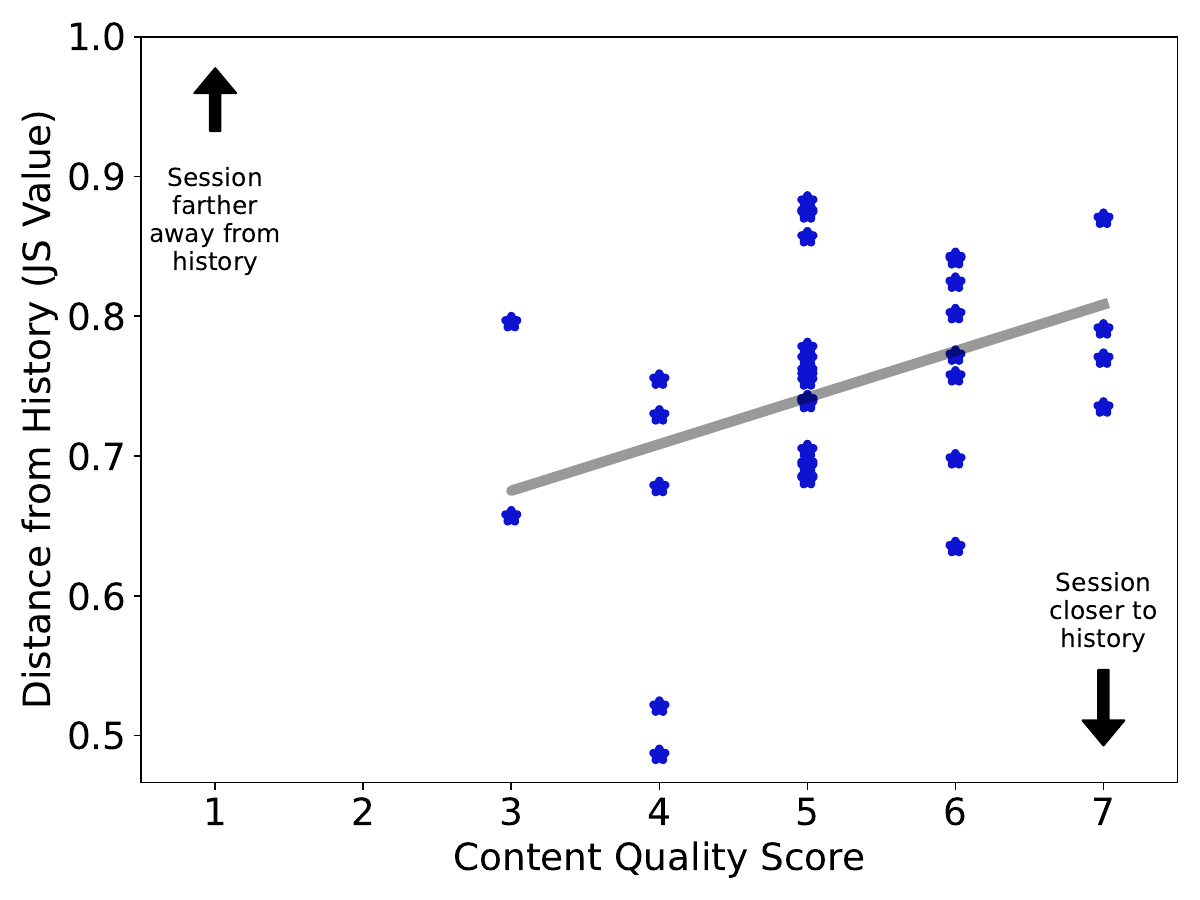}
        
    \end{subfigure}\hfill
    \Description{The X axis represents the content quality scores, ranging from 1 to 7. The Y axis represents the Jensen-Shannon divergence, ranging from 0.5 to 1.0. Each dot on the scatterplot represents one session. The line of best fit is upward sloping.}
    
    \caption{\centering Relationship between the content quality scores a session received and the similarity of the session to historical recommendations. Linear regression analysis estimates the slope to be 0.033, with $p=0.02$, which is less than 0.05 indicating statistical significance.}
    \label{fig:js_session}
\end{figure}

\subsubsection{Method}
Our method of quantifying user engagement is using scores provided by the users in the study described in \S\ref{sec:study}. In our study, each participant appeared for 2 sessions that they rated after: unmodified and another session where a percentage of videos was dropped. For the analysis of this RQ, we only use data from the unmodified session of each participant - this is to avoid any biases resulting from dropping of videos. %One of the two sessions (randomly ordered) dropped a percentage of videos. 
 We compare the content quality score, which was a response to the question \textit{Rate session X based on the content quality of the videos that you saw. Note that we are entirely referring to the content of the videos.} The score was provided on a Likert Scale of 1 to 7 with 1 being Very Dissatisfied and 7 being Very Satisfied. 

To quantify the deviation from historical recommendations in the browsing history vs. the videos seen in both the sessions of the study, we use Jensen-Shannon (JS) divergence  \cite{menendez1997jensen}. JS divergence measures the similarity between the probability distributions of the number of videos in the 100 abstract classes that exist in a user's browsing history and both of their sessions. A smaller JS value means the distribution of videos in the user's history and  the session are similar. 

\subsubsection{Results}
One would expect that the session that is the most similar to the browsing history distribution is more preferred and so, would receive a higher score (smaller JS value aligns with a higher score). However, Fig. \ref{fig:js_session} has an upward sloping trend that goes against this expectation. We find this positive relation between the distance from the history and the content quality score to be statistically significant using linear regression (p = 0.02). {We conclude that users report a better perceived experience when they see content {\it different} from content they have seen in the past.}
% Users like change.

\subsubsection{HCI Implications}
{Our analysis implies that recommending familiar videos to users can disrupt their experience on TikTok. This is in line with existing HCI work \cite{topicdiversification, diversityrec, userpeceptiondiversity}  that argues for diversity in recommendations. It also suggests that "good" personalization is not about replaying a user’s past, but instead, is about creating a mix of familiar and novel content. Moreover, rather than viewing “different from your usual” suggestions solely as a fairness or exploration mechanism, our findings indicate that such recommendations can also enhance users’ overall experience. } 

\subsection{{RQ4: Beyond the videos recommended, does the continuity of the recommended video sequence matter (in terms of user satisfaction)?}}

\label{sec:rq4}

\begin{figure}[h]
    \centering
    \begin{subfigure}{0.45\textwidth}
        \centering
        \includegraphics[width=\textwidth]{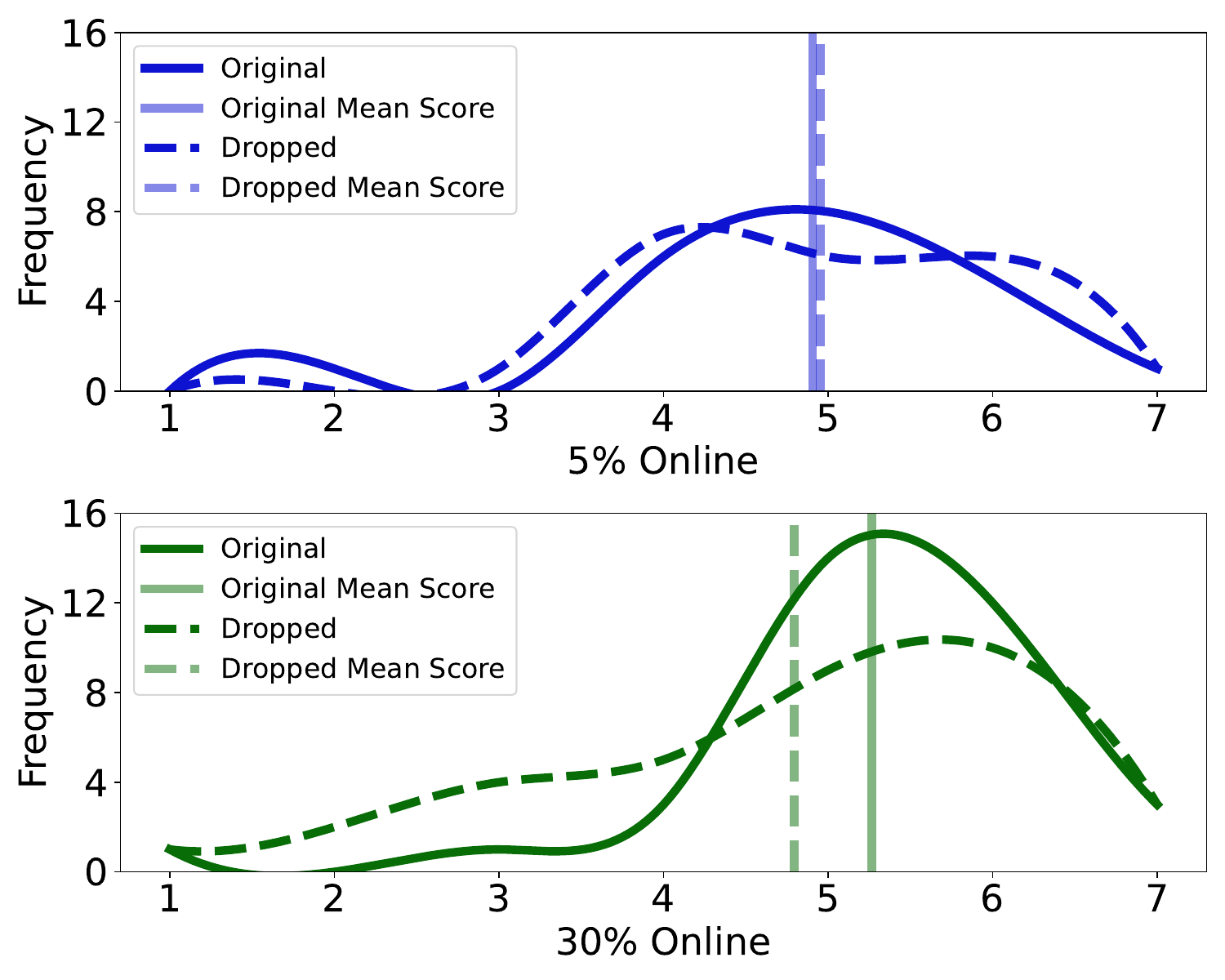}
        
        \label{fig:engagement}
    \end{subfigure}
    
    \Description{In this figure, we compare the engagement scores. For each variation, the X axis represents the engagement scores ranging from 1 to 7 and the Y axis represents the frequency of the scores ranging from 0 to 16. We also have two curves, one representing the original session and one representing the dropped session. For the 5\% dropped session, the original and dropped mean engagement score are at approximately 4.9. For the 30\% dropped session, the mean score for the original session is above 5, whereas for the dropped session it is below 5. }
    \caption{\centering Comparison of engagement scores across original and dropped sessions and the different drop percentages used. A higher number on the Likert scale means a better rating. Top = 5\% Drop Session, Bottom = 30\% Drop Session.}
    \label{fig:scores}
\end{figure}

TikTok recommends not just {\it individual} videos, but also optimizes the {\it sequence} in which the videos are recommended (again, both parts are proprietary). We attempt to understand: how much does altering the {\it sequence} of recommended videos by dropping some from the sequence, affect user experience?

\noindent {\bf Key Takeaway:} 
We find that while (randomly) dropping a small fraction of videos from the recommended sequence ($5\%$) does not affect user experience metrics, but \textbf{\textit{dropping a moderate}} ($30\%$) \textbf{\textit{of the sequence affects user experience}}. This suggests that the recommendation cares about not just the content being recommended but the order of videos (even though consecutive videos may appear to be unrelated).

\subsubsection{Method} 
To remind the reader, in our user study of \S\ref{sec:study}, users watched two TikTok sessions of ten minutes each. One of these sessions was the unmodified FYP (original session). For the other session, two variants of modifications existed (dropped session): 1. In the first kind, 5\% of the videos recommended on the FYP were randomly dropped. 2. The other variant had 30\% of the videos randomly dropped. At the end of these two sessions, participants rated both of the sessions based on their engagement with each of the session. Engagement was rated on a Likert Scale of 1 to 7 with a higher score meaning a favorable experience. %We report the results  in Fig. \ref{fig:scores}. 

We chose dropping videos as our method of disrupting a sequence of videos because it is a realistic scenario. Network issues could lead to buffering of videos that then get skipped. Dropping videos was the most feasible for us to do as it did not involve modifying the recommendation system (which is beyond our control).

The in-person study involved a statistically small number of participants and also lacked  the 5\% drop variant. Therefore, we focus our analysis on the online version conducted via Prolific.

To statistically validate the differences observed in the modified and original sessions, we perform the Wilcoxon signed-rank test \cite{wilcoxontest}. This
non-parametric test does not assume a normal distribution and is appropriate for the comparative ordinal data we collected through our 7-point Likert scale questionnaire. The Wilcoxon test tests whether the median difference between original and modified session scores is zero, and so, effectively determines whether participants exhibited a preference for one over the other. Past works also use the test for similar evaluations \cite{aideation}. 

\subsubsection{Results}
The engagement scores are summarized in Fig.~\ref{fig:scores}. We first discuss the 30\% dropped experiments. We note that users find the modified version with 30\% dropped videos less engaging. The mean score, and the distribution of scores, plotted in the bottom plot in Fig.~\ref{fig:scores} validate this point - a Wilcoxon signed-rank test revealed a statistically significant difference (W=0.0,  p<0.001) between the engagement scores of the original ($\mu=5.26$, IQR=[5,6]) and 30\% dropped sessions ($\mu=4.79$, IQR=[4,6]). This is a surprising finding---users are not aware that one of these sessions has been modified and can browse an infinite sequence of videos. Therefore, we expected users to have an identical experience.

In sessions with just 5\% video drops (top graph in Fig. \ref{fig:scores}), we do not observe the same difference in engagement scores. These sessions compared to the unmodified sessions had statistically insignificant difference (W=2.0, p=0.6) in engagement scores (Original $\mu=4.90$, IQR=[4,6], Dropped $\mu=4.95$, IQR=[4,6]) 

indicating that it is harder for users to perceive very small changes in the recommendations and their sequence.

We hypothesize a possible reason for this phenomenon: TikTok's recommendation algorithm applies temporal logic in placement of videos, e.g., to not place many similar videos next to each other. Such temporal logic can break due to random drops. With a 30\% drop rate, the sequence of videos gets impacted more significantly and so, we note a greater drop in engagement.

\subsubsection{HCI Implications} {

Our analysis identifies a tolerance band within which external changes to the sequence of recommendations remain effectively invisible to users. 
For HCI, this relates directly to how much designers can safely modify or experiment with sequential recommendations (for exploration, fairness, etc.) without harming perceived quality.}
Furthermore, even though consecutive videos in a recommended sequence often appear to be unrelated, our findings suggest that the recommendation algorithm creates a medium-term or long-term ``continuity'' across a sequence. This  warrants  further study by the research community (especially IR/information retrieval), as it opens a new understudied dimension of recommender output.

\subsection{RQ5: To what extent can we predict whether a user will watch a recommended video using our video analysis pipeline?}

\label{sec:rq5}

{Finally, we ask whether the video-content representations produced by VCA offer predictive value for user engagement. Specifically, we examine the extent and limits of predictable structure in short video viewing behavior.}

\noindent {\bf Key Takeaway:} We find that the use of VCA vectors provides additional, but limited, predictability in user watch behavior. The maximum accuracy is 69.65\%, demonstrating that user behavior is hard to predict.

\subsubsection{Method}
Past works have shown that user preference and behaviors in commercial short video streaming platforms are extremely hard to predict due to the complexities of the human decision-making and large number of factors influencing user interactions ~\cite{li2023dashlet, ng2021will, zhan2022deconfounding}. We use content vectors derived from our VCA technique to capture the semantic and contextual nuances of the videos,{ and explore the benefit that rich video‑content representations can provide, if any.}

\begin{table}[h]
\begin{tabular}{|c|c|}
\hline
\textbf{Feature Group} & \textbf{List of Features}                                                                \\ \hline
VCA Vectors (V)        & Embeddings Vectors (PCA)                                                                 \\ \hline
Time (T)               & \begin{tabular}[c]{@{}c@{}}Hour of the Day + Day of the Week + \\ Month\end{tabular}     \\ \hline
User Status (U)        & \begin{tabular}[c]{@{}c@{}}User Name + No. Videos Watched + \\ Time Watched\end{tabular} \\ \hline
\end{tabular}
\vspace{5pt}
\caption{\centering Input features of user prediction include three feature groups.}
\label{table:features}
\end{table}

\begin{table*}[t]
\centering
\begin{tabular}{|c|c|c|c|c|}
\hline
\textbf{Feature Groups}
& \textbf{Accuracy (\%)}
& \textbf{Accuracy Delta (\%)}
& \textbf{F1 Score}
& \textbf{AUC-ROC} \\
\hline
V + T     & 69.65 & +0.04  & 0.703 & 0.771 \\
\hline
V + U     & 69.53 & -0.08  & 0.711 & 0.785 \\
\hline
T + U     & 66.48 & \textbf{-3.13} & 0.681 & 0.757 \\
\hline
V + T + U & 69.61 & 0      & 0.724 & 0.795 \\
\hline
\end{tabular}
\vspace{5pt}
\caption{\centering Classification performance under different feature groups.}
\label{table:ablation}
\end{table*}

We model the problem of whether a user will stay on a video and see or just swipe as a binary classification problem. Our dataset contains 3,870,540 data points and we split the training and testing dataset with a ratio of 8:2. To train our model, we obtain the ground truth label by checking the ratio of the video that the user watched and set the threshold to be 10\%, meaning that if a video is watched by this user for less than 10\%, we classify this video as not interesting to the user. We choose a threshold of 10\% because it yields a balanced classification problem, i.e., 50\% of videos in our dataset were watched over 10\%.

We leverage the random forest algorithm for our task. The input features are shown in Table ~\ref{table:features}. We compared several standard classifiers on a validation set: logistic regression, a linear SVM, neural network, and a random forest. 

All models used the same features as input. The random forest achieved the best performance and, we therefore report these results only. We set the random forest algorithm to have 120 estimators based on validation set accuracy. 

We embed the user names using one-hot encoding and include the number of videos and the time that user has been watching so far during that video scrolling session. {Our choice of features reflects both the constraints of the available data and insights from publicly reported factors that affect short video watch time \cite{reelsrec}.} 

\subsubsection{Results}
The prediction results with different feature group input combinations are shown in Table~\ref{table:ablation}. With all feature groups included, the model achieves the highest accuracy of 69.61\%, along with the best F1 score (0.724) and AUC-ROC (0.795). To evaluate the contribution of each feature group, we systematically removed one feature group at a time and analyzed the impact on performance. We find that removing the VCA vectors results in the most significant decline, with a 3.13\% drop in accuracy, as well as noticeable reductions in F1 score and AUC-ROC, making it the most crucial feature group. In contrast, the User Status and Time features have a less pronounced effect compared to the VCA vectors.

\subsubsection{HCI Implications}

{The modest but consistent gains from VCA vectors suggest that semantic representations of videos contribute to engagement prediction, yet they leave a substantial portion of user behavior unexplained. This points to two implications: first, such content-based representations can provide researchers with a valuable lens into an otherwise opaque recommendation system. Second, content-based explanations (“you are seeing this because it is similar to videos you watched”) of algorithmic experience capture only part of the story and user interactions exhibit randomness beyond these features.}

\section{Discussion and Implications}
\label{sec:discussion}

{Apart from the ``HCI Implications'' embedded into each of the respective RQ subsections above, we now note a few additional implications for users, for algorithm design, for long video in general, and for researchers who may want to extend VCA.}

\paragraph{Implications for Users}  A common fear among online users (including TikTok users) is that interacting with a video will change their set of seen videos and take them into unknown topics or topics that they would like to avoid. But our finding from RQ2 (\S\ref{sec:rq2}) of long-term impact on the algorithm, indicates that such interaction does not affect upcoming recommendations, but instead is a {\it result} of past recommendations. This indicates that users who do not want to interact with TikTok content but feel compelled to do so, may be able to curate their interaction urges by introspecting about recent ``similar'' videos they may have seen. Further, our findings indicate that users need not feel helpless at the mercy of the algorithm---the slow reaction from TikTok algorithm shows that users still have time to ``steer'' the algorithm. Even after a negative interaction with a video, users are not ``stuck'' into topics due to our finding from RQ1 that TikTok's recommended topics to a user are highly dynamic.

\paragraph{Implications for Recommendation Algorithm Design - I}  A common misconception among designers of personalized recommender algorithms is that the goal of the algorithm's learning is to  converge to what the user prefers. This is evident in recommender algorithms for music apps (e.g., Pandora, YouTube Music, etc.), where the set of recommended audios converges (and stops recommending new audios, except in corners of the feed). Our finding that users like change (RQ3, \S\ref{sec:rq3}) indicates that the goal of algorithms must be to ``continuously learn'' as opposed to converge. TikTok's churning recommendations (RQ1, \S\ref{sec:rq1}) is indicative that its algorithms are  cognizant of  this fact. %Designers of other systems (e.g., online music delivery) appear to be lagging in this manner, and may need to re-design their algorithms.

\paragraph{Implications for Recommendation Algorithm Design - II}  While available documentation appears to indicate that TikTok's production algorithm largely relies on hashtags, our findings (especially RQ5 in (\S\ref{sec:rq5})) indicate that deeper video content-based recommendations may be closer to user preferences. Short video systems should likely consider recommender algorithms that account for video content. {While this implies additional computational cost, the cost is limited---the embedding for a video needs to be computed only once at upload time, similar to how a codec already generates multiple resolutions of the uploaded video for different network qualities and screen types.}

\paragraph{Implications for Long Form Video} Designers of long-term multimedia systems (e.g., movie streaming, TV content, YouTube, Netflix, etc.) rarely focus on the sequences of video delivery (which video is watched after which). %While little is known about how TikTok creates the sequence out of a set of recommended vidoes, 
Our finding from RQ4 indicates that  sequence matters to the user. This implies that designers of all video services (not just short videos) including long videos, movie channels on TV, etc., will need to pay attention to the sequence of content delivered. It is common, for instance, for television channel hosts to ``handover'' to the next host, or for ``lead in'' programs (e.g., a live talk show right after a game or  awards ceremony, or a pilot episode of a new series premieres right after a popular established series' episode). More study is needed whether {\it longer} sequences (more than just two back-to-back videos) of such content affect user experience in long form videos. Understanding the implications of this question in both pull-based systems (e.g., YouTube, Netflix) as well as push-based systems (e.g., TV, movie streaming) is an open direction.

\paragraph{Implications for Researchers} VCA can be used as a new measurement instrument to analyze video content beyond short video systems. The recent growth of multimodal Vision Language Models further enriches the VCA pipeline. Our research demonstrates that VCA allows us to reason about video content, rather than just video statistics, at scale. Researchers can use this approach for answering research questions related to recommendation systems in both short and long video domains. More broadly, this research can also be used for new applications such as content moderation, content querying (e.g., to surface similar or different videos), and popularity analysis. 

{\paragraph{VCA as a generalizable HCI Tool} We envision our VCA analysis pipeline to be a reusable HCI tool that can benefit other researchers working on users interacting with multimodal data. VCA allows us to convert a sequence of multimodal data (videos in our case) into human-interpretable clusters. By doing so, VCA converts a large sequence of data into manageable units that can be visualized across time allowing both sequence-level and temporal analysis of multimodal data. This can power dashboards showcasing dominant user interests over time and allow analysis of cluster transitions relation to user engagement. It also allows researchers to study multimodal data at the granularity of a sequence.}

\section{Limitations}
\label{sec:limitations}

In our HCI measurement approach, we use our VCA pipeline in combination with our user study and data donation effort to validate our five research questions. We note the following limitations within our user studies, VCA technique and analysis.

\subsection{User Study}
\begin{itemize}
\item Most of our user study participants were aged 18 to 34. While this age range does significantly dominate on TikTok ~\cite{tiktokdemo}, older users do exist on the app. The skew in age of our population likely arose from two factors: a) recruitment on a university campus, and b) age of the typical survey taker on Prolific. 

\item In our user studies, participants browsed TikTok in a web browser on macOS computers. The default mode for users is the TikTok mobile application. Given our inability to modify the mobile app or programmatically collect statistics, we limit our study to a web browser. Additionally, due to the way our extension was designed, our user study was limited to participants with macOS laptops. {Our use of both the Chrome extension and web browser  to simulate TikTok interactions potentially reduces the ecological realism of a TikTok user's experience and may limit the generalizability of our findings.} Some of our respondents reported in their comments that the form factor (on browser) was unfamiliar and may have affected their overall level of engagement---however we note that such users reported the effect on both their 10-minute browsing sessions. 
\item While we conducted a user study to validate that VCA clusters group together meaningfully similar videos, the evaluation is based on a moderate sample size consisting mostly of students and subjective judgments. A more diverse study may yield different results.

\end{itemize}

\subsection{VCA Technique}
\begin{itemize}

\item A component of our algorithmic analysis relies on clustering of videos into abstract clusters. The unsupervised clustering approach removes any labeling biases. However, the cluster definitions are abstract and data-driven. Future work could  exploit other supervised clustering approaches.

\item 
Our choice of 100 clusters is driven by our manual pre-analysis of meaningful clusters and verification using external volunteers. Our analysis also assigns only one cluster to each video. In principle, soft clustering (i.e., assigning a video to multiple clusters) or a deeper or different manual pre-analysis may yield different clusters, although we do not expect it to change our overall conclusions. 

\item We utilized the pretrained Video-LLaMA model for generating embeddings from videos. A large generative model trained specifically on TikTok videos may yield different, and perhaps more informative, set of embeddings.

\item  We focus on audio and visual embeddings to represent TikTok videos and capture their content. Text transcripts can sometimes be error-prone (noisy environment, unclear voice-overs, music lyrics) and absent in TikTok videos, and their contribution is largely redundant with the audio modality.

\end{itemize}

\subsection{Analysis}
\begin{itemize}

\item Participants in our user study joined TikTok at different times, which may influence their usage patterns or platform engagement and in turn, impact our analysis. To address this limitation, we standardized timelines by measuring activity relative to each participant’s study date.

\item Our analysis is generally conducted at the group level to examine the overall trend. Individual level analysis can investigate heterogeneities in people's TikTok usage behavior.

\end{itemize}

\section{Conclusion}

{Short video platforms are now central to everyday media consumption, yet the mechanisms by which their algorithms shape user experience remain difficult for researchers to observe and evaluate. In this work, we presented a new HCI measurement approach which combines a new tool for automated Video Content Analysis (VCA), along with large-scale data donation, and controlled user studies. This trifecta  offers a multi-faceted account of how a recommendation system interacts with user history, behavior, and perception, and unearth five insightful results about the algorithm-user interaction.} 

Some of our conclusions run contrary to folk theory (e.g., user interaction is more correlated with {\it recently seen} videos, and  less to {\it upcoming}  recommended videos), while other results of ours point to the need to rethink fundamental assumptions of personalization (e.g., we find users like change, and a user's interests never converge), while  yet other findings of ours provide the first quantitative evidence of phenomena (e.g., the sequence of videos recommended matters; users like change). {Across these contributions, we demonstrate how integrating multimodal content representations with user-centered methodologies can surface new insights about algorithm–user interaction. We hope this work opens pathways for future research that uses VCA not only as an analytic lens but also as a basis for developing more transparent, controllable, and user-aligned recommendation experiences.}

\begin{acks}
We thank the anonymous reviewers and Associate Chairs for their valuable feedback. The work in this paper was funded in part by NSF Grant CNS 2504595, and in part by a gift from Microsoft. 
\end{acks}

%%%% conclusion.tex ends here %%%%

\bibliographystyle{ACM-Reference-Format}

%%%% bibliography starts here %%%%

%%% -*-BibTeX-*-
%%% Do NOT edit. File created by BibTeX with style
%%% ACM-Reference-Format-Journals [18-Jan-2012].

%\bibliography{reference}%% Commented by merge tool

%%%% appendix.tex starts here %%%%

\section*{Appendix}
\label{sec:appendix}

\subsection*{Additional User Study Details} 
\label{appendix:userstudy}

% \vspace{-10pt}
\begin{figure}[h]
    \centering
    \includegraphics[width=0.6\linewidth]{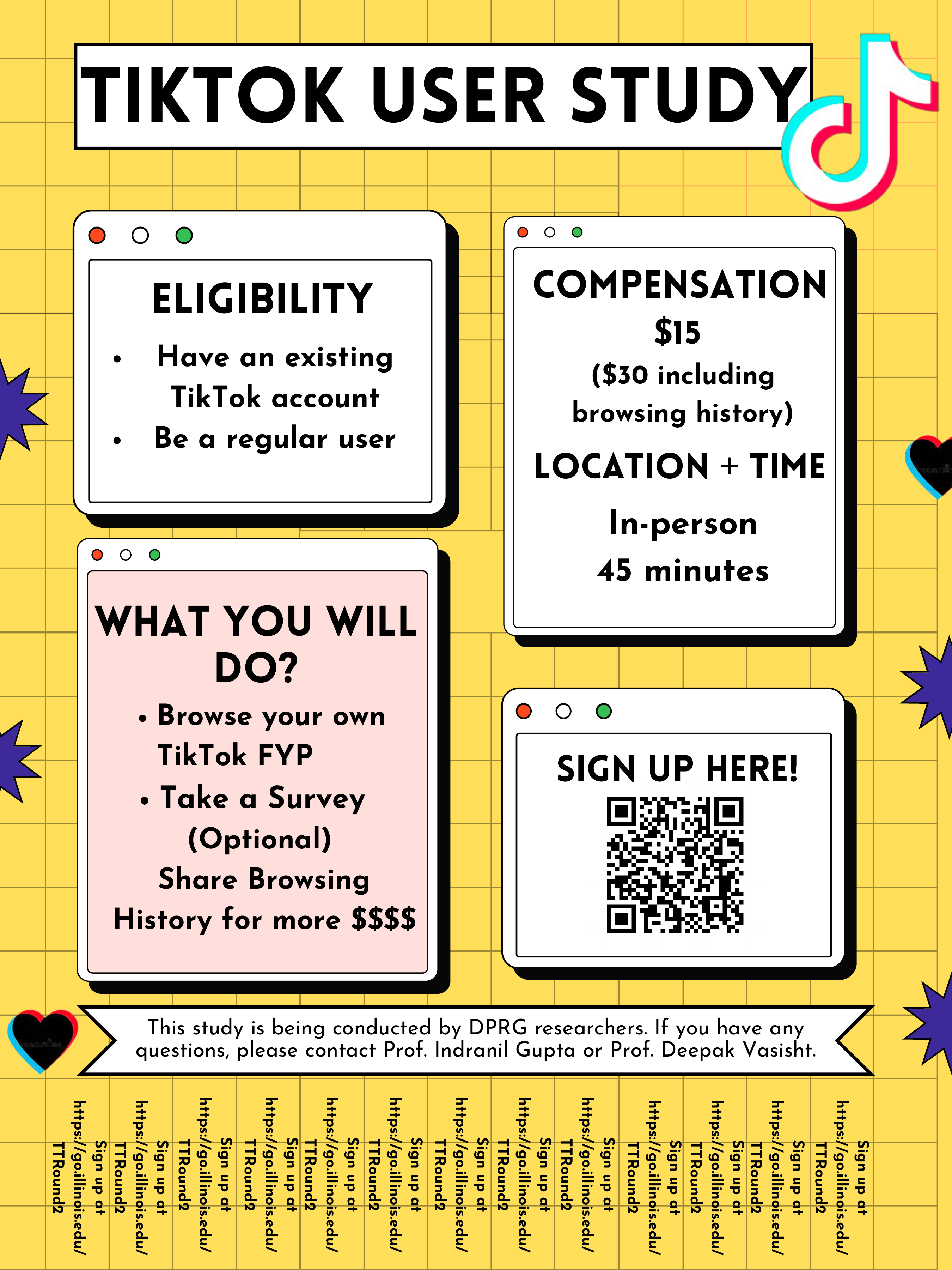}
    \Description{In this figure, we illustrate the poster that we utilized to recruit participants for our in-person user study. The poster contains details such as the eligibility, compensation, location, duration of study, description of activities done by user as well as details of research group conducting the study.}
    \caption{\centering \centering The poster we utilized to recruit participants for our in-person user study. This poster was placed in various buildings, coffee shops and restaurants on our university campus.}
    \label{fig:tiktok_us_poster}
\end{figure}

% \begin{figure}[H]
% \centering

% \begin{subfigure}{\columnwidth}
%   \centering
%   \includegraphics[width=\linewidth]{figures/tiktok_us_part1.png}
%   \Description{Screenshot of the Prolific study page for Part~1, where participants browse TikTok in two sessions. The page shows the study description, compensation, duration, eligibility requirements, and a button to begin the study.}
%   \caption{\centering Part 1: Browsing sessions.}
%   \label{fig:tiktok_us_part1}
% \end{subfigure}

% \vspace{6pt}

% \begin{subfigure}{\columnwidth}
%   \centering
%   \includegraphics[width=\linewidth]{figures/tiktok_us_part2.png}
%   \Description{Screenshot of the Prolific data donation page for Part~2, where participants donate their activity data. The page lists the task description, compensation, duration, requirements, and a button to proceed with data donation.}
%   \caption{\centering Part 2: Activity data donation.}
%   \label{fig:tiktok_us_part2}
% \end{subfigure}
% 
% \Description{Screenshots of the two parts of our online user study conducted through Prolific.}
% \caption{\centering Snapshot of our online user study conducted through Prolific.}
% \label{fig:tiktok_prolific_userstudy}
% \end{figure}

\begin{figure}[t]
  \centering

  \begin{subfigure}[t]{0.45\textwidth}
    \centering
    \includegraphics[width=\linewidth]{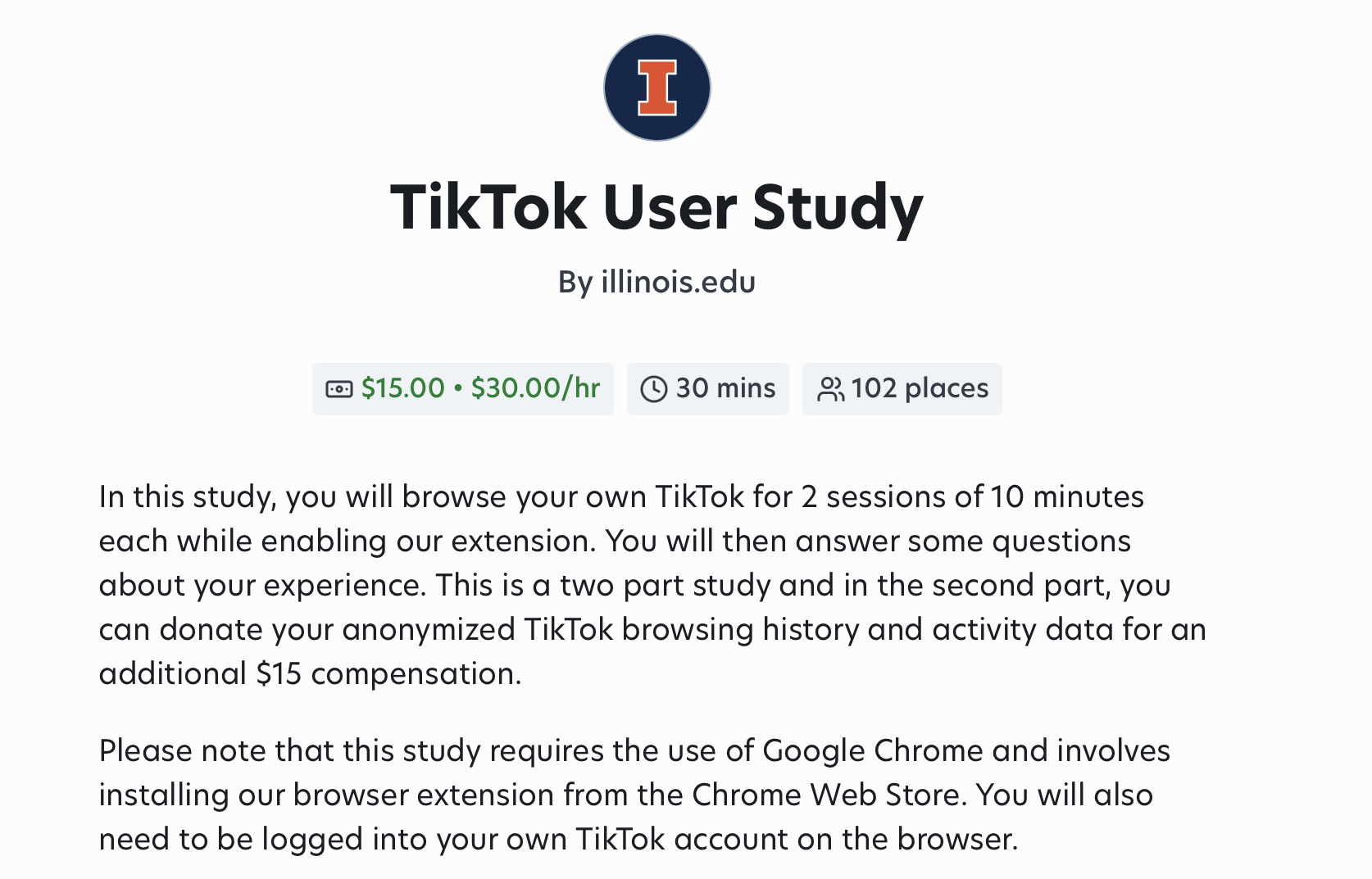}
    \caption{\centering Part 1: Browsing sessions.}
    \label{fig:tiktok_us_part1}
  \end{subfigure}
  \vspace{8pt}
  \begin{subfigure}[t]{0.45\textwidth}
    \centering
    \includegraphics[width=\linewidth]{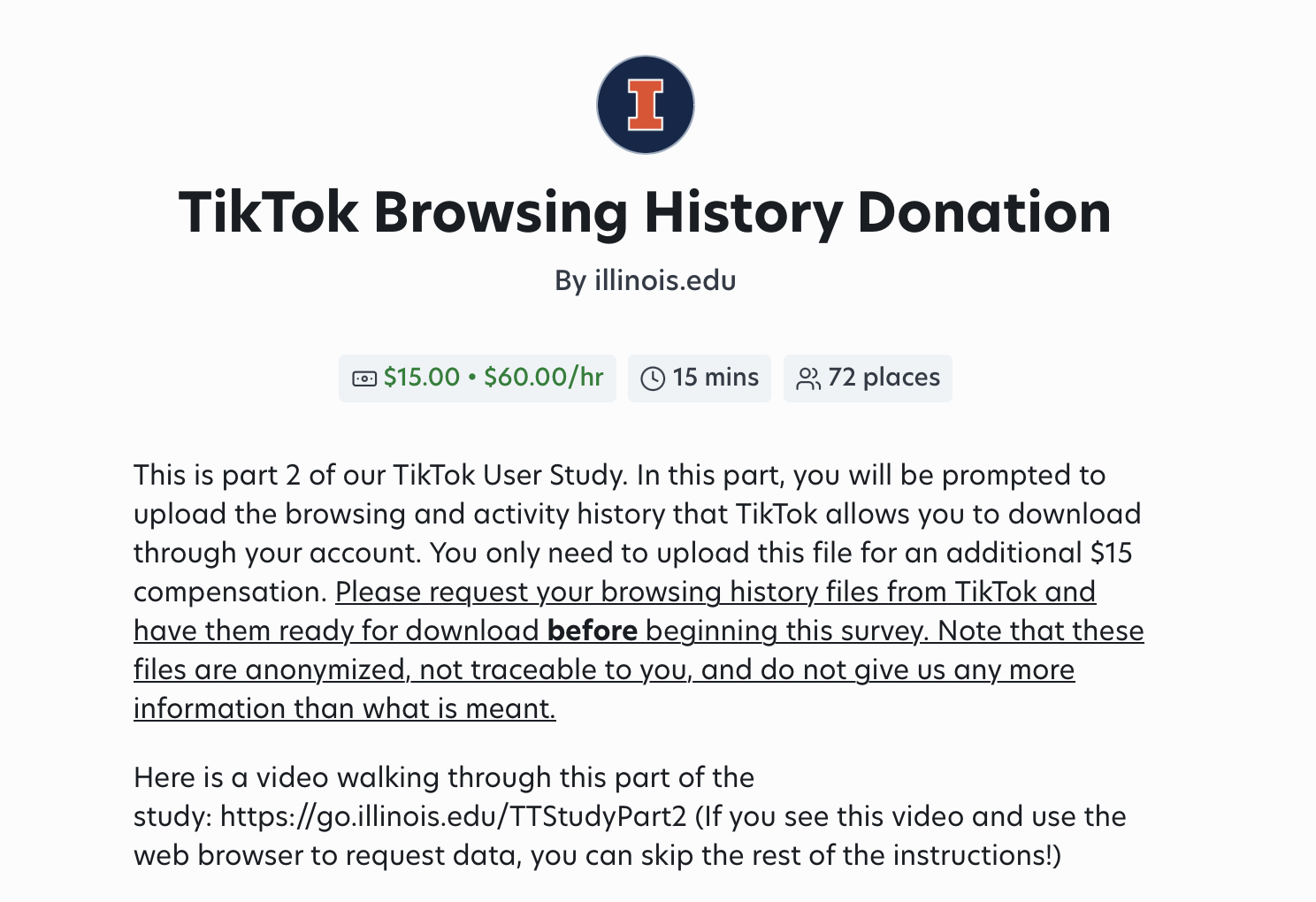}
    \caption{\centering Part 2: Activity data donation.}
    \label{fig:tiktok_us_part2}
  \end{subfigure}

  \Description{Prolific study pages for our two-part online user study: (a) TikTok browsing sessions and (b) optional donation of TikTok browsing/activity history.}
  \caption{\centering Snapshot of our two-part online user study on Prolific.}
  \label{fig:tiktok_prolific_userstudy}
\end{figure}

% \begin{figure}[h]
%     \centering
%     \begin{subfigure}{0.85\textwidth} 
%         \centering
%         \includegraphics[width=\textwidth]{figures/tiktok_us_part1.pdf}
%         \Description{We have the snapshot of our Prolific user study in which participants browse through TikTok in two sessions. Prolific provides a description of the user study, compensation, duration, requirements to take the study, and a button that guides the users to start the user study.}
%         \caption{\centering Part 1}
%         \label{fig:tiktok_us_part1}
%     \end{subfigure}
%     \hfill  % 
%     \begin{subfigure}{0.85\textwidth}  
%         \centering
%         \includegraphics[width=\textwidth]{figures/tiktok_us_part2.pdf}
%         \Description{We have the snapshot of our data collection facilitated by Prolific in which participants donate their activity data. Prolific provides a description of the data donation, compensation, duration, requirements to take the study, and a button that guides the users to donate data.}
%         \caption{\centering Part 2}
%         \label{fig:tiktok_us_part2}
%     \end{subfigure}
%     \Description{In this figure, we provide two subfigures with present snapshorts of our online user study conducted through Prolific.}
%     \caption{\centering Snapshot of our online user study conducted through Prolific.}
%     \label{fig:tiktok_prolific_userstudy}
% \end{figure}

Snapshots of materials from the study are placed in Figs. \ref{fig:tiktok_us_poster} and \ref{fig:tiktok_prolific_userstudy}. In Table~\ref{tab:tiktokstudy}, we list the different questions that we asked our participants at the end of both sessions along with the corresponding Likert scales used to record responses. 

\begin{table*}[t]
\centering
\small
\begin{tabular}{|l|c|}
\hline
\textbf{Question} & \textbf{Scale} \\
\hline

I was more engaged by session \_\_\_\_. &
1 -- Session 1;\; 7 -- Session 2 \\ \hline

In session \_\_\_\_, I saw videos that were closer to the kinds of videos I normally saw in my past TikTok viewing. &
1 -- Session 1;\; 7 -- Session 2 \\ \hline

I was more satisfied by session \_\_\_\_. &
1 -- Session 1;\; 7 -- Session 2 \\ \hline

I was more absorbed in session \_\_\_\_. &
1 -- Session 1;\; 7 -- Session 2 \\ \hline

I felt more frustrated in session \_\_\_\_. &
1 -- Session 1;\; 7 -- Session 2 \\ \hline

I found my experience more rewarding in session \_\_\_\_. &
1 -- Session 1;\; 7 -- Session 2 \\ \hline

I felt more interested in session \_\_\_\_. &
1 -- Session 1;\; 7 -- Session 2 \\ \hline

Rate session 1 based on the content quality of the videos you saw. &
1 -- Very dissatisfied;\; 7 -- Satisfied \\ \hline

Rate session 2 based on the content quality of the videos you saw. &
1 -- Very dissatisfied;\; 7 -- Satisfied \\ \hline

Rate session 1 based on network conditions or technical issues. &
1 -- Very poor;\; 7 -- Excellent \\ \hline

Rate session 2 based on network conditions or technical issues. &
1 -- Very poor;\; 7 -- Excellent \\ \hline

How engaging was session 1 for you? &
1 -- Very boring;\; 7 -- Very engaging \\ \hline

How engaging was session 2 for you? &
1 -- Very boring;\; 7 -- Very engaging \\

\hline
\end{tabular}
\vspace{5pt}
\caption{\centering User study questions and corresponding response scales for Part~1.}

\label{tab:tiktokstudy}
\end{table*}

\section{Cluster Descriptions}
The descriptions of the 100 abstract clusters (when we are reasonably able to infer) that we created via KMeans are listed below:

% \clearpage
\begin{enumerate}
\item Videos of people gathering at events such as sports matches and concerts
\item Videos of influencers interviewing other people on the streets
\item Videos of influencers asking questions to their audience with text displayed on screen
\item Videos of influencers trying new clothes and reviewing outfits
\item Advertisements from companies
\item A single male influencer trying a new product and explaining his experience
\item Videos of people sharing emotional moments with their family
\item Videos of people using appliances
\item People displaying their daily routines
\item Videos of people reacting to news articles and other videos on the internet
\item Videos of people at restaurants serving food
\item Videos of people traveling to forests and beaches
\item Videos of a single female influencer giving advice
\item ``Get Ready With Me'' videos by female influencers
\item Clippings from reality shows and movies
\item Female influencers working out at the gym
\item Clippings from movies with dramatic background music
\item Videos of cats and kittens
\item Videos of people unboxing products and reviewing them, with no person on screen
\item Videos of people at music concerts
\item Videos of people singing songs with lyrics emphasized on screen
\item TV show clips where the video does not occupy the full frame
\item Videos of women talking about their relationships and friendships
\item Clippings featuring popular celebrities
\item Videos of people commenting on celebrities' outfits or posts
\item Videos featuring the sky
\item Videos of male influencers sharing realizations
\item \textit{Unclear}
\item Videos of people building products from scratch
\item Videos of female influencers reacting to comments
\item Videos of friends hanging out together
\item Videos of people making dishes and sharing recipes
\item Videos of pop artists hanging out together
\item ``POV'' videos explaining a situation
\item Videos with dramatic audio and bold white text displayed on screen
\item \textit{Unclear}
\item Videos titled ``Pop culture moments we don't talk enough about''
\item Videos featuring K-pop artists
\item Clippings from K-pop music videos
\item Videos of influencers talking about their favorite food
\item Women performing trends and dances in corridors
\item Videos from basketball matches
\item Videos of female influencers stating preferences with white text displayed on screen
\item Videos of football matches where the video does not occupy the full frame
\item Make-up and hair tutorials by female influencers
\item Videos of one person holding a microphone and interviewing others
\item Screen recordings of YouTube videos or messages
\item Podcast clips
\item Videos of cats and dogs with their pet parents
\item Videos of female influencers giving relationship advice
\item Videos of women looking at the camera as if it were a mirror and fixing their hair
\item Videos of women sitting in cars and talking about topics
\item \textit{Unclear}
\item Videos of male influencers reacting to other websites and news articles
\item Videos of one or more girls performing a popular trend
\item Advertisements from food companies
\item Minecraft gameplay videos
\item \textit{Unclear}
\item Videos of people exploring cafés and restaurants
\item Videos of male influencers talking about a specific topic
\item Videos featuring clothing products
\item Videos featuring babies and toddlers
\item ``POV'' and realization videos featuring movie scenes
\item Romantic videos of couples
\item Make-up and hair tutorials featuring specific products
\item \textit{Unclear}
\item Clippings of cartoons and animated movies
\item Aesthetic cooking videos with no visible chef
\item Videos of female influencers reacting to comments in foreign languages
\item Videos sharing Reddit stories
\item Videos labeled as ``Stitch''
\item Advertisements by technology companies
\item Clippings from talk shows and newsrooms
\item Movie clips occupying only a small region of the frame
\item Videos of people during their vacations
\item Minecraft videos combined with reactions to Reddit posts
\item \textit{Unclear}
\item Videos of people showcasing achievements and important moments
\item Videos of crowds enjoying concerts
\item Videos from TikTok Research
\item Videos of women discussing controversial topics while in their cars
\item Split-screen videos with unrelated content
\item \textit{Unclear}
\item Videos where people enact skits
\item Celebrity interview clips not occupying the full frame
\item Videos in foreign languages
\item Videos of people pranking others
\item Videos of male pop artists performing
\item Cartoon clips with white text displayed over them
\item \textit{Unclear}
\item DIY videos where people organize their homes
\item Videos where influencers question their actions
\item Funny videos of cats and dogs with their pet parents
\item \textit{Unclear}
\item Clippings from very famous music videos
\item ``Day in the life'' videos
\item \textit{Unclear}
\item Videos with vertically split screens showing unrelated content
\item Videos of people reacting to controversial user comments
\item Videos filmed in homes or bedrooms
\end{enumerate}

%%%% appendix.tex ends here %%%%

\end{document}